\documentclass[a4paper,11pt]{article}
\pdfoutput=1

\usepackage{jheppub} 

\usepackage[T1]{fontenc} 

\usepackage{amssymb}\usepackage[normalem]{ulem} 
\usepackage{ulem}

\usepackage{graphicx} 
\usepackage{graphics}
\usepackage{epsfig}
\usepackage{color}
\usepackage{amssymb}
\usepackage{amsmath}
\usepackage{doi}
\usepackage{dsfont}
\usepackage{setspace}
\usepackage{float}
\usepackage{caption}
\usepackage{subcaption}

\usepackage{slashed}
\usepackage{booktabs}
\usepackage{multirow}
\usepackage{soul}

\def\WWW{e^+\nu_e\, \mu^-\bar{\nu}_\mu\, \tau^+\nu_\tau\, b\bar{b}}

\title{Multi-scale improved  predictions for $\boldsymbol{pp \to t\bar{t}W^+ +X}$ }

\author{Nikolaos Dimitrakopoulos}
\author{and Malgorzata Worek}

\affiliation{Institute for Theoretical Particle Physics and Cosmology, RWTH Aachen University, D-52056 Aachen, Germany}

\emailAdd{ndimitrak@physik.rwth-aachen.de}
\emailAdd{worek@physik.rwth-aachen.de}

\abstract{We compare standard fixed-order NLO QCD predictions for the full off-shell $pp \to \WWW \, (j) + X$ processes  in the multi-lepton decay channel with the results obtained using the \textsc{MiNLO} approach. The \textsc{MiNLO} method, now also implemented in the \textsc{Helac-NLO} framework, provides a dynamic determination of the renormalization and factorization scale settings by identifying the most likely branching histories of the additional jets through an inverse $k_T$-clustering algorithm. The inclusion of Sudakov form factors accounts for the large logarithms that arise in the presence of  widely separated scales. We perform a dedicated comparison of the two approaches at both the integrated and differential (fiducial) cross-section levels for the LHC
Run II energy of $\sqrt{s}=13$ TeV. Finally, we present the merging of the full off-shell  predictions for $pp \to \WWW \, + X$, $pp \to \WWW \, j +X$ and $pp \to \WWW \, jj$, with the aim of improving the description of the underlying $pp \to \WWW + X$ process.}

\dedicated{\rm TTK-26-21, P3H-26-057}

\keywords{Higher-Order Perturbative Calculations, Specific QCD Phenomenology, Top Quark}

\begin{document} 

\maketitle
\flushbottom

%
\section{Introduction}
\label{sec:introduction}
%

Perturbative QCD calculations form the theoretical backbone for the high‑precision predictions that are needed to describe hard‑scattering processes at the Large Hadron Collider (LHC). At any fixed order in the perturbative expansion, however, such predictions depend on unphysical parameters, most notably the renormalization scale $\mu_R$ and the factorization scale $\mu_F$. The renormalization scale enters through the scale dependence of the strong coupling, $\alpha_s$, whereas the factorization scale separates the short-distance partonic cross section from the long-distance parton distribution functions (PDFs). Although physical observables are independent of these scales when computed to all orders, a residual scale dependence remains at any finite order in perturbation theory and is commonly used to estimate the size of missing higher-order corrections.  Scale uncertainties in perturbative calculations are conventionally estimated by varying the renormalization and factorization scales independently or simultaneously by a factor of two around a chosen central value. Consequently, the choice of central scale is an important ingredient of any fixed-order prediction. In next-to-leading order (NLO) calculations, this scale is often determined a posteriori by requiring that the NLO corrections should be small or that the scale sensitivity would be minimal. These criteria are motivated by the expectation that inappropriate scale choices can generate large logarithms of ratios of the renormalization or factorization scale to the characteristic scale of the process, leading to sizable higher-order corrections. 

An alternative approach is to determine the relevant scales dynamically from the kinematics and underlying dynamics of the event, rather than relying on an a posteriori stability criterion. This idea was introduced several years ago and is commonly referred to as the CKKW procedure (named after S. Catani, F. Krauss, R. Kuhn, and B. Webber)  \cite{Catani:2001cc,Krauss:2002up}. In this method, the kinematic configuration of a given event is associated with the most probable branching history using an exclusive jet-clustering algorithm. The transverse momentum associated with each branching is then used as the renormalization scale for the corresponding power of $\alpha_s$, at that vertex. In addition, Sudakov form factors are included to account for the large double logarithms arising when the reconstructed branching history contains widely separated scales.

The multi-scale improved NLO (\textsc{MiNLO})  method \cite{Hamilton:2012np} extends this philosophy to NLO calculations. The goal is to construct an NLO computation in which the Born, virtual, and real-emission contributions are evaluated with scales and Sudakov form factors assigned according to the CKKW procedure, while retaining the formal NLO accuracy of the calculation. In this way, the method can yield more reliable predictions for full inclusive quantities, even when they are obtained from calculations that explicitly include additional QCD radiation. It is also particularly valuable for matching NLO calculations to parton showers. A scale-setting prescription that reflects the branching structure of the event and includes the appropriate Sudakov suppression provides a natural bridge between the fixed-order calculation and the parton-shower description.  

Nevertheless, the \textsc{MiNLO} method has also been applied in purely fixed-order computations, i.e. without matching to parton-shower programs, see, e.g. Ref.~\cite{Hoche:2016elu}. In this context, its purpose was to provide an alternative scale-setting prescription and to quantify the impact of Sudakov form factors.  In the current work, we adopt the same strategy and employ the \textsc{MiNLO} approach to obtain NLO QCD predictions for the full off-shell $pp \to t\bar{t}W^+ +X$ process in association with up to two additional jets. We focus on multi-lepton final states arising from the decays of the $t\bar{t}$ pair and the associated $W^+$ boson. This is the first application of the \textsc{MiNLO} method to a final state of such complexity, with full off-shell effects consistently taken into account.

The simultaneous production of a top-quark pair together with a $W^\pm$ gauge boson is a rather unique process to study at the LHC.  This is mainly due to the importance of the $gq$- and $gg$-initiated subprocesses, which first appear at higher orders in QCD. Given the sizable contributions from gluon-induced channels at the LHC, higher-order corrections are essential for an accurate description of this process. Beyond these perturbative aspects, $pp\to t\bar{t}W^\pm$ production is also of considerable phenomenological interest, see e.g. Refs.~\cite{Maltoni:2014zpa,Bevilacqua:2020srb}.  In addition,  the $t\bar t W^\pm$ production process plays an important role in global SMEFT analyses \cite{Brivio:2019ius}, especially in constraining four-quark operators. Its rich variety of final-state signatures also makes it an important background for  Higgs boson measurements and searches for physics beyond the Standard Model. In this respect, the $pp\to t\bar{t}W^\pm$  process in the multi-lepton decay channel constitutes an important background for many new physics searches targeting same-sign lepton signatures \cite{Contino:2008hi,DeSimone:2012fs, Almeida:1997em,vonBuddenbrock:2016rmr,vonBuddenbrock:2017gvy,vonBuddenbrock:2018xar,Buddenbrock:2019tua}. 

On the experimental side, the $pp\to t\bar t W^\pm +X$  process  has been measured  by both the ATLAS \cite{ATLAS:2016wgc,ATLAS:2019fwo,ATLAS:2023xay,ATLAS:2024moy} and CMS collaborations \cite{CMS:2017ugv,CMS:2022tkv,CMS:2025iwa}. The measured $pp\to t\bar{t}W^\pm$ cross sections generally exceed the corresponding Standard Model predictions.  The resulting tension remains moderate, typically at the level of about $2\sigma$, and does not constitute evidence for new physics. In addition to the discrepancy in the overall normalization, tensions have also been observed in the modeling of final-state kinematics. These observations highlight the need for increasingly precise and accurate theoretical predictions to ensure a reliable interpretation of the LHC data.

On the theoretical side, substantial effort has been devoted to improving the precision of predictions for this process. For on-shell $t\bar{t}W^\pm$ production, theoretical predictions are available including NLO QCD and NLO electroweak corrections, as well as subleading LO and NLO contributions and matching to parton-shower programs 
\cite{Hirschi:2011pa,Garzelli:2012bn,Maltoni:2015ena,Frixione:2015zaa,Frederix:2018nkq,Frederix:2020jzp,FebresCordero:2021kcc,Bevilacqua:2021tzp}. To provide a more accurate description of the associated jet activity, merged simulations based on the \textsc{FxFx} method \cite{Frederix:2012ps} have also been developed \cite{vonBuddenbrock:2020ter,Frederix:2021agh}. On the other hand, soft gluon resummation effects with the next-to-next-to-leading logarithmic (NNLL) accuracy have been studied in Refs. \cite{Broggio:2016zgg,Kulesza:2018tqz}. In recent years, significant progress has been made towards calculations of the two-loop virtual corrections for the $pp\to t\bar{t}W^\pm$ process \cite{Becchetti:2025osw,Becchetti:2025qlu}. Owing to the complexity of these calculations, previous next-to-next-to-leading order (NNLO) QCD predictions for $pp\to t\bar{t}W^\pm$ have relied on dynamical approximations for the two-loop contribution \cite{Buonocore:2023ljm}. The NNLO QCD predictions based on a direct computation of the required two-loop amplitudes in the leading-color approximation have only recently become available in the literature \cite{Becchetti:2026awn}. 

Improving the precision and accuracy of theoretical predictions for $pp \to t\bar{t}W^\pm$ requires going beyond the stable-top approximation and incorporating top-quark and $W$-boson decays already at the matrix-element level. Besides calculations in the on-shell approximation \cite{Campbell:2012dh}, full off-shell effects have been studied at NLO in QCD and with combined NLO QCD and electroweak corrections, including double-resonant top-quark contributions as well as single-resonant and non-resonant configurations \cite{Bevilacqua:2020pzy,Denner:2020hgg,Denner:2021hqi}. Such calculations are crucial for achieving an accurate description of differential observables, especially in phase-space regions where off-shell effects are enhanced. 
They have been studied in the multi-lepton decay channel and, more recently, for the $pp \to t\bar{t}W^+ j + X$ process at NLO in QCD \cite{Bi:2023ucp}, where the additional light jet is described with NLO accuracy. As shown in the latter study, jet activity has a significant impact on this process, with the fiducial $pp\to t\bar{t}W^+j+X$ cross section accounting for a substantial fraction of the corresponding $pp\to t\bar{t}W^+ +X$ cross section, namely up to $50\%-55\%$. This sizable contribution highlights the importance of accurately describing additional QCD radiation in the $pp\to t\bar{t}W^+ +X$ process. In particular, the scale choice associated with the extra jet can significantly affect the perturbative behavior of the calculation, especially when compared with the underlying $pp\to t\bar{t}W^+ +X$ process, which does not involve this additional hard jet. This motivates a comparison between the standard fixed-order NLO prediction and the \textsc{MiNLO}  approach, in which the scale of the additional emission is assigned dynamically according to the reconstructed branching kinematics. This comparison allows us to assess whether the standard NLO calculation captures the expected behavior of the process in the presence of hard QCD radiation, or whether the use of a dynamically assigned scale leads to noticeable differences.

The paper is organized as follows. Section \ref{sec:MINLOmethod}  describes the \textsc{MiNLO} method and  its implementation in the \textsc{Helac-NLO}  framework. The computational setup and input parameters are presented in Section \ref{sec:setup}. In Section \ref{sec:integrated_results}, we compare the standard NLO and \textsc{MiNLO} predictions at the integrated (fiducial) cross-section level for the full off-shell $pp \to t\bar{t}W^+$ processes with up to two additional jets. The impact of the \textsc{MiNLO}  method on various differential cross-section distributions is discussed in Section \ref{sec:differential}. Finally, Section  \ref{sec:merging} presents merged predictions designed to improve the overall description of the $pp \to t\bar{t}W^+ +X$ process in the multi-lepton decay channel.

%
\section{Description of the \textsc{MiNLO} method}
\label{sec:MINLOmethod}
%

We begin with a brief overview of the multi-scale improved NLO  method first introduced in Ref. \cite{Hamilton:2012np}, highlighting the differences between the original method and our implementation. The \textsc{MiNLO} method aims to reduce the ambiguity associated with the choice of renormalization  $(\mu_R)$ and factorization $(\mu_F)$ scales in perturbative calculations. This is achieved by assigning these scales according to the most probable branching history of the additionally emitted jets relative to a given set of particles that defines the so-called \textit{core process}. In our study, the core process comprises 
$pp\to \WWW $, which is the $pp \to t\bar{t}W^+$ process in the multi-lepton  decay channel. The scales associated with the clustering of extra jets relative to that system will be extracted from the \textsc{MiNLO} method, however, the definition of the scale setting for the core process, 
the so-called \textit{core scale}, denoted as $q_{core}$, remains arbitrary. This scale choice generally depends on the kinematics of the particles in the final state after all clusterings of the extra jets are performed. In addition to the scale handling, the \textsc{MiNLO} method dresses up the calculation with Sudakov form factors. The inclusion of these factors improves the accuracy of the computation by resumming large logarithms arising from events with widely separated scales. 
\begin{figure}[t!]
        \centering        
        \includegraphics[width=1.0\linewidth]{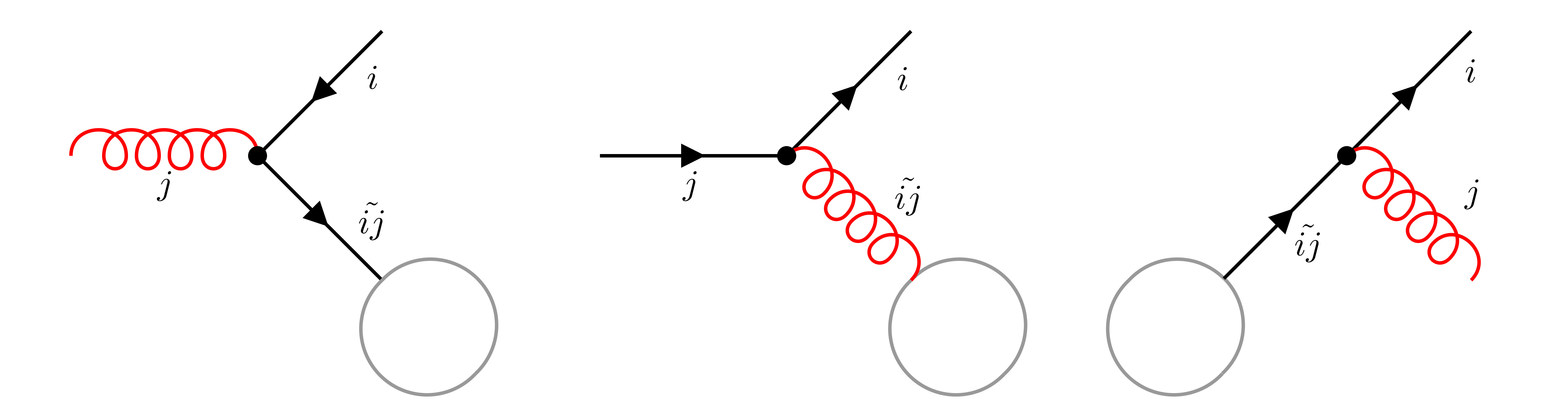}
        \caption{\textit{Examples of $i+j \rightarrow \tilde{ij}$ branchings. The $\tilde{ij}$ represents the pseudoparton created after clustering partons $i$ and $j$. The gray blobs represent the rest of the process, capturing also the core interaction. The first two clusterings involve an incoming parton where $\tilde{p}_{ij} = p_j - p_i$, while the third one comprises the case where two final-state partons are clustered together with $\tilde{p}_{ij} = p_j + p_i$. Graphs are generated with the \textsc{FeynGame} program \cite{Harlander:2020cyh,Harlander:2024qbn,Bundgen:2025utt}}.}
        \label{fig:minlo_branchings}
\end{figure}

First, we focus on the LO case, denoted as \textsc{MiLO}, which is essentially the same as the CKKW procedure introduced in Refs. \cite{Catani:2001cc,Krauss:2002up}.  To reconstruct the branching history, \textsc{MiLO} applies an inverse $k_T$-clustering algorithm \cite{Ellis:1993tq,Catani:1993hr} defined according to 
\begin{equation}
\label{yij}
d_{ij} = 
 \dfrac{\min(p_{T,\, i}^2\,, \,p_{T,\, j}^2)\, \Delta R_{ij}^2}{R^2} \,,\quad \quad \quad \quad \quad \quad \quad \quad 
d_{iB} = p_{T,\,i}^2  \,, 
\end{equation}
where $\Delta R_{ij}^2=(y_i-y_j)^2+(\phi_i-\phi_j)^2$, while $R$, $p_{T, \,i}$, $y_i$ and $\phi_i$ are the jet-resolution parameter,  transverse momentum, rapidity and azimuthal angle of parton $i$, respectively. We use all final‑state QCD jets to compute all distances $d_{ij}$ and beam distances $d_{iB}$ in order to determine the minimum distance $d_{min} = \min\left\{d_{ij},d_{iB}\right\}$.  If a pair distance $d_{ij}$ is the minimum one, we cluster $i+j \to \tilde{ij}$. On the other hand, if the beam distance $d_{iB}$ is the smallest one, we cluster parton $i$ into the incoming leg. At each stage of clustering, only those splittings that are compatible with flavor conservation are permissible. Consequently, we consider the following branchings $gg \to g$, $g\bar{q}\to \bar{q}$, $gq \to q$ and $q\bar{q}\to g$, where $q = u,d,c, s,b$. Some illustrative examples are depicted in Figure \ref{fig:minlo_branchings}. For each clustering step $k$, where $k=1,\dots,N$ and $N$ is the number of extra jets, a \textit{nodal scale}, denoted as $q_k$, is assigned, corresponding to the square root of the minimal distance at that step, i.e. $q_k =\left[\sqrt{d_{min}} \,\right]_k$. For the final‑state clustering, we use the relative $k_T$ of the pair (and not simply the $p_T$ of one parton)
\begin{equation}
q_k=\left[\sqrt{d_{ij}}\, \right]_k=\left[\min(p_{T,\, i}\,, \,p_{T,\, j}) \frac{\Delta R_{ij}}{R}\right]_k\,, 
\end{equation}
whereas for the initial‑state clustering the transverse momentum of the emitted parton is used instead 
\begin{equation}
q_k=\left[\sqrt{d_{iB}} \, \right]_k = \left[p_{T,\,i}\right]_k\,.
\end{equation}
In the formulas above, the $[...]_k$ notation denotes that the quantities inside the brackets are computed for the $k$-th clustering. From Eq. \eqref{yij} it follows that an ambiguity arises whenever an incoming parton is involved in the clustering algorithm. In such a case, the $k_T$ measure is given by the transverse momentum of the final-state parton, regardless of the flavor of the incoming one. Thus, when a final-state parton can be clustered with either of the two initial-state partons, neither of the branchings is kinematically favored. To resolve this ambiguity, we require that clustering with incoming partons is allowed only if
\begin{align} \label{rapidity_condition}
    \begin{split}
    y_f > 0 \quad & \textrm{and} \quad p_{i,z} > 0\,,   \\
    &\; \textrm{or} \\
    y_f < 0 \quad & \textrm{and} \quad p_{i,z} < 0 \,,
    \end{split}\;
\end{align}
where $y_f$ denotes the rapidity of the final-state parton and $p_{i,z}$ is the $z$-component of the momentum of the incoming parton in the centre-of-mass frame of the incoming beams, which are along the $z$-axis. This indicates that a Lorentz boost to the center-of-mass frame of the initial-state partons is necessary, as both partons may acquire momenta along the $x$ and $y$ axes after absorbing the kinematics of the final-state partons.  Furthermore, the condition specified in Eq. \eqref{rapidity_condition} must be satisfied even when only a single clustering involving the initial-state parton is possible. This implies, for instance, that in the case of large-angle emission in the direction opposite to the incoming beam, the \textsc{MiLO} method is not applied and no nodal scale is determined. The \textsc{MiLO} procedure is repeated until the core process is reached, i.e.
when all extra jets are clustered away, and $q_{core}$ can be evaluated based on the momenta of the available particles. Because the $k_T$-clustering algorithm always clusters the smallest distances first, the nodal scales are ordered as follows 
\begin{equation} \label{orderd_qi}
    q_1 < ... < q_n \leq q_{core}\,,
\end{equation}
where $n \leq N$. A few comments are in order here. Firstly, whenever the nodal scale $q_{i+1}$ turns out to be smaller than the previous scale $q_i$, the clustering is terminated, and all remaining partons are considered part of the core system. The same applies to the last nodal scale $q_n$, which must be smaller than $q_{core}$. If this is not the case, we set $q_{core} = q_n$. In both cases, the number of extracted nodal scales is smaller than the number of additional jets, which necessitates the condition $n < N$. Conversely, in situations where no unordered branching occurs, the number of nodal scales is exactly equal to $N$. Secondly, the above ordering of the nodal scales is imposed because only ordered emissions can be consistently resummed with the Sudakov form factors. Thus, in the clustering sequence, the first merged pair corresponds to the softest or most collinear emission, whereas the last merged pair is the hardest one. Finally, in the described procedure, $R$ is an input parameter, which we set to $R=0.4$. The effect of varying the value of $R$ is also going to be investigated and identified as a source of systematic uncertainty.

Based on the description above, we can construct the so-called \textit{skeleton} of the event, consisting of \textit{external} and \textit{internal lines}. The internal lines connect two successive branchings $q_i$ and $q_{i+1}$ in the branching history, while the external lines, which do not branch further, are associated with a single nodal scale $q_i$. Once the skeleton of the event has been constructed, the event is dressed with Sudakov form factors. For all lines (internal and external) we use the same type of the Sudakov form factors but with different lower limits. Following the prescription outlined in Ref. \cite{Hoche:2016elu,Anger:2017nkq} we employ
\begin{equation} \label{sudakov_int}
\Delta^{\rm ext}_{f_j}(q_{res}^2;q_j^2) = \dfrac{\Delta_{f_j} (q_{res}^2,q_j^2)}{\Delta_{f_j} (q_{res}^2,q_1^2)}, \quad \quad \quad 
    \Delta^{\rm int}_{f_{ij}}(q_{res}^2;q_i^2,q_j^2) = \dfrac{\Delta_{f_{ij}} (q_{res}^2,q_j^2)}{\Delta_{f_{ij}} (q_{res}^2,q_i^2)},  
\end{equation}
where the \textit{resolution scale}, denoted as $q_{res}$, is defined according to
\begin{equation}
q_{res} = 
\begin{cases} \label{qres}
q_1\,,  \quad \quad &\textrm{for the final-state line} \,,\\
\xi_F q_1\,, \quad \quad&\textrm{for the initial-state line} \,.
\end{cases}
\end{equation}
In Eq. \eqref{qres}, the parameter $\xi_F$ denotes the factorization-scale factor and $q_1$ is the smallest extracted nodal scale, i.e. the scale below which all radiation is considered unresolved. Therefore, for the final-state lines, the probabilities of no emission between the corresponding nodal scale and $q_1$ are taken into account. For  the incoming lines, however, the quantity $q_{res}$ is modified to account for the dependence on the factorization scale. We always set $\mu_F = \xi_F q_1$ in order to match the evolution in the PDFs. As a result, QCD radiation up to the scale $\mu_F$ is resummed within the PDFs, while above that value, it is accounted for via Sudakov form factors. To examine how changing the resolution scale may affect the logarithmic resummation, we also consider the case where $q_{res}$ is fixed at $q_{res} = q_1$, that is, the Sudakov form factors remain unchanged by scale variations. This approach, employed in the original formulation of the \textsc{MiNLO} method  \cite{Hamilton:2012np},  will henceforth be referred to as \textsc{Mi(N)LO}$_{\rm no\xi_F}$.

The form of the Sudakov form factors used in our implementation is consistent with that presented in  Ref. \cite{Hoche:2016elu,Anger:2017nkq}
\begin{equation} \label{sudakov_form}
    \Delta_f (t_0,t_1) = \exp\left\{ -\int_{t_0}^{t_1} \dfrac{dt}{t}\dfrac{\alpha_s(t)}{2\pi} \sum_{b=q,g}  \int_{0}^{1-\sqrt{t/t_1}} dz \left( zP_{bf}(z) + \delta_{bf} \dfrac{2C_f}{1-z}\Lambda_2(t)\right)\right\}\,,
\end{equation}
where $C_f = C_F, C_A$ for quarks and gluons, respectively. In addition, $P_{bf}(z)$ are the $f\to bc$ Dokshitzer-Gribov-Lipatov-Altarelli-Parisi (DGLAP) splitting functions \cite{Altarelli:1977zs, Gribov:1972ri, Dokshitzer:1977sg}, and $\Lambda_2(t)$ is the two-loop correction to the splitting functions \cite{Catani:1990rr} given by
\begin{equation}
    \Lambda_2 (t) = \dfrac{\alpha_s(t)}{2\pi}\left[\left(\dfrac{67}{18} - \dfrac{\pi^2}{6}\right)C_A - \dfrac{10}{9}T_R N_f(t)\right] \,.
\end{equation}
The number of active flavors $N_f(t)$ is determined dynamically while integrating the exponent in Eq. \eqref{sudakov_form} according to
\begin{equation}
    N_f(t) = 
    \begin{cases} 
    5\,,  \quad \quad t < m_t^2 \,, \\
    6\,, \quad \quad t \geq m_t^2 \,,
    \end{cases}
\end{equation}
where $m_t$ is the top-quark mass. By construction, the above definition of the Sudakov form factor does not exceed unity, allowing it to be interpreted as the probability of no emission between two fixed scales $t_0$ and $t_1$. Furthermore, it accounts for the resummation of soft and collinear effects to next-to-leading-log (NLL) accuracy. In our case, the Sudakov form factors are calculated numerically for each event using the Gauss-Legendre quadrature method. The precise formulas used for numerical integration can be found in Ref. \cite{Moretti:2016jnv}. The final form of the \textsc{MiLO} weight included in the matrix element is given by the formula
\begin{equation} \label{miloweight}
    [\alpha_s(q_{core})]^m \times \prod_{i=1}^n \alpha_s(q_i) \times \prod_{i}\Delta_{f_i}^{\rm ext}(q_{res}^2;q_i^2) \times \prod_{ij} \Delta_{f_{ij}}^{\rm int} (q_{res}^2;q_i^2,q_j^2)\,,
\end{equation}
where $m$ denotes the powers of $\alpha_s$ associated with the core process, while the $n$ nodal scales serve as arguments for the remaining powers of the strong coupling.

When extending the \textsc{MiLO} method to the NLO level, referred to as \textsc{MiNLO}, certain modifications are necessary. In particular, special care must be taken to avoid double-counting of NLO terms. The inclusion of Sudakov form factors in the Born-level contribution already accounts for higher-order effects, which can be interpreted as no-branching probabilities. At NLO, however, the exact virtual corrections are included explicitly, providing an equivalent description of no-emission probabilities. On top of that, the real-emission term already contains contributions where the extra emission is unresolved. Therefore, an appropriate subtraction procedure is necessary to prevent overlap between the resummed and fixed-order contributions. This is done by expanding the Sudakov form factors in terms of  $\alpha_s$ and then subtracting the contribution proportional to $\alpha_s$. Thus, the Born-level term $\mathcal{B}$ is replaced by 
\begin{equation} \label{subtracted_delta1}
    \mathcal{B} \rightarrow \mathcal{B} \times \left( 1 - \sum_i \left[\Delta_{f_i}^{(1)}(q_{res}^2,q_i^2) - \Delta_{f_i}^{(1)}(q_{res}^2,q_1^2)\right] - \sum_{ij} \left[\Delta_{f_{ij}}^{(1)}(q_{res}^2,q_i^2) - \Delta_{f_{ij}}^{(1)}(q_{res}^2,q_j^2)\right]\right),
\end{equation}
where the $\Delta^{(1)}$ terms correspond to the NLO part of the Sudakov expansion. Regarding the scale settings and Sudakov form factors, the Born-level and one-loop contributions, as well as all infrared subtraction terms, are handled exactly the same as in the LO case. In the case of real-emission contributions, the additional QCD emission leads to the extraction of an additional nodal scale, denoted as $q_0$. The requirement of ordered clusterings is then extended to
\begin{equation} \label{orderd_qi_nlo}
    q_0 < q_1 < ... < q_n \leq q_{core} \,.
\end{equation}
Again, the scale $q_{core}$ is used as the argument for all powers of $\alpha_s$ in the core process. Instead, if $q_0 > q_1$ the algorithm is terminated, the clustering is performed, and the real-radiation kinematics are absorbed into the Born-level system. In this case, the $q_{core}$ scale is calculated based on the kinematics corresponding to this Born level. We emphasize here that the $q_0$ scale is used neither as the argument of the strong coupling nor in the evaluation of the Sudakov form factors. In other words, the softest branching at scale $q_0$ is considered unresolved and is simply excluded from the \textsc{MiNLO} procedure. This step aims to ensure the proper cancellation of infrared poles between the real-emission contributions and the virtual terms. 

For the additional power of the strong coupling appearing in the NLO calculation and in the Born-level substitution given by Eq. \eqref{subtracted_delta1}, we use the original \textsc{MiNLO} prescription provided by 
\begin{equation}
    \alpha_s^{(n+m+1)} = \dfrac{1}{n+m}\left(\sum_{i=1}^n\alpha_s(q_i) + m\alpha_s(q_{core}) \right),
\end{equation}
where $m$ is the power of $\alpha_s$ associated with the core process and $n$ is the number of nodal scales. For the renormalization scale appearing in the virtual contributions, we adopt the following geometric mean
\begin{equation} \label{geometricmean}
    \mu_R = \left((q_{core})^m \times \prod_{i=1}^n q_i\right)^{\frac{1}{m+n}} \,.
\end{equation}
The above choice guarantees that, even in the presence of multiple scales, a change in the renormalization scale will be a higher-order effect. Finally, the scale variation is performed by rescaling all relevant scales except those appearing in the Sudakov form factors, according to
\begin{equation} \label{var1}
    q_i \to \xi_R \,q_i \quad\quad \textrm{and} \quad\quad q_{core} \to \xi_R \,q_{core} \,,
\end{equation}
where $\xi_R$ is the renormalization scale factor. As previously mentioned, the factorization scale is defined as $\mu_F = \xi_F \,q_1$.

We have also implemented an alternative prescription for $\alpha_s^{(n+m+1)}$ and $\mu_R$ \cite{Hoche:2016elu, Anger:2017nkq}. In particular,  the renormalization scale $\mu_R$ is determined by considering the following effective $\alpha_s$ coupling
\begin{equation} \label{var2}
    \alpha_s(\mu_{R}) = \left( [\alpha_s(q_{core})]^m \times \prod_{i=1}^n \alpha_s(q_i) \right)^{\frac{1}{m+n}}.
\end{equation}
The additional power of $\alpha_s$ is given by $\alpha_s^{(n+m+1)} = \alpha_s(\mu_R)$, while the scale variation is accounted for in the standard way by $\mu_R \to \xi_R\, \mu_R$.  The difference between these two approaches is treated as an additional source of systematic uncertainty.

The \textsc{Mi(N)LO} method described above has been implemented in the \textsc{Helac-NLO} framework \cite{Bevilacqua:2011xh}, which includes \textsc{Helac-1Loop} \cite{Ossola:2006us,Ossola:2007ax,vanHameren:2009dr,Draggiotis:2009yb,vanHameren:2010cp} and \textsc{Helac-Dipoles} \cite{Czakon:2009ss,Bevilacqua:2013iha}. In practice, we use the in-house program \textsc{HEPlot} \cite{Bevilacqua:HEPlot} to process the obtained (N)LO results. Drawing on concepts presented in  Ref. \cite{Bern:2013zja}, we store events,  along with their corresponding matrix elements and PDF information, in the modified \textsc{Les Houches} event files \cite{Alwall:2006yp} and ROOT \textsc{Ntuples} \cite{Antcheva:2009zz}. This enables us to obtain results for various scale settings and PDF choices simply by reweighting the original results.  This approach not only offers significant advantages when different observables and more exclusive cuts are needed, but it is also a natural framework for implementing the \textsc{MiNLO} method in \textsc{Helac-NLO}. 

To verify the correctness of our implementation, we performed several cross-checks using the Monte Carlo program \textsc{Sherpa} \cite{Sherpa:2019gpd,Sherpa:2024mfk,Gleisberg:2003xi,Gleisberg:2008ta}, where \textsc{MiNLO} is implemented and can also be employed for fixed-order calculations.  Specifically, we validated the \textsc{MiNLO} approach for the processes $pp \to t\bar{t}$, $pp \to t\bar{t}j$, $pp \to t\bar{t}jj$, and $pp \to t\bar{t}jjj$ with stable top quarks, achieving in each case perfect agreement between the results.

%
\section{Computational setup}
\label{sec:setup}
%

We consider full off-shell predictions for  $pp \to \WWW + X$ and $pp \to \WWW \,j + X$ at LO and NLO. In the latter case, we calculate $\alpha_s$ corrections to the Born-level processes at perturbative orders $\mathcal{O}(\alpha_s^2 \alpha^6)$ and $\mathcal{O}(\alpha_s^3 \alpha^6)$, respectively. In addition, we provide full off-shell results for the $pp \to \WWW \, j j$ process at $\mathcal{O}(\alpha_s^4 \alpha^6)$. For brevity, we will sometimes refer to these processes as  $pp \to t\bar{t}W^+ +X$, $pp \to t\bar{t}W^+ \, j+X$, and $pp \to t\bar{t}W^+ \, jj$, respectively. We present standard (N)LO predictions alongside the \textsc{Mi(N)LO} results.

The computational setup is described in detail in Ref. \cite{Bi:2023ucp}. For the sake of completeness, however, we briefly summarize the key input parameters, PDF sets, renormalization and factorization scale settings as well as fiducial phase-space cuts employed  in the calculations. Specifically, for the (N)LO calculations, we use the (N)LO NNPDF3.1 PDF set \cite{NNPDF:2017mvq} provided by the LHAPDF library \cite{Buckley:2014ana} and adopt the following SM input parameters

\begin{equation}
\label{eq:input_parameters}
\begin{aligned}
    &G_\mu = 1.166378 \times 10^{-5}\; \rm GeV^{-2}\,, \quad \quad & m_t = 172.5 \; \rm GeV \,, \\[0.2cm]
    &m_W = 80.379 \; \rm GeV\,, &\Gamma_W^{\rm NLO} = 2.0972 \; \rm GeV \,, \\[0.2cm]
    &m_Z = 91.1876 \; \rm GeV\,, &\Gamma_Z^{\rm NLO} = 2.5074 \; \rm GeV \,, \\[0.2cm]
    & \Gamma_{t,\rm \, off-shell}^{\rm LO} = 1.45766\, \rm GeV, &\Gamma_{t,\rm off-shell}^{\rm NLO} = 1.33254 \; \rm GeV\,. 
\end{aligned}
\end{equation}
All final-state partons with pseudorapidity $|\eta|<5$ are recombined into jets via the infrared-safe  anti-$k_T$ jet algorithm  \cite{Cacciari:2008gp} with a separation parameter of $R= 0.4$.  The presence of at least two $b$-jets and exactly three charged leptons is required. Depending on the process, a specific number of light jets is also required. Final-state jets and charged leptons must satisfy the following phase-space selection criteria (fiducial cuts)
\begin{equation}
\label{eq:cuts}
\begin{aligned}
    &p_{T,\,\ell} > 25~\mathrm{GeV}, \quad\quad \quad&& p_{T,\,b} > 25~\mathrm{GeV}\,, \quad \quad\quad&& p_{T,\,j} > 25~\mathrm{GeV}\,, \\[0.2cm]
    &|y_\ell| < 2.5\,, \quad\quad\quad && |y_b| < 2.5\,, \quad \quad\quad&& |y_j| < 2.5\,, \\[0.2cm]
    &\Delta R_{\ell\ell} > 0.4\,, \quad\quad\quad && \Delta R_{bb} > 0.4\,, \quad \quad\quad&& \Delta R_{b\ell} > 0.4\,, \\[0.2cm]
    &\Delta R_{j\ell} > 0.4\,, \quad\quad \quad&& \Delta R_{bj} > 0.4\,, \quad\quad\quad && \Delta R_{jj} > 0.4 \,,
\end{aligned}
\end{equation}
where $b$ and $j$ stand for the flavored and light jet, respectively, while $\ell=e, \mu,\tau$.

To investigate the impact of scale dependence on NLO QCD results and \textsc{MiNLO} predictions, it is useful to present results for various scale settings. To this end, we first consider the commonly used fixed-scale setting
\begin{equation}
    \mu_0 = m_t + \dfrac{m_W}{2}\,.
\end{equation}
For the dynamic scale setting, we focus on the scalar sum of the transverse momenta of all decay products, $H_T$, defined according to
\begin{equation} \label{HTscale}
    H_T = p_{T,\,e^+} + p_{T,\,\mu^-} + p_{T,\,\tau^+} + p_T^{miss} + p_{T,\,b_1} + p_{T,\,b_2} + \sum_{j}p_{T,\,j}\,,
\end{equation}
where the $p_T^{miss}$ is the total missing transverse momentum from the three escaping neutrinos and the sum $\sum_j$  involves zero, one and two light jets for the full off-shell $pp \to t\bar{t}W^+$, $pp \to t\bar{t}W^+j$ and $pp \to t\bar{t}W^+jj$ processes, respectively. Furthermore, we consider the two specific cases 
\begin{equation}
    \mu_0 =  \dfrac{H_T}{2} \quad \quad \quad \quad \textrm{and} \quad\quad  \quad \quad \mu_0 = \dfrac{H_T}{3}\,.
\end{equation}
This scale choice is blind to the fact that in the $pp\to t\bar{t}W^+(jj)$ process top-quark and $W$ gauge boson resonances might appear. In the next step, however, information regarding the resonant nature of the process is taken into account. To this end, we employ the scalar sum of the transverse energies of all final-state particles, $E_T$, defined according to 
\begin{equation}
    \mu_0 =  \dfrac{E_T}{2} \quad \quad \quad \quad \textrm{and} \quad\quad  \quad \quad \mu_0 = \dfrac{E_T}{3}\,,
\end{equation}
where $E_T$ is provided by 
\begin{equation}\label{ETscale}
    E_T = \sqrt{m_t^2 + p^2_{T,\,t}} + \sqrt{m_t^2 + p^2_{T,\,\bar{t}}} + \sqrt{m_W^2 + p^2_{T,\,W}} + \sum_{j}p_{T,\,j} \,.
\end{equation}
The top-quark and $W$-gauge boson momenta are reconstructed from their decay products by finding the decay history that minimizes the ${\cal Q}$ value defined by the formula 
\begin{equation}
    {\cal Q} = | M_t - m_t | + | M_{\bar{t}} - m_t | + | M_W - m_W | \,,
\end{equation}
where $M_t$, $M_{\bar{t}}$, and $M_W$ are the reconstructed invariant masses of the top quark, the antitop quark, and the $W$ gauge boson, respectively. A detailed list of all possible resonance histories that must be taken into account can be found in Refs. \cite{Bevilacqua:2020srb,Bi:2023ucp}.

We are reiterating a few details here. Firstly, even though the above dynamical scale settings are used for the (N)LO calculations, for the \textsc{Mi(N)LO} results, the core scale  $q_{core}$ is not directly given by the above formulas. After the inverse $k_T$-clustering algorithm is terminated, the kinematics of the light jets are absorbed into the core system. As a result, the transverse momenta of the light jets are not included in the definitions of Eq. \eqref{HTscale} and \eqref{ETscale}. The only exception is unordered clusterings, where the light jets can be treated as part of the core system and as elements of the $q_{core}$ scale definition. Secondly, the lepton kinematics do not differ between the (N)LO and \textsc{Mi(N)LO} results, as leptons do not participate in the clustering algorithm. Finally, the kinematics of light and flavored jets generally differ,  because in the \textsc{MiNLO} method, pseudo-jets with combined momenta are formed after each successful clustering step.

For fixed-order (N)LO calculations, the scale uncertainties for the integrated and differential fiducial cross-section results are determined using the standard 7-point scale variation method. We identify the maximum and minimum cross-section values by considering the following combinations of $(\xi_R , \xi_F )$
\begin{equation} \label{7point}
    (\xi_R, \xi_F) = \{(1,1),(2,1),(1,2),(1,0.5),(0.5,1),(2,2),(0.5,0.5)\} \,.
\end{equation}
For the \textsc{Mi(N)LO} results, the renormalization scale is varied according to Eq. \eqref{geometricmean} and Eq. \eqref{var1}, whereas for the factorization scale, $\mu_F = \xi_F q_1$ is adopted. 

%
\section{Integrated fiducial cross-section results}
\label{sec:integrated_results}
%

%
\begin{table}[!t]
\centering
\scalebox{1.0}{
\begin{tabular}{cccccccc}
\midrule\midrule
$\mu_0$  & $\sigma^{\textrm{LO}}$ [ab] &  $\sigma^{\textsc{MiLO}}$ [ab] & $\sigma^{\textrm{NLO}}$ [ab]  &  $\sigma^{\textsc{MiNLO}}$ [ab] &  $\dfrac{\sigma^{\textrm{NLO}}}{\sigma^{\textrm{LO}}}$ &  $\dfrac{\sigma^{\textsc{MiNLO}}}{\sigma^{\textsc{MiLO}}}$ & $\dfrac{\sigma^{\textsc{MiNLO}}}{\sigma^{\rm NLO}}$\\
\midrule\midrule
  &  &  &  $pp \to t\bar{t}W^+$ &  &  & \\
\midrule\midrule
 $m_t + m_W/2$ & $202.4^{+24\%}_{-18\%}$ &  $202.4^{+24\%}_{-18\%}$ & $252.5^{+5\%}_{-7\%}$  &  $252.5^{+5\%}_{-7\%}$ & $1.25$ & $1.25$ & $1.00$\\
\midrule
 $E_T/2$ & $177.0^{+22\%}_{-17\%}$ &  $177.0^{+22\%}_{-17\%}$ & $240.4^{+6\%}_{-8\%}$  &  $236.3^{+6\%}_{-7\%}$ & $1.36$ & $1.34$ & $0.98$\\
\midrule
 $E_T/3$ & $198.8^{+23\%}_{-18\%}$ &  $198.8^{+23\%}_{-18\%}$ & $249.9^{+5\%}_{-7\%}$  &  $244.6^{+3\%}_{-6\%}$ & $1.26$ & $1.23$ & $0.98$\\
\midrule
 $H_T/2$ & $192.1^{+23\%}_{-18\%}$ &  $192.1^{+23\%}_{-18\%}$ & $246.6^{+5\%}_{-7\%}$  &  $240.4^{+4\%}_{-6\%}$ & $1.28$ & $1.25$ & $0.97$\\
\midrule
 $H_T/3$ & $216.4^{+24\%}_{-18\%}$ &  $216.4^{+24\%}_{-18\%}$ & $254.8^{+3\%}_{-6\%}$  &  $246.7^{+3\%}_{-5\%}$ & $1.18$ & $1.14$ & $0.97$\\

\midrule\midrule
  &  &  &  $pp \to t\bar{t}W^+ j$ &  &  & \\
\midrule\midrule
 $m_t + m_W/2$ & $141.1^{+41\%}_{-27\%}$ &  $133.3^{+34\%}_{-23\%}$ & $144.7^{\,\,\,+0\%}_{-14\%}$  &  $140.9^{\,\,\,+2\%}_{-11\%}$ & $1.03$ & $1.06$ & $0.97$\\
\midrule
 $E_T/2$ & $104.0^{+37\%}_{-25\%}$ &  $\,\,\, 94.7^{+32\%}_{-22\%}$ & $140.1^{\,\,\,+4\%}_{-10\%}$  &  $136.5^{\,\,\,+6\%}_{-10\%}$ & $1.35$ & $1.44$ & $0.97$\\
\midrule
 $E_T/3$ & $124.5^{+39\%}_{-26\%}$ &  $121.4^{+34\%}_{-23\%}$ & $144.9^{+1\%}_{-7\%}$  &  $141.7^{+1\%}_{-6\%}$ & $1.16$ & $1.17$ & $0.98$\\
\midrule
 $H_T/2$ & $115.8^{+38\%}_{-26\%}$ &  $112.2^{+33\%}_{-23\%}$ & $142.6^{+2\%}_{-8\%}$  &  $139.5^{+2\%}_{-7\%}$ & $1.23$ & $1.24$ & $0.98$\\
\midrule
 $H_T/3$ & $139.3^{+40\%}_{-27\%}$ &  $141.3^{+35\%}_{-24\%}$ & $144.3^{\,\,\,+1\%}_{-13\%}$  &  $138.4^{\,\,\,+4\%}_{-18\%}$ & $1.04$ & $0.98$ & $0.96$\\
\midrule
\end{tabular}}
\caption{\textit{Integrated fiducial cross-section results for the full off-shell $pp \to \WWW + X$ and $pp \to \WWW \, j + X$ processes at the LHC with $\sqrt{s} = 13 \; \rm TeV$. The fixed-order (N)LO  and  \textsc{Mi(N)LO} predictions are presented for several scale choices. Also provided  are  the corresponding scale uncertainties,  $\mathcal{K}$-factors defined as \textsc{(Mi)NLO}/\textsc{(Mi)LO} and  \textsc{MiNLO/NLO} ratios. }}
\label{tab:ttwj} 
\end{table}
Below, we compare the fixed-order (N)LO  predictions with \textsc{Mi(N)LO} results for the full off-shell $pp\to t\bar{t}W^+ (jj)$ process in the multi-lepton decay channel for the LHC Run II energy of  $\sqrt{s} = 13$  TeV. Specifically,  Table \ref{tab:ttwj} presents the integrated (fiducial) cross sections for $pp\to t\bar{t}W^+ +X$ and $pp\to t\bar{t}W^+j +X$. For the fixed-order predictions, we set $\mu_0=\mu_R=\mu_F$, whereas for the \textsc{Mi(N)LO} results, the parameter $\mu_0$ corresponds to the scale $q_{core}$.

It follows from the construction itself that the LO and \textsc{MiLO} predictions for the  $pp \to t\bar{t}W^+$ process are identical, since the \textsc{MiLO} procedure is not applied at all. The difference between the NLO and \textsc{MiNLO} predictions is at most $3\%$. The scale dependence is very similar in both approaches. The theoretical error ranges from $5\%$ to $8\%$, depending on the scale choice. For the $pp \to t\bar{t}W^+ j+X$ process, the comparison of the two methods reveals a moderate reduction in the theoretical uncertainties, amounting to up to $7$ percentage points at LO, i.e. from $41\%$ to $34\%$, and $3$ percentage points at NLO, from $14\%$ to $11\%$. The only exception is observed for the $\mu_0 = H_T/3$ scale choice, where the \textsc{MiNLO} uncertainties exceed those obtained in the standard NLO QCD calculation. Focusing on the $\mathcal{K}$-factor, defined as the  $\textsc{(Mi)NLO}/\textsc{(Mi)LO}$ ratio, we observe a wide range of values $(0.98-1.44)$ and $(1.14-1.34)$ for $pp\to t\bar{t}W^+j$ and $pp\to t\bar{t}W^+$, respectively, in the case of \textsc{MiNLO}. For the fixed-order predictions we obtain  $(1.03-1.35)$ and $(1.18-1.36)$ instead. These differences stem from the spread of the $\textsc{(Mi)LO}$ results, as the $\textsc{(Mi)NLO}$ predictions are very close to one another, differing by at most $4\%$, a fact confirmed by the last column of Table \ref{tab:ttwj}.  Overall, for all scale choices considered, the integrated cross-section results at (N)LO and \textsc{Mi(N)LO} predictions show excellent agreement within their respective theoretical uncertainties.
\begin{table}[!t]
\centering
\renewcommand{\arraystretch}{2.5}   
\scalebox{1.0}{
\begin{tabular}{c|cccccc}
\hline\hline
$\mu_0$  & $\sigma^{\textsc{MiNLO}}$ [ab] &  $\sigma^{\textsc{MiNLO}}_{R=1.0}$ [ab] & 
$\sigma^{\textsc{MiNLO}}_{\rm var_2}$ [ab] &  $\sigma^{\textsc{MiNLO}}_{\rm no\xi_F}$ [ab] & 
$\dfrac{\sigma^{\textsc{MiNLO}}_{R=1.0}}{\sigma^{\textsc{MiNLO}}}$ & 
$\dfrac{\sigma^{\textsc{MiNLO}}_{\rm var_2}}{\sigma^{\textsc{MiNLO}}}$  \\
\hline\hline
$E_T/2$ & $136.5^{\,\,\,+6\%}_{-10\%}$ &  $135.3^{\,\,\,+7\%}_{-11\%}$ & $135.0^{+6\%}_{-9\%}$ &  
$136.5^{\,\,\,+7\%}_{-12\%}$ & 0.99 & 0.99 \\
\hline
$H_T/2$ & $139.5^{+2\%}_{-7\%}$ &  $137.9^{+3\%}_{-8\%}$ & $138.1^{+2\%}_{-7\%}$ &  
$139.5^{+3\%}_{-9\%}$ & 0.99 & 0.99 \\
\hline
\end{tabular}}
\caption{\textit{Integrated fiducial cross-section results for the full off-shell  $pp \to \WWW \, j + X$ process at the LHC with $\sqrt{s} = 13$ TeV. The \textsc{MiNLO} predictions are presented for two dynamical scale choices with the default setup. Furthermore, the case with the parameter $R=1.0$ and the case where the scale variation is performed according to Eq. \eqref{var2} (denoted as $\rm var_2$) are shown. A variant without scale variation in the Sudakov form factors, denoted as $\rm no\xi_F$, is also included. Corresponding scale uncertainties and ratios relative to the results obtained in the default configuration are also provided.}}
\label{tab:different_prescriptions}
\end{table}

Next, we analyze the impact of using alternative variants when generating predictions via the \textsc{MiNLO} method. Specifically, we adopt a different value for the resolution parameter, $R = 1.0$, in Eq. \eqref{yij} and use Eq.  \eqref{var2} to calculate $\mu_R$,  which serves as the argument for the additional power of the strong coupling, $\alpha_s^{(n+m+1)}$. We refer to these approaches as $\sigma^\textsc{MiNLO}_{R=1.0}$ and $\sigma^\textsc{MiNLO}_{\rm var_2}$, respectively. Furthermore, we present results for the case where scale variations in the Sudakov form factors are not taken into account, and the resolution scale is determined by the lowest nodal scale $q_1$ rather than by Eq. \eqref{qres}. This approach, proposed in the original formulation of the \textsc{MiNLO} method, is referred to as $\sigma^{\textsc{MiNLO}}_{\rm no\xi_F}$. Differences between these variants and the default setting are treated as an additional source of systematic uncertainty for the $\textsc{MiNLO}$ method. Ultimately, we present results only for the $pp \to t\bar{t}W^+j+X$ process, where the application of the $\textsc{MiNLO}$ method is better justified due to the presence of additional radiation already at the LO level. Results for $\mu_0 = E_T/2$ and $\mu_0 = H_T/2$ are summarized in Table \ref{tab:different_prescriptions}. It should be noted, however, that similar conclusions have also been drawn for other scale choices listed in Table \ref{tab:ttwj}.

All alternative prescriptions differ by only $1\%$ from the default $\textsc{MiNLO}$ configuration. The magnitude of the scale uncertainties is the same for the $\rm var_2$ case and our default setup. For the $\textsc{MiNLO}_{\rm no\xi_F}$ case, a slight increase in the size of the scale uncertainties, up to 1 to 2 percentage points, can be observed. We also verified that these differences increase significantly when the parameters $\xi_R$ and $\xi_F$ are set to very large ($\xi_R=\xi_F=8$) or very small ($\xi_R=\xi_F=0.125$) values. Indeed, in such cases, the default \textsc{MiNLO} prescription yields smaller theoretical uncertainties. To summarize the section on additional systematic uncertainties, the results obtained with our chosen scale settings, where $\xi_R, \xi_F \in \{0.5, 1, 2\}$, indicate that, although the $\textsc{MiNLO}$ procedure allows for several possible implementations, all variants lead to mutually consistent predictions for the $pp\to t\bar{t}W^+(j)+X$ process.
\begin{figure}[t!]
        \centering        
        \includegraphics[width=0.49\linewidth]{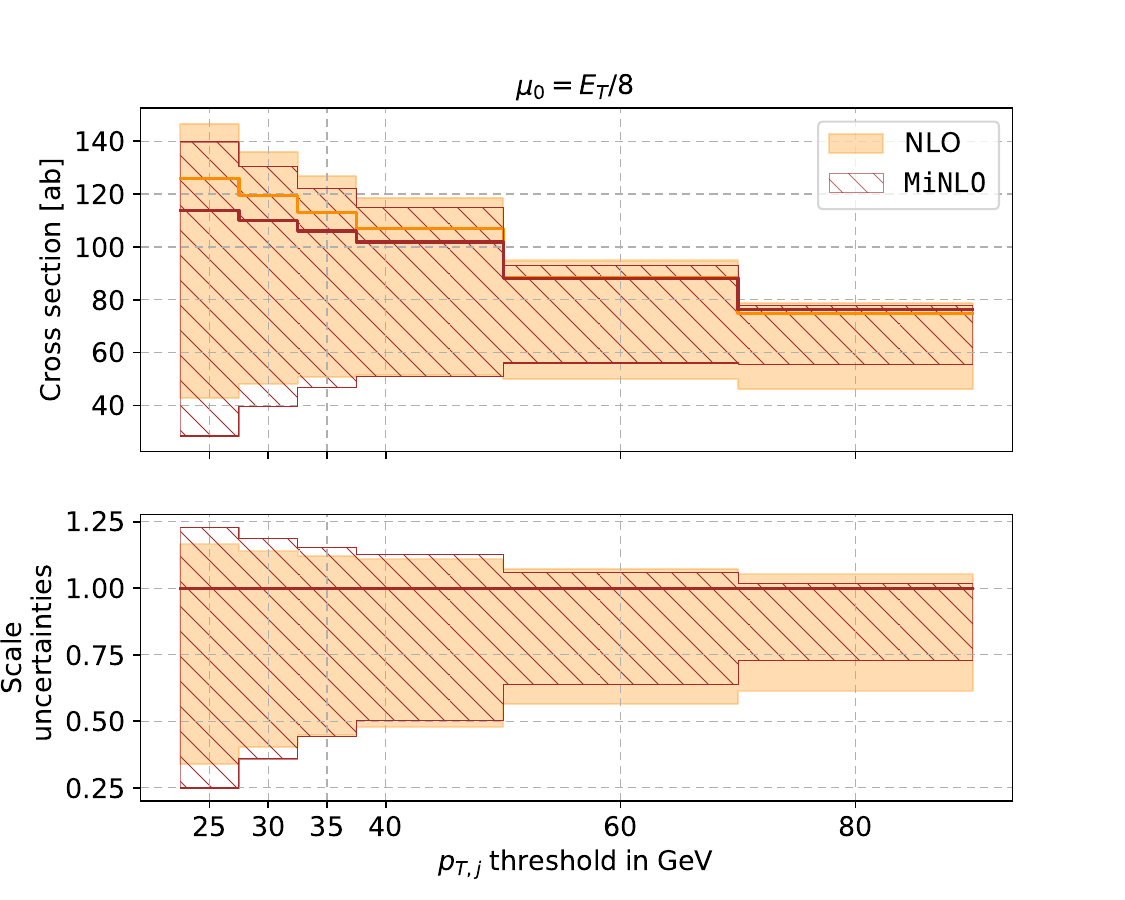}
        \includegraphics[width=0.49\linewidth]{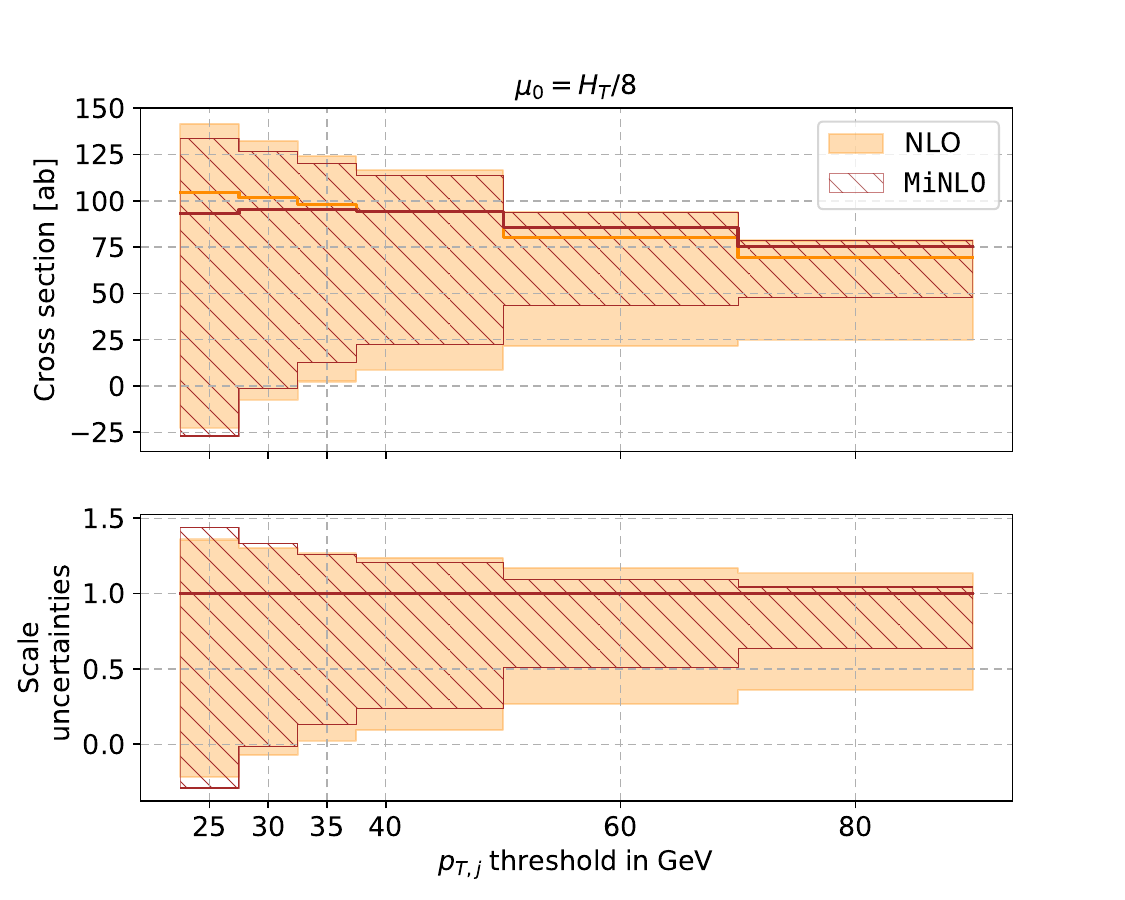}
        \caption{\textit{Integrated fiducial cross-section results for the full off-shell $pp \to \WWW\, j + X$ process at the LHC with $\sqrt{s} = 13$ TeV. The NLO and \textsc{MiNLO} predictions are shown for the two scale settings $\mu_0 = E_T/8$ (left) and $\mu_0 = H_T/8$ (right). We display results for different values of the $p_{T,\, j} > p_{T,\, j}^{min}$ cut, where $p_{T,\, j}^{min}\in \left\{ 25,30,35,40,60,80\right\}$ GeV. Corresponding scale uncertainties are also shown.}}
         \label{fig:ptj_thresholds}
\end{figure}

So far, we have focused exclusively on the scale choices already employed in our previous studies of the $pp\to t\bar{t}W(j)+X$ process   \cite{Bevilacqua:2020pzy,Bevilacqua:2020srb,Bevilacqua:2021tzp,Bi:2023ucp}. It would also be worthwhile to investigate less optimal or even unfavorable scale settings for $q_{core}$ to see how well the \textsc{MiNLO} method performs in such cases compared to fixed-order NLO QCD calculations.  One might assume that the result obtained using the $\textsc{MiNLO}$ method will always be more reliable, since a poorly chosen $q_{core}$  is replaced by the nodal scale $q_i$, determined dynamically based on the process kinematics, thereby improving the overall accuracy of the calculations. Indeed, when $q_{core}$ is actually too small, the ordering condition specified in Eq. \eqref{orderd_qi_nlo} should ensure that the largest nodal scale is used in the calculations instead, which in practice eliminates the problematic core scale from the analysis. To this end, in Figure \ref{fig:ptj_thresholds} we present NLO and \textsc{MiNLO} predictions for two scale choices: $\mu_0 = E_T/8$ and $\mu_0 = H_T/8$, for various values of $p_{T,\, j}^{min}\in\left\{25,30,35,40,60,80\right\}$ GeV, where $p_{T,\,j} > p_{T,\,j}^{min}$. The upper panels show the absolute NLO QCD and \textsc{MiNLO} predictions, while the lower panels display the magnitude of scale uncertainties, normalized to the corresponding predictions for the default configuration. Such scale settings are obviously a poor choice, as they entail large NLO  uncertainties. For example, for $p_{T,\,j} > 25$ GeV, these uncertainties are of the order of $66\%$ and $122\%$ for $\mu_0=E_T/8$ and $\mu_0=H_T/8$, respectively. However, we observe that the corresponding \textsc{MiNLO} uncertainties are also large, amounting to $75\%$ (for $\mu_0=E_T/8$) and $129\%$ (for $\mu_0=H_T/8$). It is only above a value of approximately $p_{T,\,j}^{min} = 60$ GeV that the situation changes significantly. As the threshold for $p_{T,\,j}$ is raised, the differences in the magnitude of scale uncertainties between the fixed-order NLO predictions and the \textsc{MiNLO} approach become more pronounced, favoring the latter. For example, for $p_{T,\,j} > 80$ GeV, the \textsc{MiNLO} uncertainties are of the order of $30\%-35\%$, compared to NLO QCD uncertainties, which fall within the $40\%-60\%$ range.
This behavior can be easily explained by analyzing the distributions of differential cross-sections for $H_T$ and $E_T$ and noting that they peak at around $H_T \approx 500$ GeV and $E_T \approx 600$ GeV, respectively. Thus, for $p_{T,\,j}^{min} \gtrsim 60$ GeV, the probability of unordered clusterings with $q_i > q_{core}$ increases significantly, because the nodal scale associated with the additional jet is determined by the kinematics of the light and flavored jets. On the other hand, for $p_{T,\,j}^{min} \lesssim 60$ GeV, we encounter (well-)ordered clusterings in the majority of cases.
\begin{table}[!t]
\centering
\scalebox{1.0}{
\begin{tabular}{ccccc}
\midrule\midrule
$p_{T,j} \geq \; [\rm GeV]$  & $\sigma^{\textrm{NLO}}_{\mu_0 =E_T/2}$ [ab] &  $\sigma^{\textsc{MiNLO}}_{\mu_0 =E_T/2}$ [ab] & $\sigma^{\textrm{NLO}}_{\mu_0 = m_t + m_W/2}$ [ab] &  $\sigma^{\textsc{MiNLO}}_{\mu_0 =m_t + m_W/2}$ [ab] \\
\midrule\midrule
 $25$ & $140.1^{\,\,\,+4\%}_{-10\%}$ &  $136.5^{\,\,\,+6\%}_{-10\%}$ & $144.7^{\,\,\,+0\%}_{-14\%}$ &  $140.9^{\,\,\,+2\%}_{-11\%}$\\
\midrule
 $30$ & $128.4^{\,\,\,+5\%}_{-11\%}$ &  $126.2^{+6\%}_{-9\%}$ & $133.7^{\,\,\,+0\%}_{-13\%}$ &  $130.3^{\,\,\,+2\%}_{-11\%}$ \\
\midrule
 $35$ & $118.6^{\,\,\,+6\%}_{-11\%}$ &  $117.3^{+6\%}_{-9\%}$ & $124.3^{\,\,\,+1\%}_{-12\%}$ &  $121.3^{\,\,\,+2\%}_{-11\%}$\\
\midrule
 $40$ & $110.2^{\,\,\,+6\%}_{-11\%}$ &  $109.5^{\,\,\,+7\%}_{-10\%}$ & $116.1^{\,\,\,+1\%}_{-11\%}$ &  $113.4^{\,\,\,+2\%}_{-10\%}$ \\
\midrule
 $60$ & $86.3^{\,\,\,+8\%}_{-12\%}$ &  $86.7^{\,\,\,+7\%}_{-10\%}$ & $92.4^{\,\,\,+1\%}_{-10\%}$ &  $90.9^{+2\%}_{-8\%}$\\
\midrule
 $80$ & $70.7^{+10\%}_{-13\%}$ &  $71.5^{\,\,\,+8\%}_{-11\%}$ & $76.7^{+1\%}_{-9\%}$ &  $75.8^{+2\%}_{-7\%}$\\
\midrule
\end{tabular}}
\caption{\textit{Integrated fiducial cross-section predictions for the full off-shell $pp \to \WWW \,j + X$ process at the LHC with $\sqrt{s} = 13 \; \rm TeV$, obtained for two scale choices $\mu_0 = E_T/2$ and $\mu_0 = m_t + m_W/2$. The results are presented for standard NLO QCD predictions and the \textsc{MiNLO} procedure. Various transverse momentum thresholds for the light jet are considered, ranging from $25$ GeV to $80$ GeV. Corresponding scale uncertainties are also shown.}}
\label{tab:ptj_thresholds} 
\end{table}

Similar results, but for well-behaved scales are also presented in Table \ref{tab:ptj_thresholds}. We present results only for the dynamic scale $\mu_0 = E_T/2$ and the fixed scale $\mu_0 = m_t + m_W/2$, however, similar conclusions can be drawn for other dynamic scales. In the case of the dynamic scale, scale uncertainties increase with the $p_{T,j}$ threshold, however, the uncertainties for the \textsc{MiNLO} method remain consistently smaller than those for standard NLO calculations. The opposite trend is observed for the fixed scale, where theoretical uncertainties decrease at higher $p_{T,j}$ thresholds. Nevertheless, even in this case, \textsc{MiNLO} predictions exhibit reduced scale dependence compared to standard NLO results across all the $p_{T,j}$ thresholds considered. We also note that the differences in scale uncertainties between \textsc{MiNLO} and standard NLO predictions are less pronounced than those seen in Figure \ref{fig:ptj_thresholds}. This is expected, as both $\mu_0 = E_T/2$ and $\mu_0 = m_t + m_W/2$ are well-chosen scales, unlike the previously analyzed $\mu_0 = E_T/8$ and $\mu_0 = H_T/8$, which are known to lead to poorer perturbative stability.

Consequently, we conclude that for the $\textsc{MiNLO}$ method to yield better results when using non-optimal scales, one must consider jets with higher transverse momentum or include a larger number of jets in the algorithm. Indeed, including more jets allows for the determination of additional scales directly from the process kinematics, thereby making the $\textsc{MiNLO}$ predictions even more reliable. 
\begin{table}[!t]
\centering
\scalebox{1.0}{
\begin{tabular}{ccc}
\midrule\midrule
$\mu_0$  & $\sigma^{\textrm{LO}}$ [ab] &  $\sigma^{\textsc{MiLO}}$ [ab] \\
\midrule\midrule
 $m_t + m_W/2$ & $80.3^{+59\%}_{-35\%}$ &  $72.0^{+49\%}_{-30\%}$ \\
\midrule
 $E_T/2$ & $47.7^{+51\%}_{-32\%}$ &  $41.3^{+47\%}_{-29\%}$ \\
\midrule
 $E_T/3$ & $60.4^{+54\%}_{-33\%}$ &  $57.9^{+48\%}_{-30\%}$ \\
\midrule
 $H_T/2$ & $53.7^{+53\%}_{-32\%}$ &  $50.8^{+48\%}_{-30\%}$ \\
\midrule
 $H_T/3$ & $68.5^{+56\%}_{-34\%}$ &  $69.2^{+49\%}_{-30\%}$ \\
\midrule
\end{tabular}}
\caption{\textit{Integrated fiducial cross-section predictions at LO for the full off-shell process $pp \to \WWW \,j j$ at the LHC with $\sqrt{s} = 13 \; \rm TeV$. Results are presented for the standard LO approach and the \textsc{MiLO} scheme using various scale choices. Corresponding scale uncertainties are also shown.}}
\label{tab:ttwjj} 
\end{table}

To investigate this, Table \ref{tab:ttwjj} presents LO predictions for the $pp \to \WWW \,j j$ process, which includes two additional jets beyond the core system. We compare various scale choices to assess the stability of the $\textsc{MiLO}$ predictions relative to the standard LO results. We find that, across all scales considered, \textsc{MiLO} consistently exhibits smaller scale uncertainties. Overall, standard LO scale uncertainties fall within the $51\%-59\%$ range, whereas the corresponding uncertainties for \textsc{MiLO} lie between $47\%$ and $49\%$. This demonstrates that the \textsc{MiLO} method is less sensitive to the choice of scale than the conventional LO approach, regardless of the scale setting considered.

%
\section{Differential cross-section predictions}
\label{sec:differential}
%

In the following, we focus on the differential cross-section predictions for the $pp \to \WWW \,j + X$ process at NLO in QCD,  comparing the \textsc{MiNLO} method with the standard NLO approach. Similar results for the process $pp \to \WWW \,j  j$ are presented in Appendix \ref{appendix:a}. We use our default  \textsc{MiNLO} setup. We present results for the dynamic scale choice $\mu_0 = \mu_R = \mu_F = E_T/2$, though we have verified that other dynamic scale settings yield similar results. We also show results for the fixed scale $\mu_0 = \mu_R = \mu_F = m_t + m_W/2$. It is well known that using a fixed scale at the differential level can result in NLO uncertainties becoming comparable to LO uncertainties in the tails of certain differential distributions, or in NLO predictions falling outside the LO uncertainty bands. Such effects have already been observed in various studies involving the top quark, see e.g. Refs. \cite{Bevilacqua:2018woc,Stremmer:2021bnk}. Our goal here is to verify whether the same conclusions can be reached when applying the \textsc{MiNLO} method to the full off-shell $pp \to t\bar{t}W^+j$ process in the multi-lepton decay channel.
The upper panels of each plot show the absolute  NLO and \textsc{MiNLO} predictions, the middle panels display the ratio to NLO QCD results, while the bottom panels illustrate the magnitude of scale uncertainties normalized to the corresponding results.
\begin{figure}[t!]
        \centering        
        \includegraphics[width=0.49\linewidth]{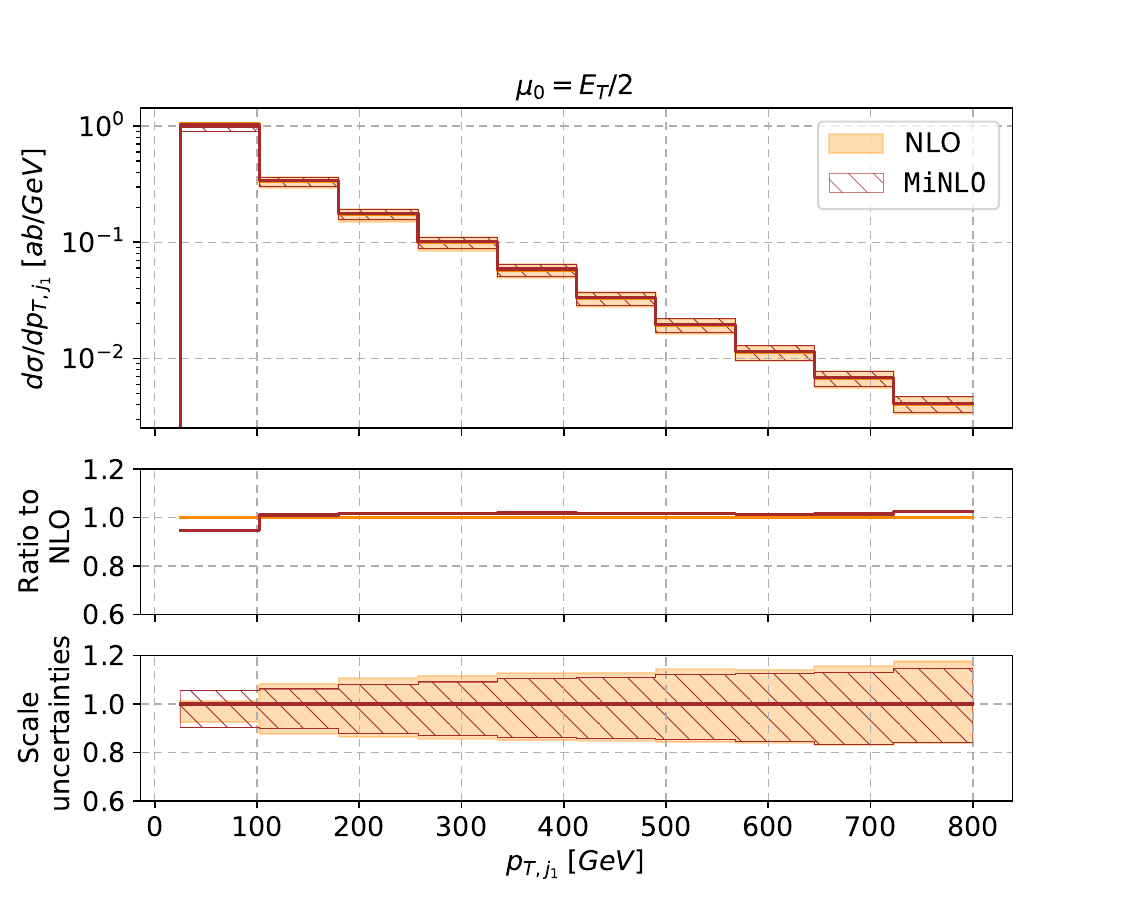}
        \includegraphics[width=0.49\linewidth]{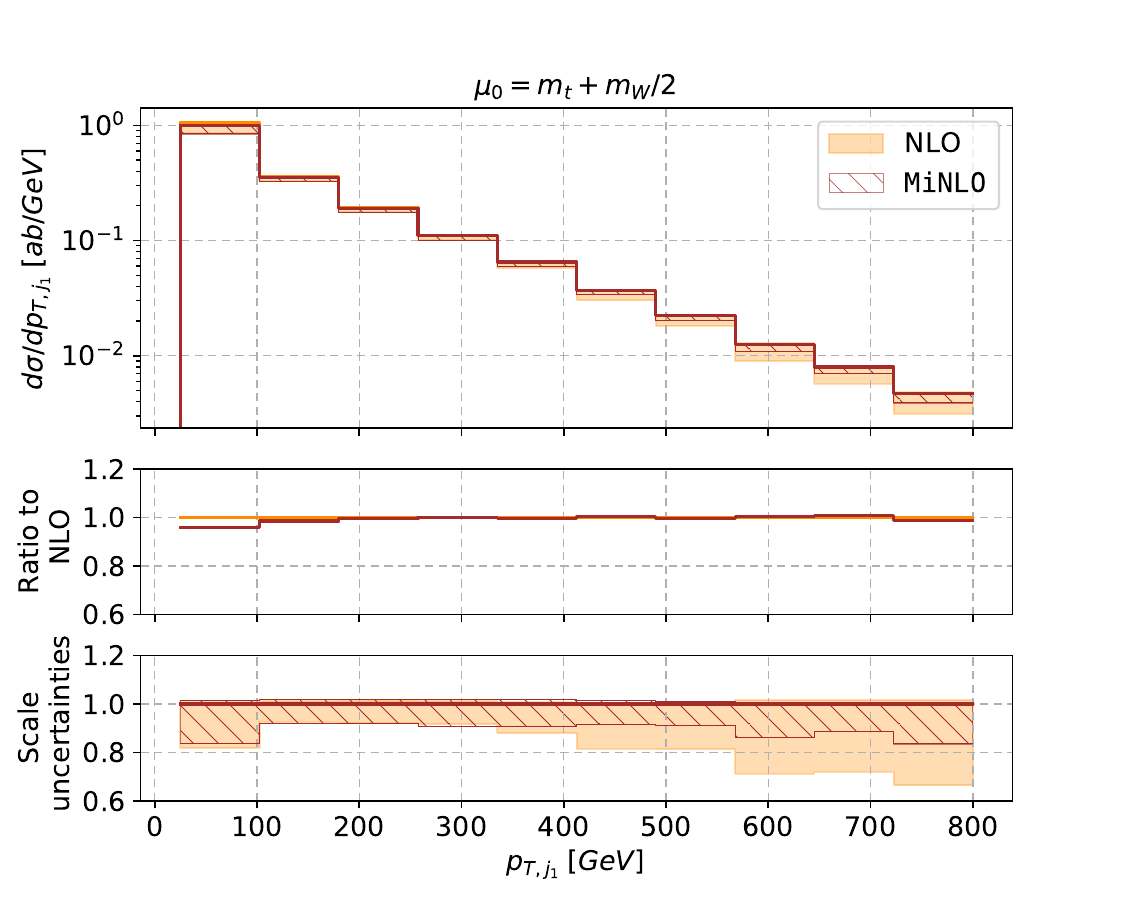}
        \includegraphics[width=0.49\linewidth]{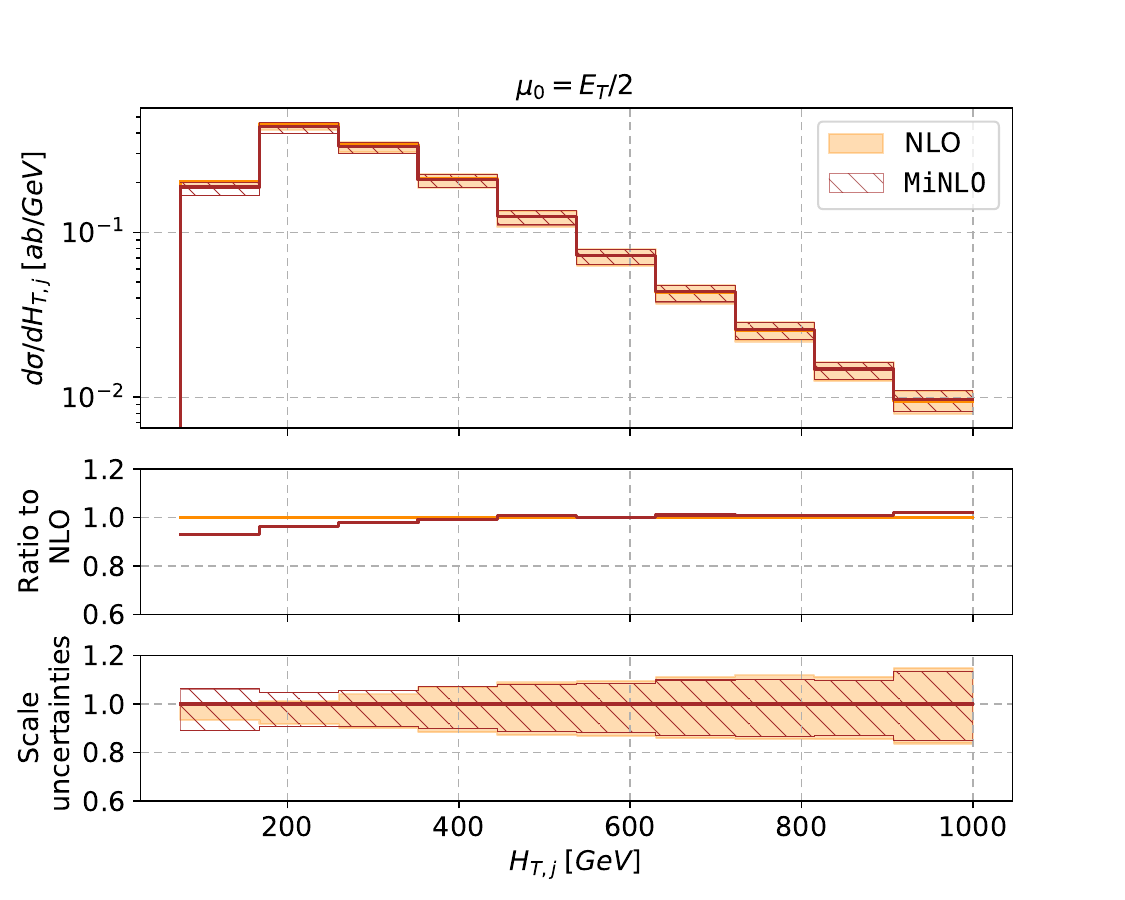}
        \includegraphics[width=0.49\linewidth]{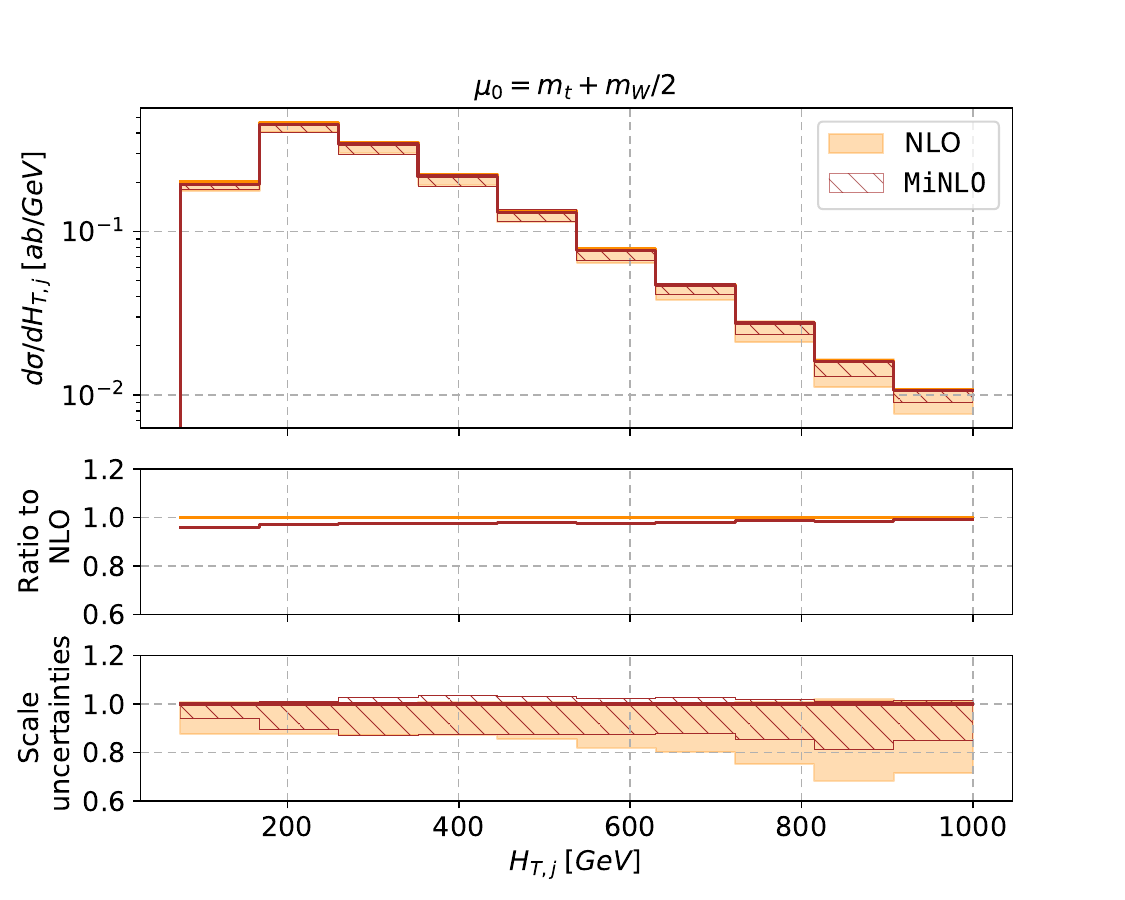}
        \caption{\textit{Differential cross-section distributions for the full off-shell $pp \rightarrow \WWW\,j $ process at the LHC with $\sqrt{s} = 13 \; \rm TeV$, for the $p_{T, \,j_1}$ and $H_{T, \, j}$ observables. Results are shown for the dynamical scale choice $\mu_0 = E_T/2$  and the fixed scale setting $\mu_0 = m_t + m_W/2$, using the standard NLO and the \textsc{MiNLO} method. The upper panels show absolute predictions, the middle panels display the ratio to NLO QCD results, while the bottom panels illustrate the magnitude of scale uncertainties normalized to the corresponding results. }}
         \label{fig:minlovsnlo_ptj}
\end{figure}
\begin{figure}[t!]
        \centering        
        \includegraphics[width=0.49\linewidth]{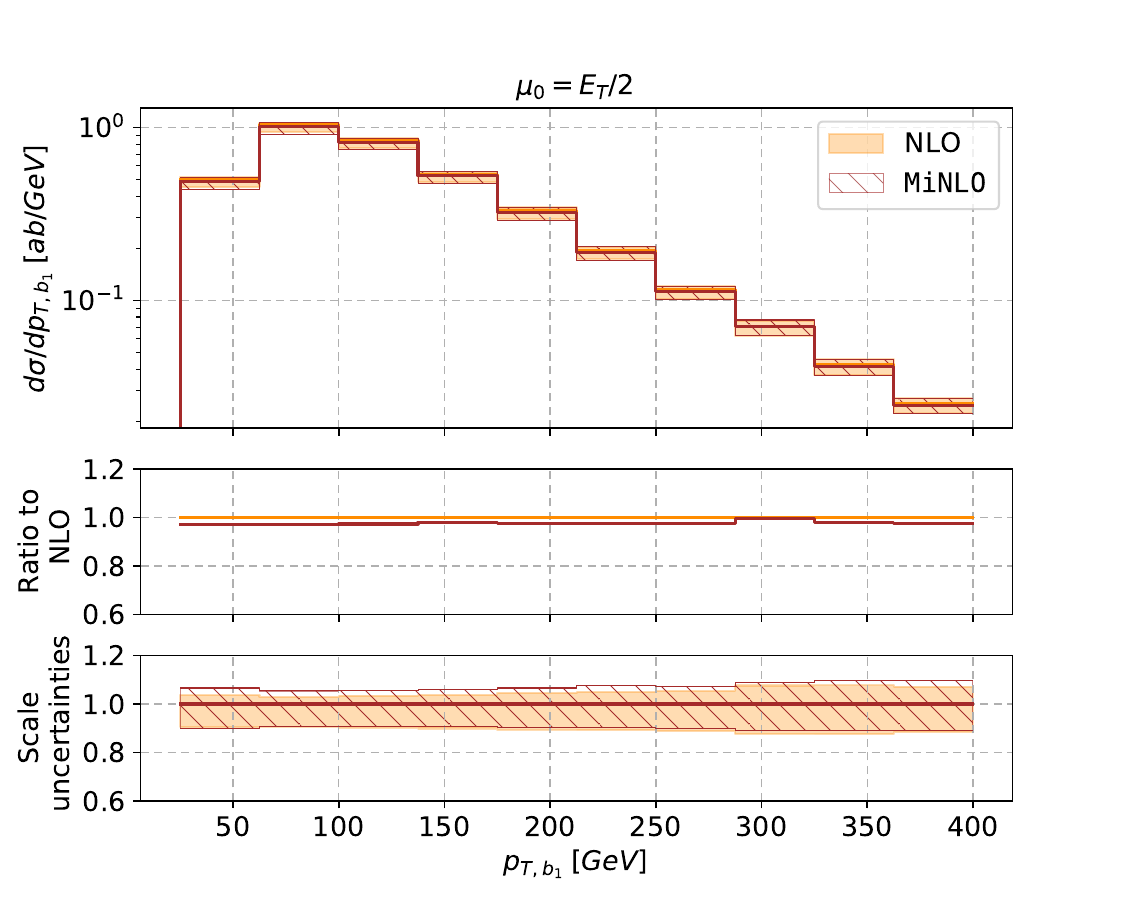}
        \includegraphics[width=0.49\linewidth]{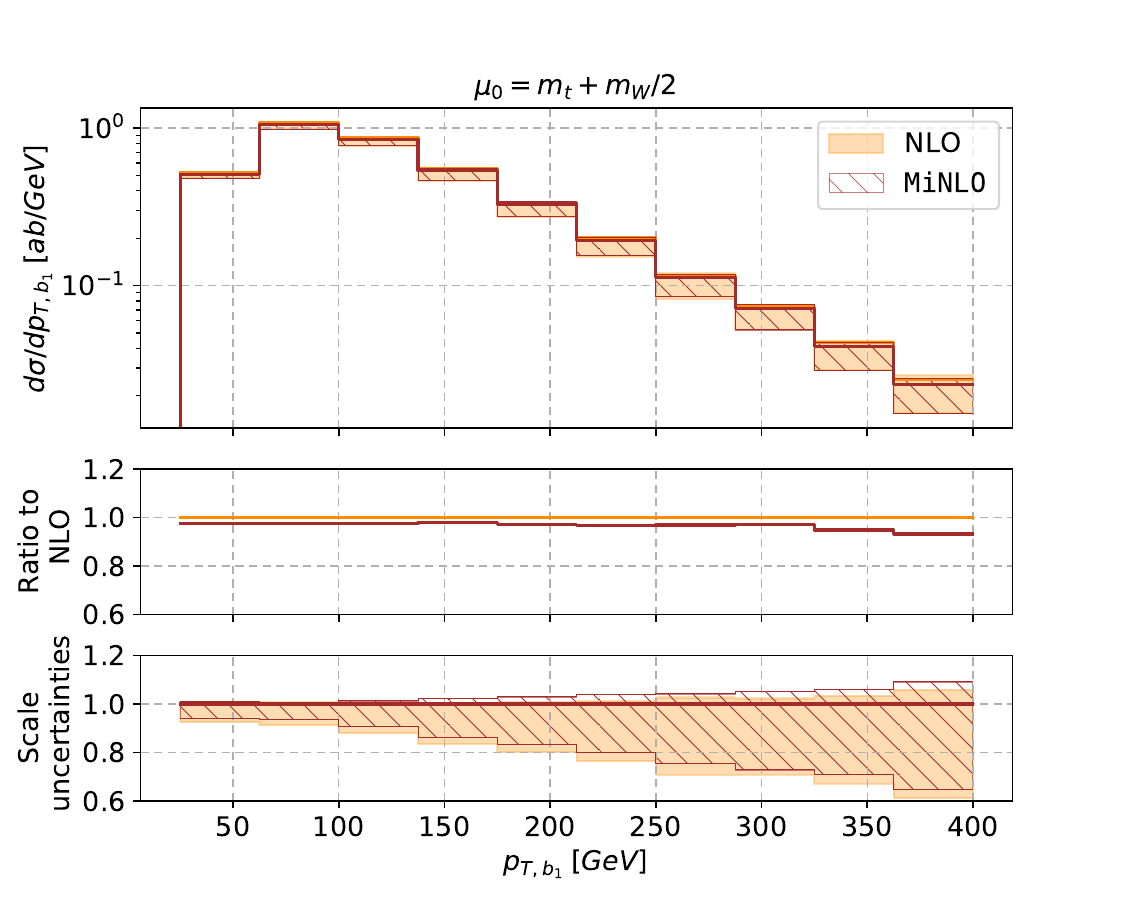}
        \includegraphics[width=0.49\linewidth]{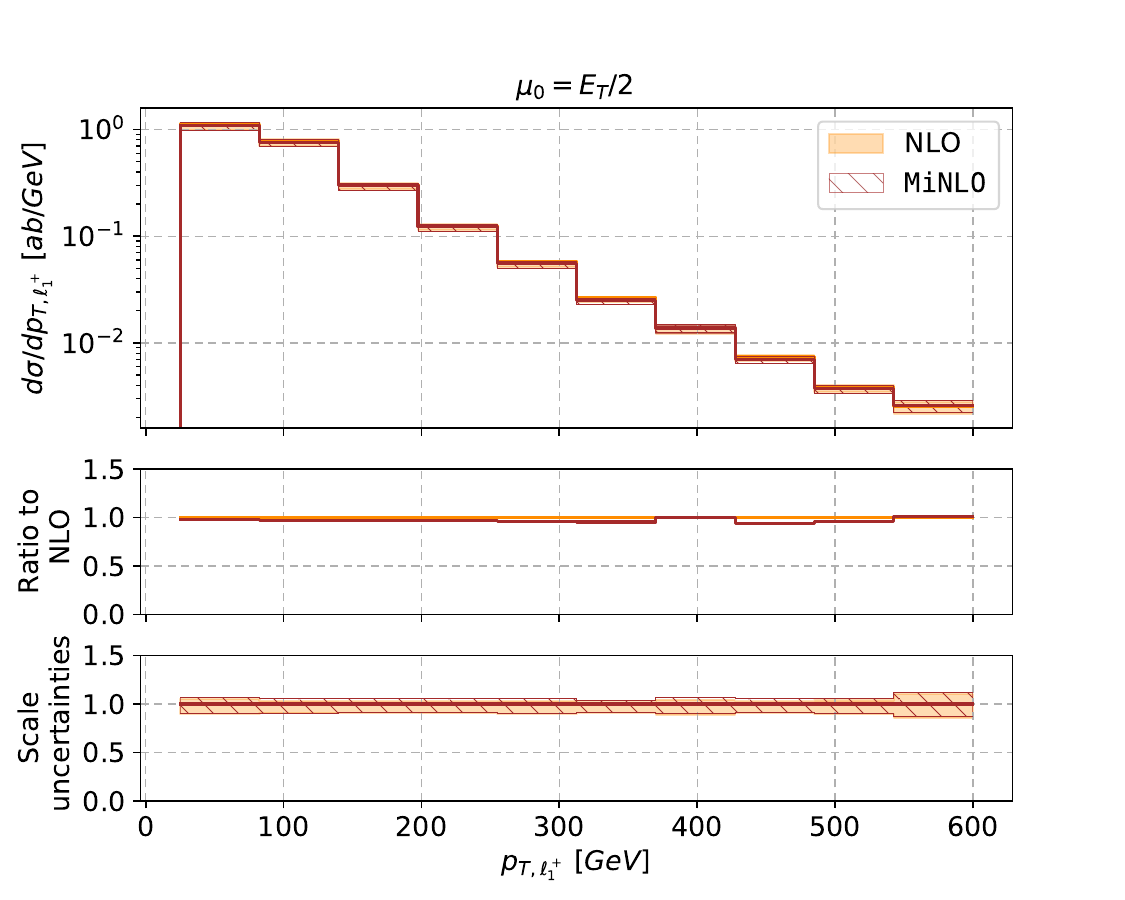}
        \includegraphics[width=0.49\linewidth]{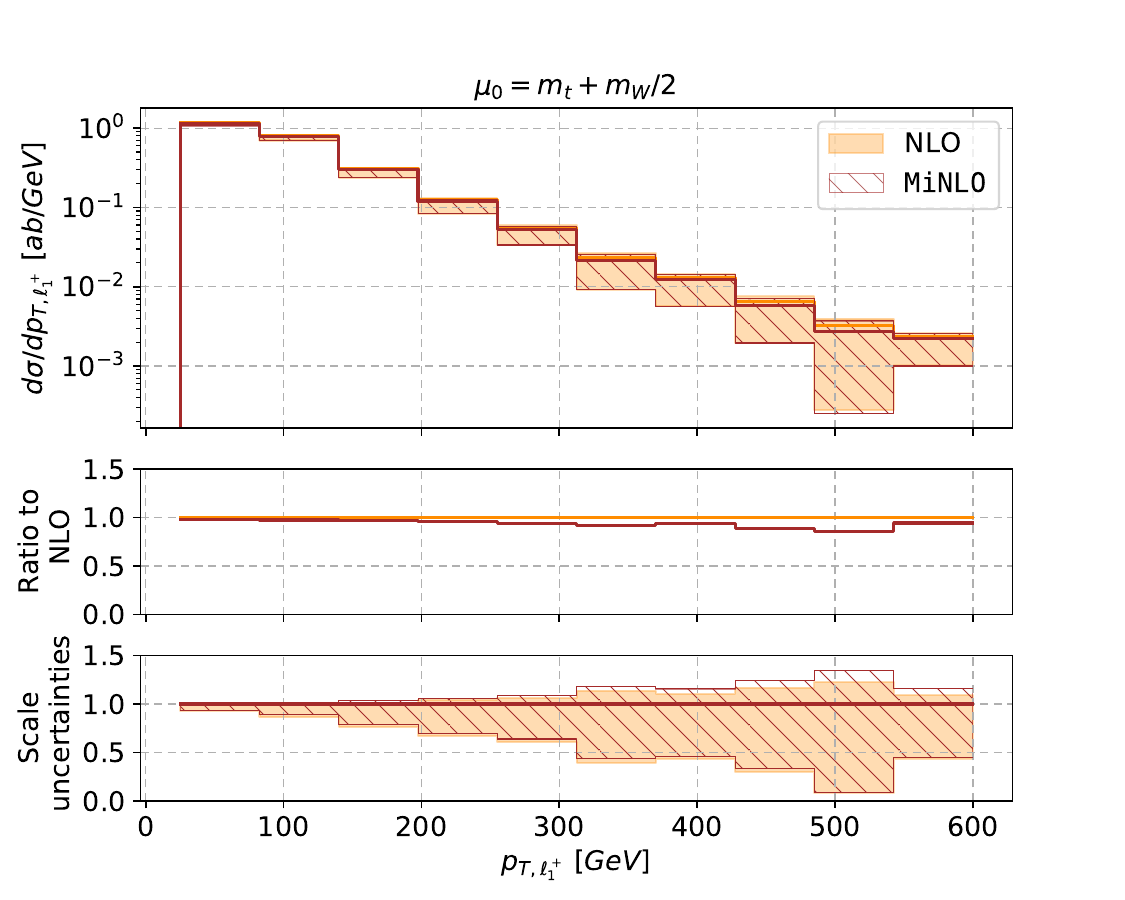}
        \caption{\textit{Same as in Figure \ref{fig:minlovsnlo_ptj}, but for the $p_{T,\, b_1}$  and $p_{T, \, \ell_1^+}$ distributions.}}
         \label{fig:minlovsnlo_ptbl}
\end{figure}
\begin{figure}[t!]
        \centering        
        \includegraphics[width=0.49\linewidth]{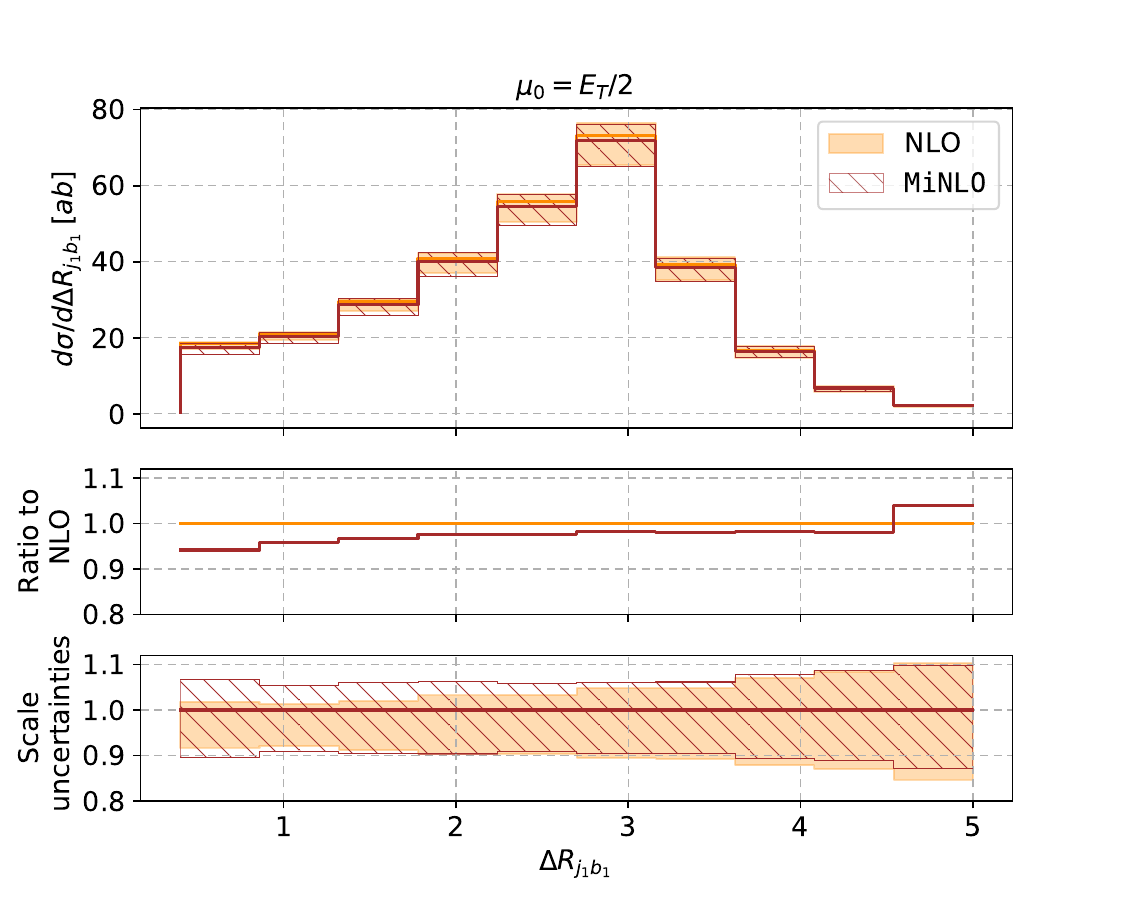}
        \includegraphics[width=0.49\linewidth]{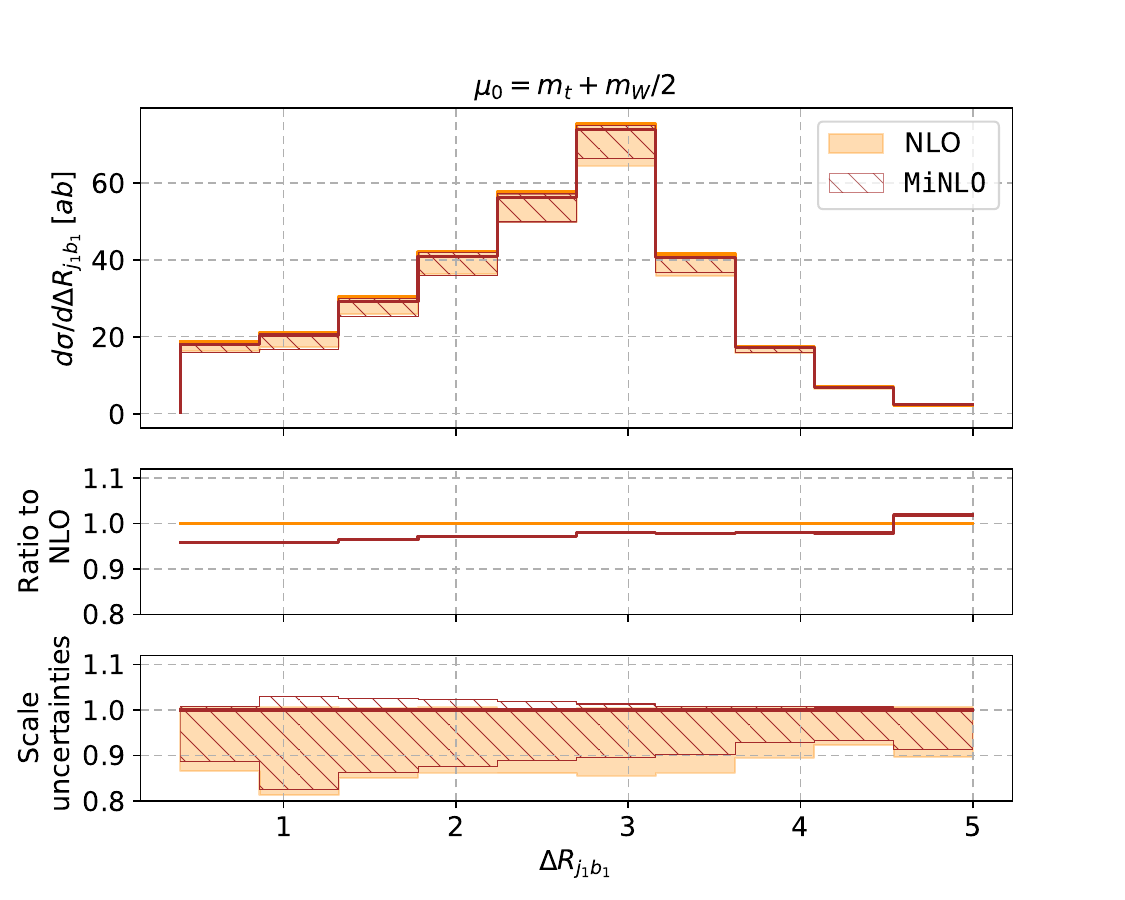}
        \includegraphics[width=0.49\linewidth]{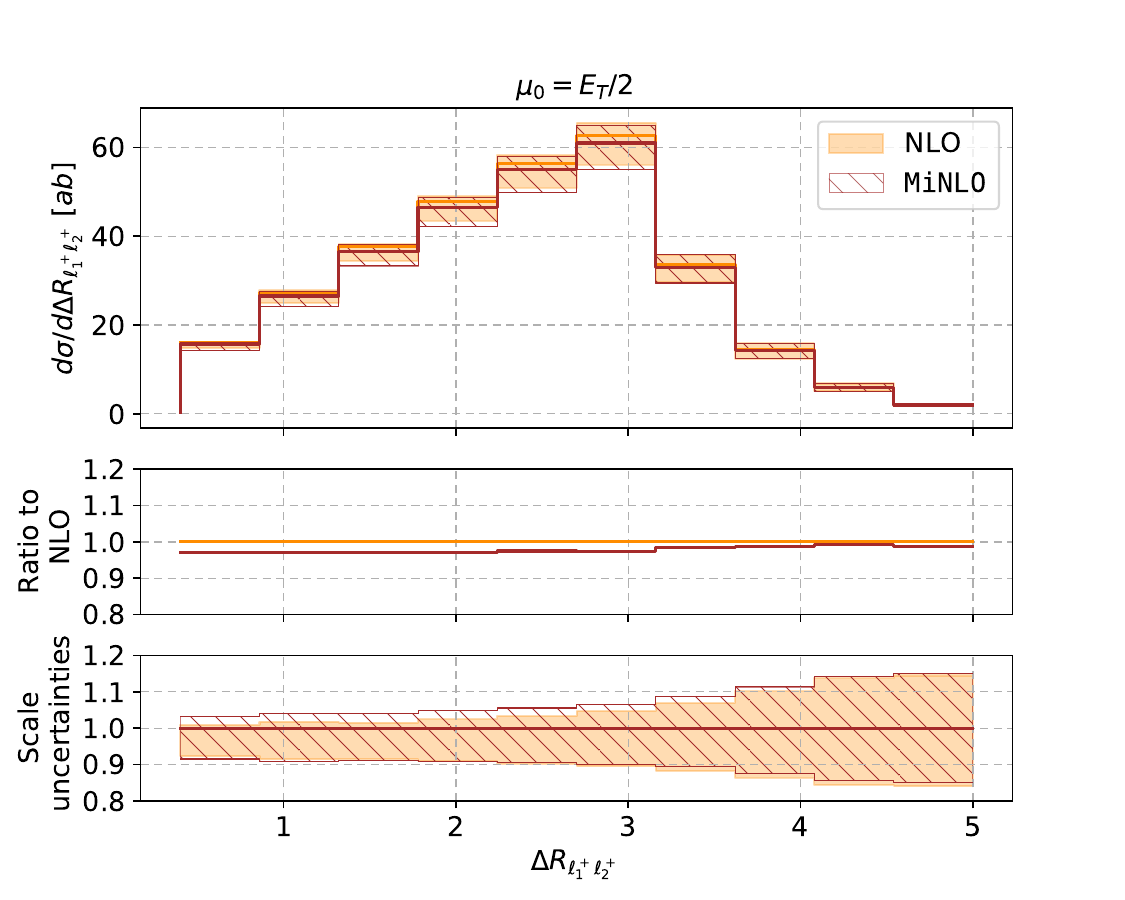}
        \includegraphics[width=0.49\linewidth]{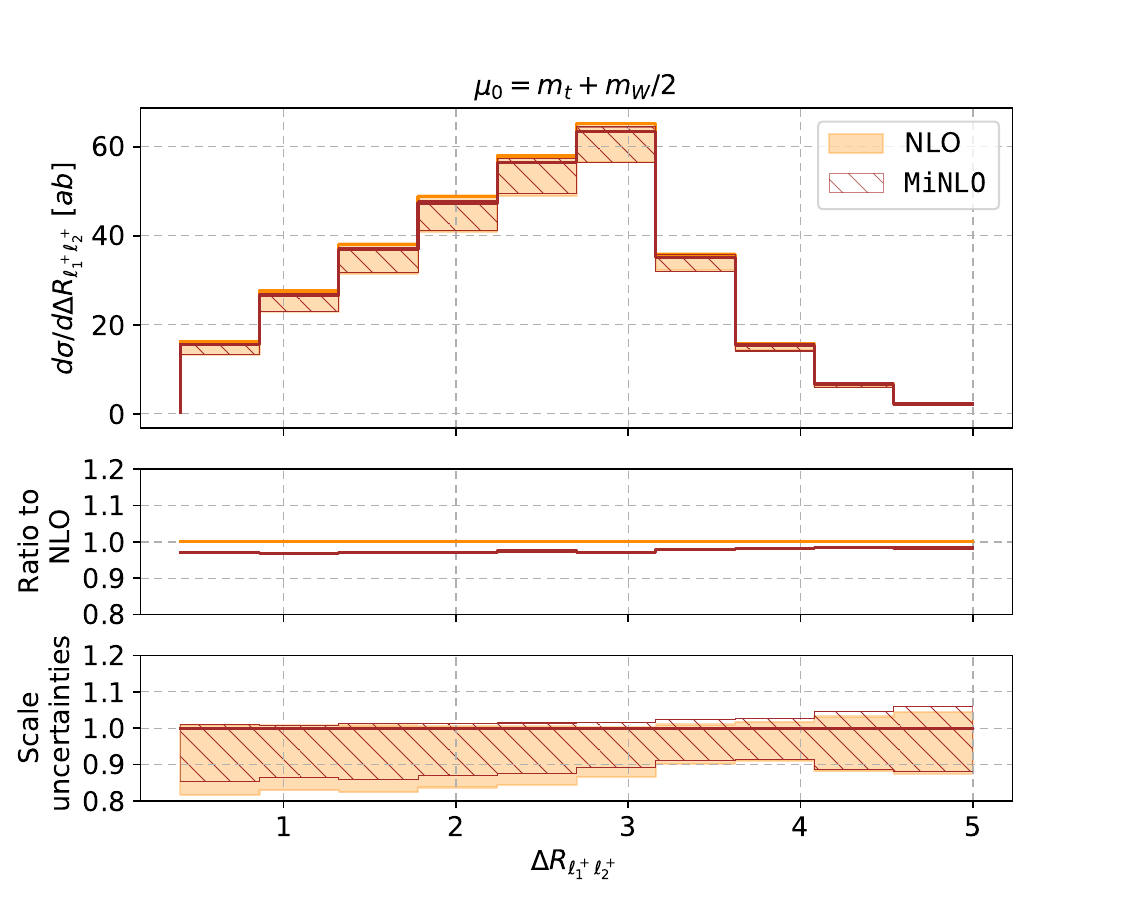}
        \caption{\textit{Same as Figure \ref{fig:minlovsnlo_ptj}, but for the $\Delta R_{j_1b_1}$ and $\Delta R_{\ell_1^+ \ell_2^+}$ distributions.}}
         \label{fig:minlovsnlo_dr}
\end{figure}

In Figure \ref{fig:minlovsnlo_ptj}, we present $p_{T,\,j_1}$ and $H_{T, \, j}$, where $H_{T, \, j}=p_{T,\,b_1}+p_{T,\,b_2}+p_{T,\,j_1}$. The focus is thus placed first on the observables related to light jets. We find that the results obtained using \textsc{MiNLO} and the standard NLO approach are fully consistent for both observables, with deviations at the beginning of the distributions amounting to at most $7\%$. Furthermore, the use of \textsc{MiNLO} leads to a significant reduction in scale uncertainties when a fixed scale is employed, particularly in the high-$p_T$ tails. In these phase-space regions, the uncertainties associated with the \textsc{MiNLO} method are $17\%$ for $p_{T, \,j_1}$ and $20\%$ for $H_{T, \, j}$, respectively. The corresponding NLO standard uncertainties are roughly twice as large for $p_{T, \,j_1}$, reaching values of up to $34\%$, while for $H_{T, \, j}$ they amount to up to $32\%$. Moreover, in these regions, the light jet becomes highly energetic, increasing the probability that the nodal scale $q_1$ obtained from the inverse $k_T$-clustering procedure exceeds the value of the fixed scale $\mu_0 = m_t + m_W/2 \approx 213 \; \rm GeV$.  This occurs, for example, when the additional jet originates from a light quark that can only be clustered with the initial-state partons. In such cases, the clustering scale $q_1$ is set to the transverse momentum of this highly energetic light jet, forcing  $q_{core} = q_1$ due to the ordered clustering condition specified in Eq. \eqref{orderd_qi_nlo}. If the additional jet is actually a gluon, which can also couple to final-state objects, a situation may still arise where $q_1 > 213 \; \rm GeV$, provided that the $b$-jets are also sufficiently energetic. If the transverse momenta of both the $b$-jet and the additional gluon exceed $p_{T,\,j} = 213 \; \rm GeV$, Eq. \eqref{yij} guarantees that the nodal scale will exceed the value $\mu_0$. In such a case, as in the previous instance, the fixed scale is omitted from the calculations, and the scale $q_1$ is used as the core scale. This behavior is reflected in the $H_{T, \, j}$ distribution, where, around $H_{T,\, j} \approx 600 \; \rm GeV$, the \textsc{MiNLO} scale uncertainties become smaller than in the standard NLO calculation, precisely because the impact of the fixed-scale setting diminishes in the computation. 

Such behavior is not observed in Figure \ref{fig:minlovsnlo_ptbl}, which shows the transverse momenta of the hardest $b$-jet $(p_{T,\,b_1})$,  and the hardest positively charged lepton $(p_{T, \, \ell_1^+})$. For both observables and both scale choices, the differences in the magnitude of scale uncertainties between the \textsc{MiNLO} and NLO QCD predictions remain negligible across the entire kinematic range. In the case of the dynamic scale, the scale uncertainties for MiNLO and NLO are essentially identical for both observables and amount to $10\%-11\%$. However, for the fixed scale setting, the theoretical uncertainties become significantly larger, reaching $40\%$ and $90\%$, respectively, in the tails of the $p_{T, \, b_1}$ and $p_{T, \, \ell_1^+}$ distributions. Since leptons do not participate in the clustering algorithm, their kinematic distributions are expected to remain largely unchanged by the \textsc{MiNLO} procedure. On the other hand, $b$-jets participate in the clustering procedure but discarding the fixed scale also requires the presence of a highly energetic light jet, since the nodal scales are determined based on the most probable branchings of the additional light jet relative to the core process. Only if this additional light jet is sufficiently energetic, $q_1 > \mu_0 = m_t+m_W/2$, and we can have $q_{core} = q_1$. However, as shown in Figure~\ref{fig:minlovsnlo_ptj}, the leading light jet is predominantly produced with a transverse momentum below $100$ GeV. Thus, the fixed scale setting may persist in the high-$p_T$ region of the $p_{T,\, b_1}$ distribution, leading to large theoretical uncertainties similar to those observed in the standard NLO QCD calculation.

Finally, Figure \ref{fig:minlovsnlo_dr} shows the angular separation between the hardest light jet and the hardest $b$-jet ($\Delta R_{j_1b_1}$), as well as the angular separation between the two positively charged leptons ($\Delta R_{\ell_1^+ \ell_2^+}$). We therefore turn our attention to dimensionless observables. For $\Delta R_{j_1 b_1}$, we observe slight distortions in the distribution shape between the \textsc{MiNLO} and NLO predictions. The obtained $\textsc{MiNLO}/\rm NLO$ ratios fall within the range of $(0.94 -1.04)$ for the dynamic scale and $(0.96- 1.02)$ for the fixed scale choice. In the phase-space regions where the back-to-back configurations occur, the uncertainty bands are comparable for $\mu_0 = E_T/2$. However, with the fixed scale choice we adopted, the scale uncertainties in the \textsc{MiNLO} method are slightly smaller, amounting to $11\%$ compared to $14\%$ for the standard NLO calculation. This reduction reflects the behavior already observed at the integrated fiducial cross-section level. For small values of $\Delta R_{j_1 b_1}$, the \textsc{MiNLO} uncertainties exceed the NLO uncertainties for $\mu_0=E_T/2$, whereas for $\mu_0=m_t+m_W/2$, both methods yield uncertainties of similar magnitude. For $\Delta R_{\ell_1^+ \ell_2^+}$ the size of the theoretical uncertainties remains comparable between the two methods over the entire kinematic range.

To sum up this part, it can be stated that the \textsc{MiNLO} method performs better given a fixed scale choice for observables associated with light jets.  For observables related to charged leptons and $b$-jets, such improvements are no longer evident.  The same applies to a judiciously chosen dynamic scale setting, for which no significant differences are observed between the \textsc{MiNLO} results and NLO QCD predictions. On the other hand, one can reverse this line of reasoning and state that the \textsc{MiNLO} method allows us to verify whether the chosen dynamic scale setting is appropriate for the process under study.

%
\section{Multi-jet merged predictions for 
$\boldsymbol{pp\to t\bar{t}W^+ +X}$}
\label{sec:merging}
%

As a bonus of this work, we combine predictions for the $pp \to t\bar{t}W^+ +X$ process with different jet multiplicities and compare the obtained results with the $pp \to t\bar{t}W^+ + X$ process, calculated at NLO in perturbative QCD. The aim of this is to increase the accuracy of full off-shell predictions for the $pp \to t\bar{t}W^+ + X$ process in the multilepton decay channel. The merged results are expected to incorporate parts of the NNLO QCD contributions and, in particular, to improve the modeling of additional jet activity. Accounting for these contributions is particularly important for processes dominated by the real-emission part of the NLO result, as is the case with the $pp\to t\bar{t}W^\pm$ process, see e.g. Ref.  \cite{Bi:2023ucp}. Although merged predictions for the process $pp \to t\bar{t}W^\pm + X$ already exist in the literature
\cite{Frederix:2021agh, vonBuddenbrock:2020ter}, they are limited to scenarios considering stable top quarks and $W^\pm$ gauge bosons. Decays are modeled within the parton shower programs.

To combine different jet multiplicities for the process $pp \to \WWW + X$ at (N)LO in QCD, i.e. to merge 
\begin{itemize}
\item 
$pp\to \WWW+X$ at NLO in QCD, 
\item $pp\to \WWW \, j +X$ at NLO in QCD,
\item $pp\to \WWW \, jj$ at LO,
\end{itemize}
the merging scale, denoted as $p_{T,\, merging}$, is introduced to veto emissions exceeding this value, thereby ensuring no overlap between different contributions and preventing double counting. 

It should be emphasized, however, that the merging with the \textsc{MiNLO} method, as described in the literature, see e.g. Refs. \cite{Hamilton:2012rf,Luisoni:2013cuh,Hamilton:2016bfu}, does not rely on the introduction of the explicit merging scale. In these studies,  NLO accuracy is preserved in observables that are both exclusive and inclusive in the additional radiation, provided $\alpha_s$ is appropriately chosen, and the Sudakov form factors are modified. Nevertheless, this method has so far been applied primarily to relatively simple processes involving the production of stable Higgs bosons and $W^\pm/Z$ gauge bosons, as well as $HW^\pm /HZ$ final states, for which the results with the NNLL accurate resummation of soft-parton emissions are available in the literature.  These results are essential for achieving NLO accuracy, in which the neglected terms can only be of the relative order of $\alpha_s^2$ or higher, see e.g. Ref. \cite{Hamilton:2012rf}. In the \textsc{MiNNLO} case \cite{Monni:2019whf} the results have also been reported for more complex processes, including $pp \to t\bar{t}$ \cite{Mazzitelli:2020jio, Mazzitelli:2021mmm}, $pp\to t\bar{t}H$ \cite{Biello:2026nhj}, and $pp\to Zb\bar{b}$ \cite{Mazzitelli:2024ura}. However, they have been calculated for stable top quarks and $Z$ bosons, that is, for processes for which the corresponding resummation predictions are also known. The general concept of the merging with \textsc{MiNLO} could, in principle, be extended to the process discussed here by lifting the restrictions on additional jet radiation. A reliable implementation that would preserve NLO accuracy, however, would require  the  NNLL result for $pp\to \WWW+X$, including full off-shell effects and fiducial phase-space cuts. These aspects are simply not yet fully understood in the context of resummation frameworks. Therefore, to preserve  NLO accuracy for $pp\to \WWW+X$, we use a direct sum instead, the main purpose of which is to combine  the available fixed-order jet multiplicities.

For all the fixed-order predictions, we apply the \textsc{MiNLO} method separately, and the merged cross-section results are obtained via the following direct sums
\begin{align}
    &\left(\sigma^{\textsc{MiNLO}}_{pp\to t\bar{t}W^+ + X}\right)_{\rm merged} = \left(\sigma^{\textsc{MiNLO}}_{pp \to t\bar{t}W^+}\right)_{\rm excl}+\left(\sigma^{\textsc{MiNLO}}_{pp \to t\bar{t}W^+ j}\right)_{\rm incl}\,, \label{merge1jet}\\
    &\left(\sigma^{\textsc{MiNLO}}_{pp\to t\bar{t}W^+ + X}\right)_{\rm merged} = \left(\sigma^{\textsc{MiNLO}}_{pp \to t\bar{t}W^+}\right)_{\rm excl}+\left(\sigma^{\textsc{MiNLO}}_{pp \to t\bar{t}W^+j}\right)_{\rm excl} + \left(\sigma^{\textsc{MiLO}}_{pp \to t\bar{t}W^+jj}\right)_{\rm incl}\,, \label{merge2jet}
\end{align}
for merging with up to 1 and 2 additional jets, respectively. The above prescription specifies which contributions are treated as exclusive and which as inclusive regarding additional jet radiation, in accordance with the definition of the merging scale $p_{T,\, merging}$. Although the $\textsc{Mi(N)LO}$ method is applied to all contributions, we have verified that using the corresponding standard (N)LO calculations led to only minor differences. Both approaches therefore yield results that are consistent and in agreement within their respective theoretical uncertainties.
\begin{figure}[t!]
        \centering        
        \includegraphics[width=0.49\linewidth]{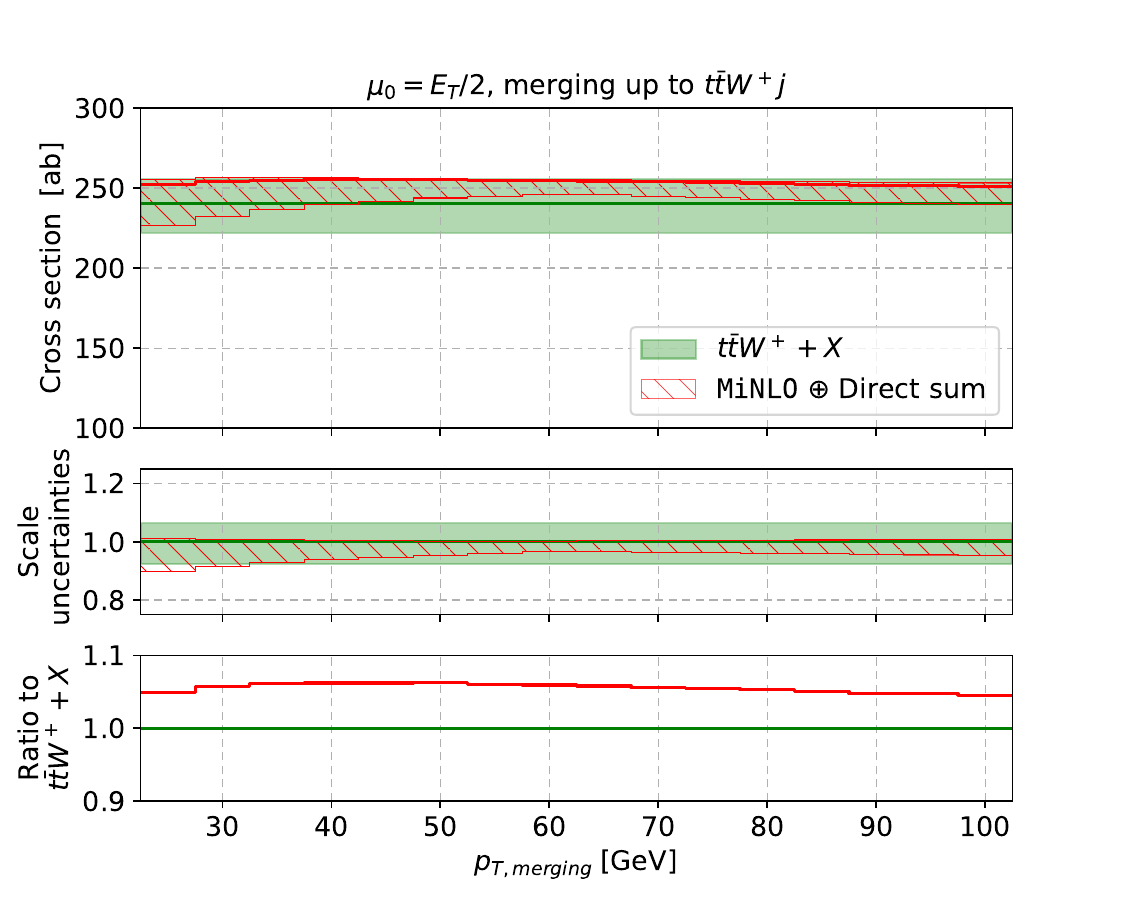}
        \includegraphics[width=0.49\linewidth]{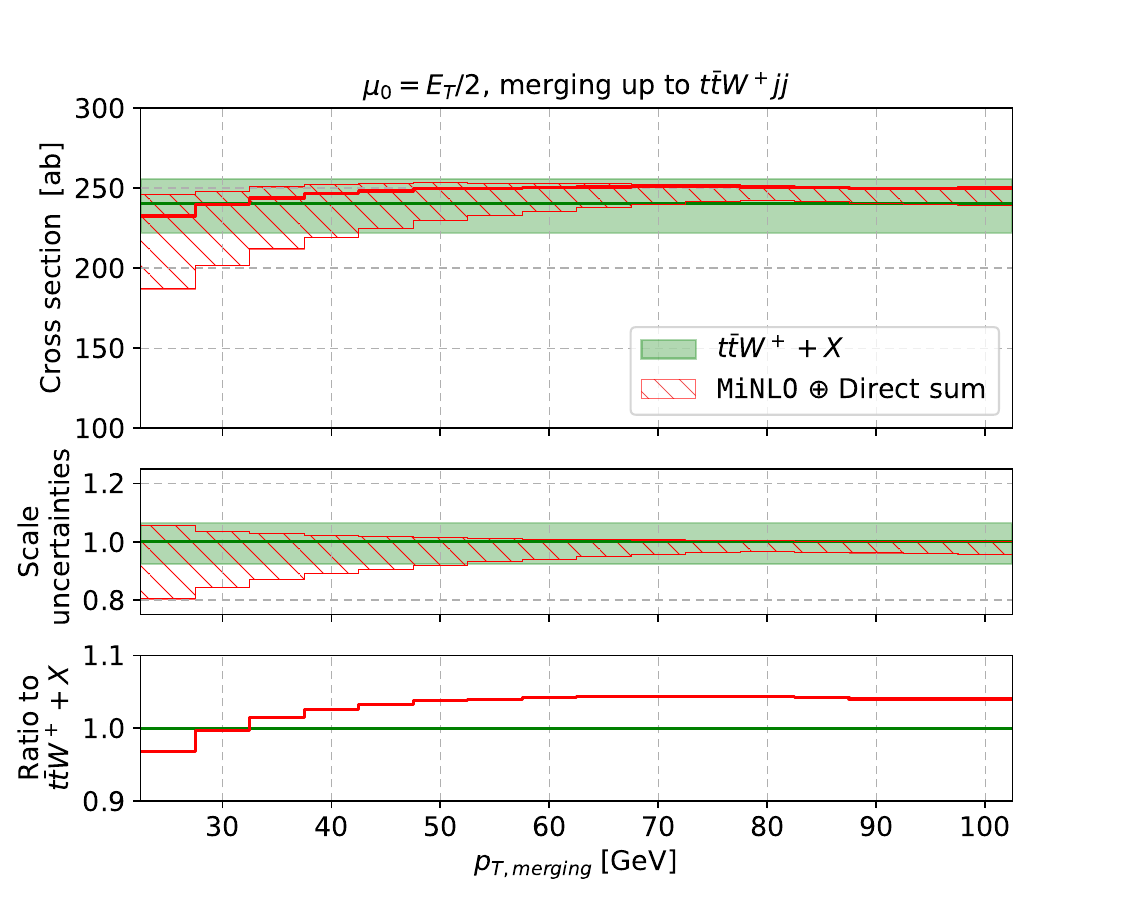}
        \caption{\textit{Merged cross-section predictions as a function of $p_{T,\,merging}$ for the full off-shell $pp \to \WWW  + X$ process at the LHC with $\sqrt{s} = 13 \; \rm TeV$. Results are shown for merging with up to 1  and 2 additional jets,  based on Eq.  \eqref{merge1jet} and Eq. \eqref{merge2jet}, respectively. The fixed-order NLO QCD predictions are also presented. Scale uncertainties for both variants are shown in the middle panels, while the bottom panels illustrate the ratio to the fixed-order NLO calculation. }}
         \label{fig:merged_integrated}
\end{figure}
\begin{figure}[t!]
        \centering        
        \includegraphics[width=0.49\linewidth]{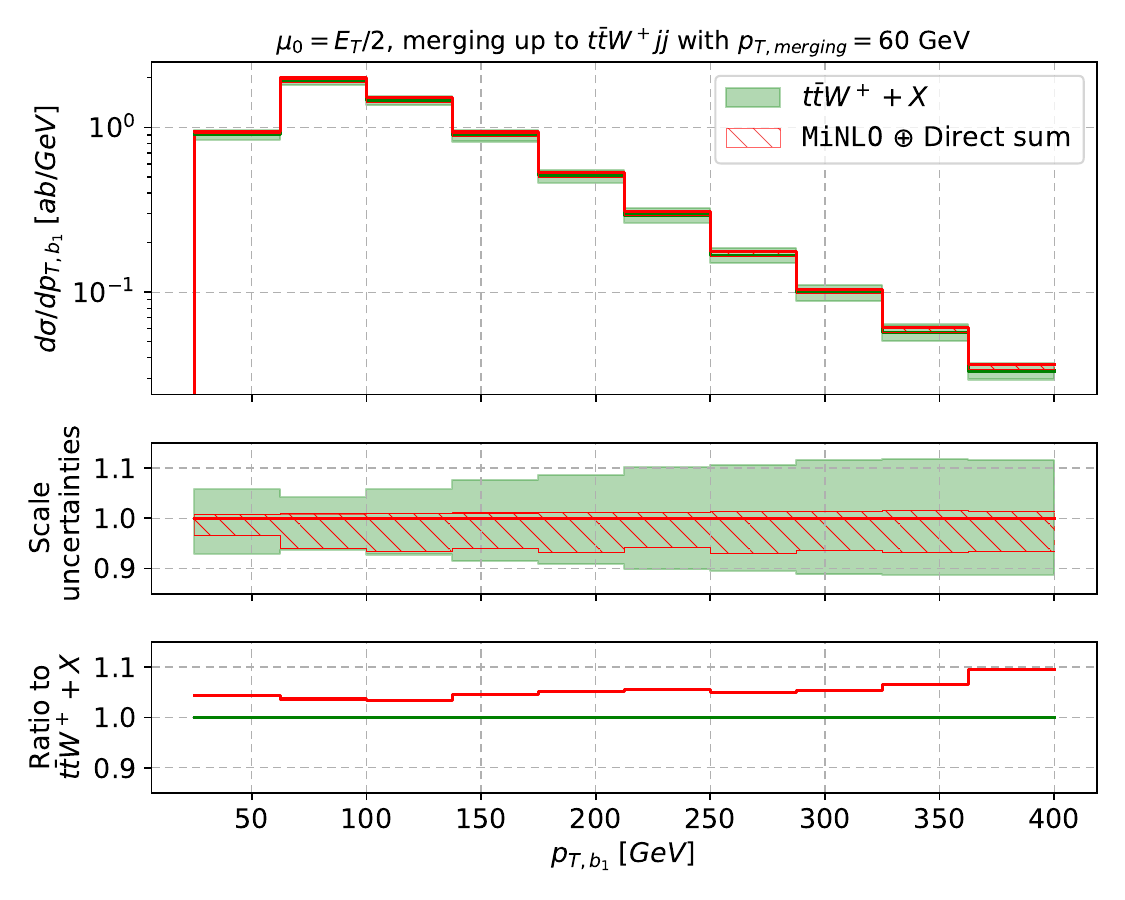}
        \includegraphics[width=0.49\linewidth]{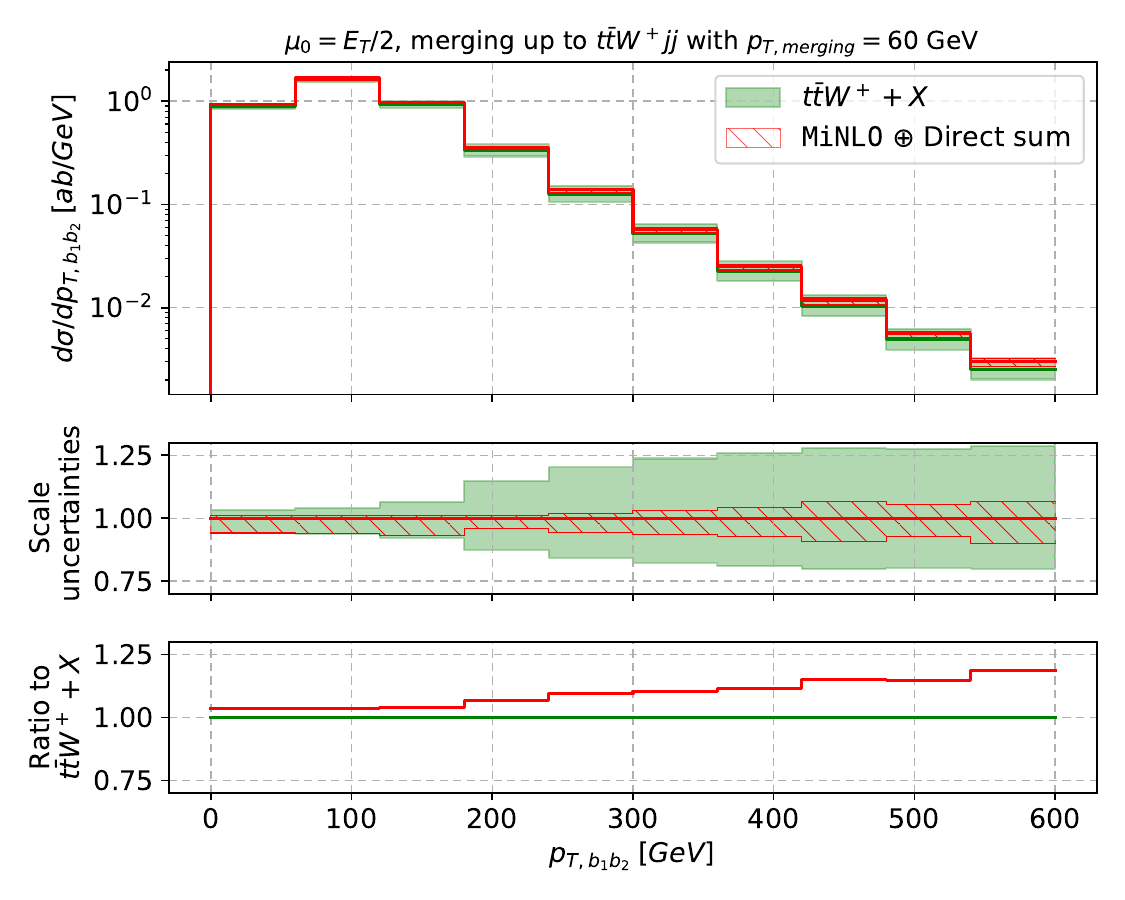}
        \includegraphics[width=0.49\linewidth]{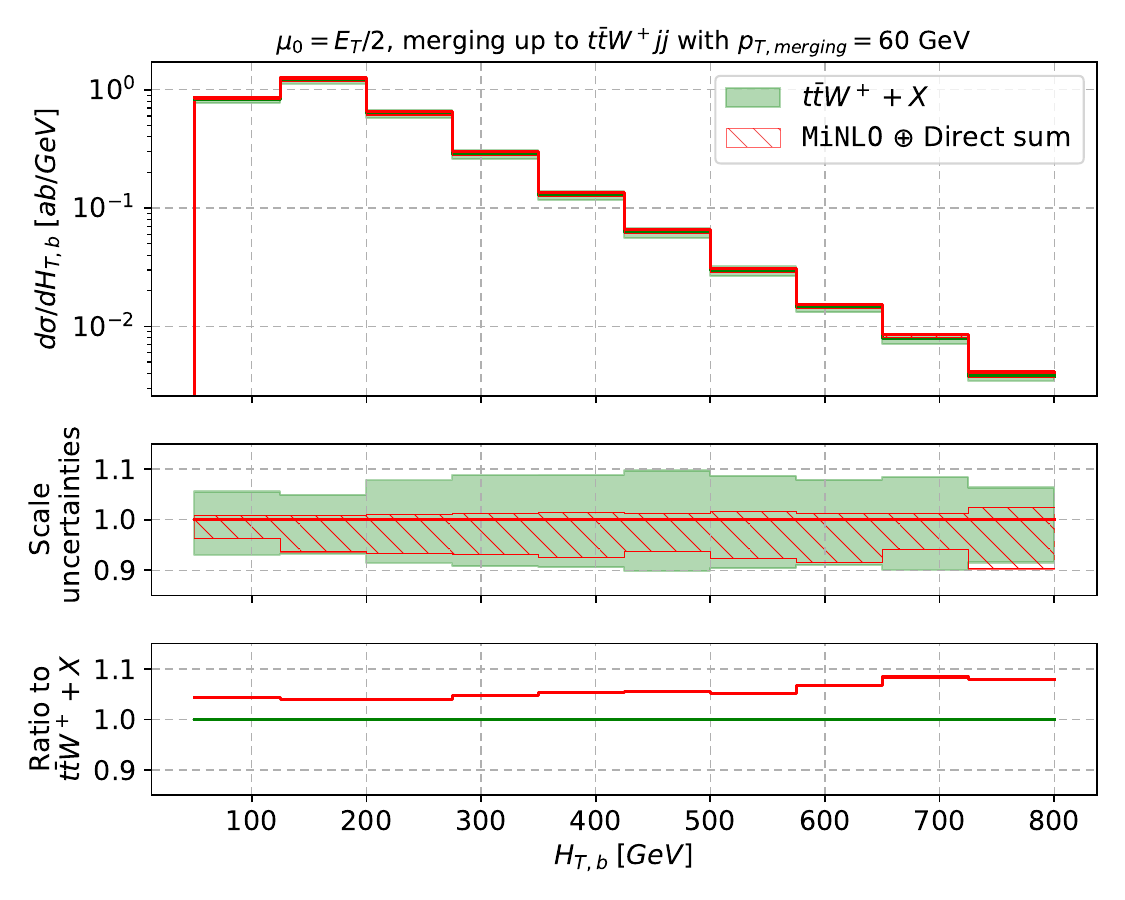}
        \includegraphics[width=0.49\linewidth]{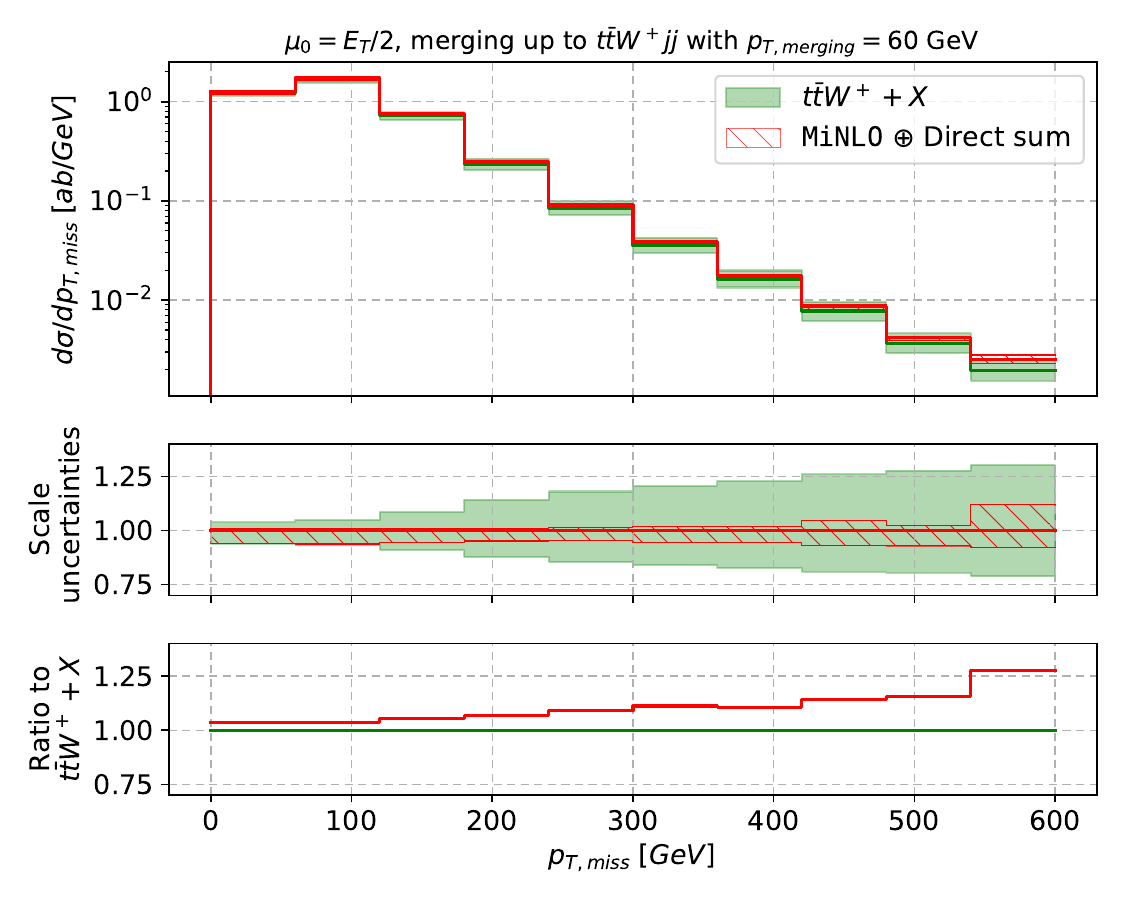}
        \caption{\textit{Differential cross-section distributions for the $pp \to \WWW + X$ process at the LHC with $\sqrt{s} = 13 \; \rm TeV$, as well as the merged prediction obtained using Eq. \eqref{merge2jet} with the  merging parameter set to $p_{T,\,merging} = 60 \; \rm GeV$ for the following observables: $p_{T,\,b_1}, p_{T,\,b_1b_2}, H_{T,\,b}$ and $p_{T,\,miss}$. Results are obtained using the dynamical scale choice $\mu_0 = E_T/2$ and the default setup comprising the NNPDF3.1 PDF set. The upper panels display the absolute predictions, the middle panels illustrate scale uncertainties, while the bottom panels present the ratio to the fixed-order prediction.}}
         \label{fig:diff_merged_1}
\end{figure}
\begin{figure}[t!]
        \centering        
        \includegraphics[width=0.49\linewidth]{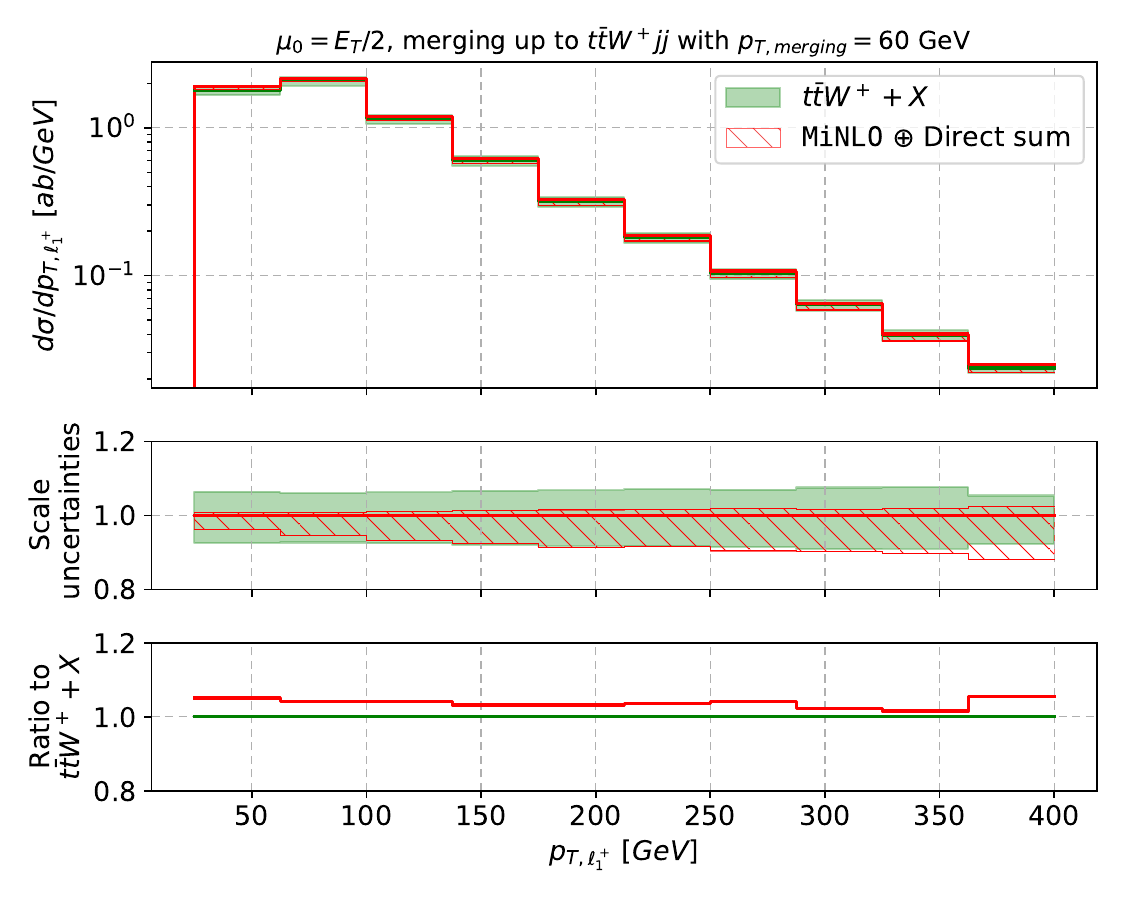}
        \includegraphics[width=0.49\linewidth]{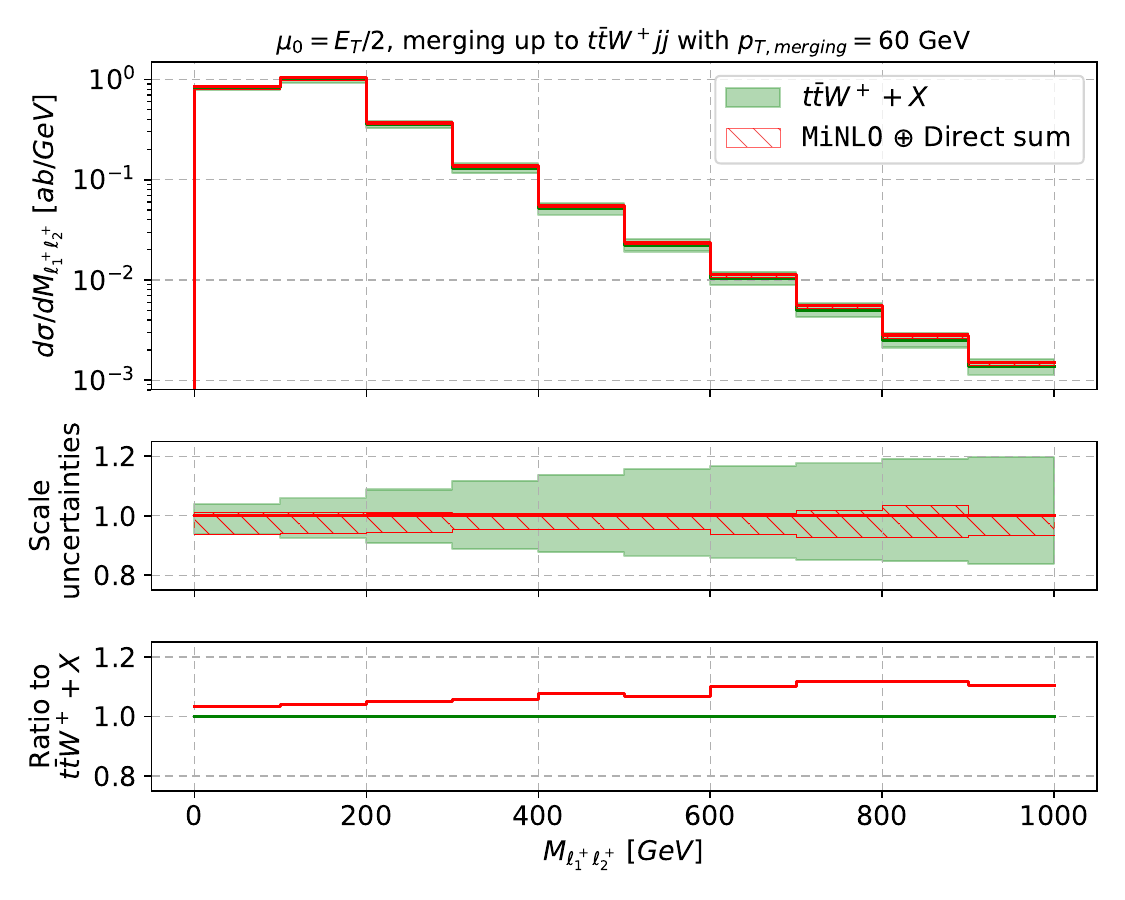}
        \includegraphics[width=0.49\linewidth]{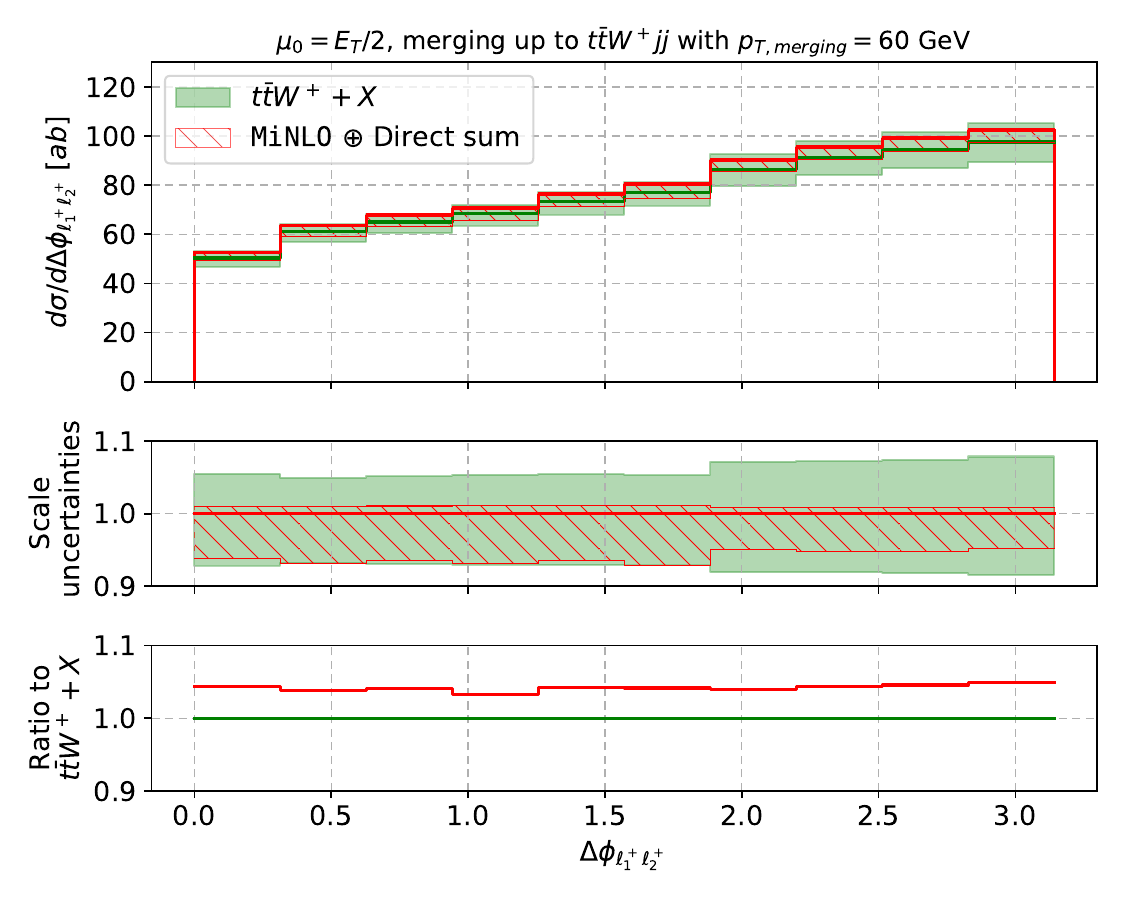}
        \includegraphics[width=0.49\linewidth]{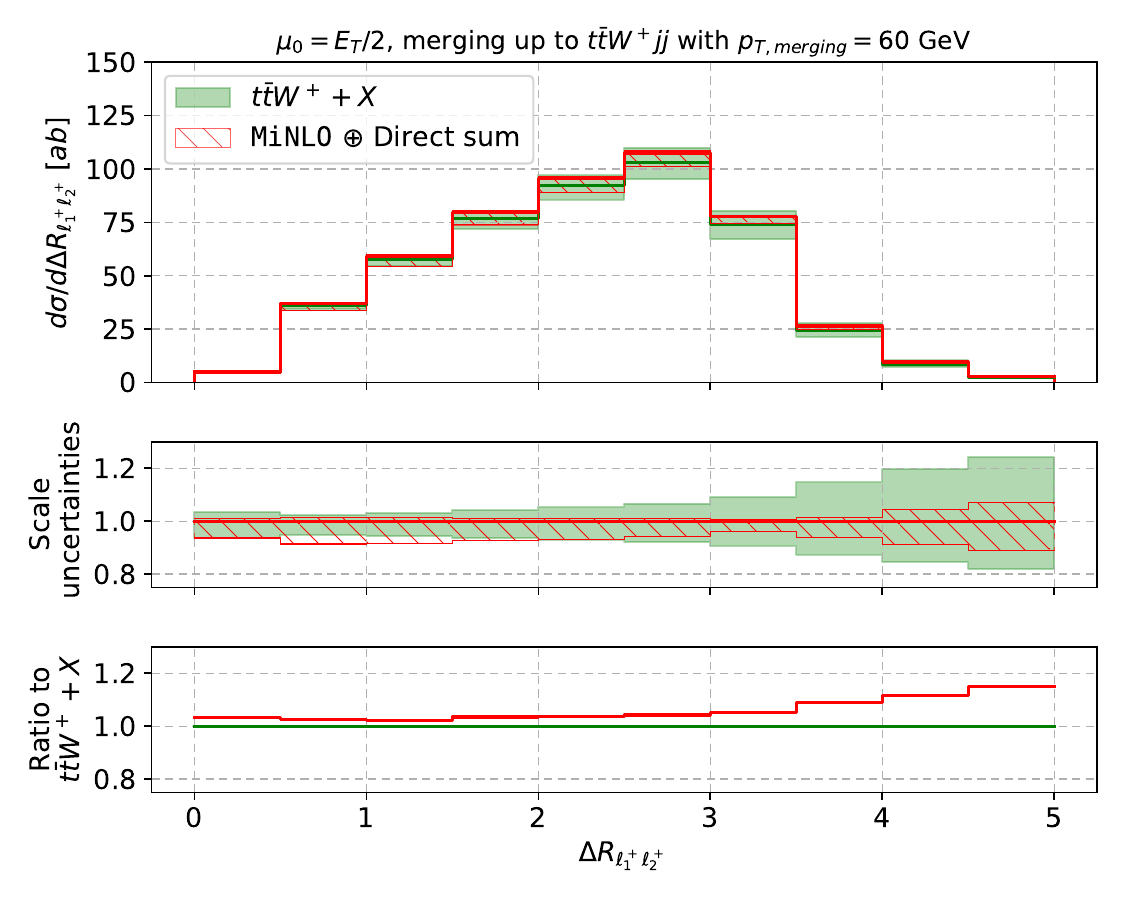}
        \caption{\textit{Same as Figure \ref{fig:diff_merged_1}, but for $p_{T,\,\ell_1^+}$, $M_{\ell_1^+ \ell_2^+}$, $\Delta \phi_{\ell_1^+ \ell_2^+}$ and $\Delta R_{\ell_1^+ \ell_2^+}$.}}
         \label{fig:diff_merged_2}
\end{figure}
\begin{figure}[t!]
        \centering        
        \includegraphics[width=0.49\linewidth]{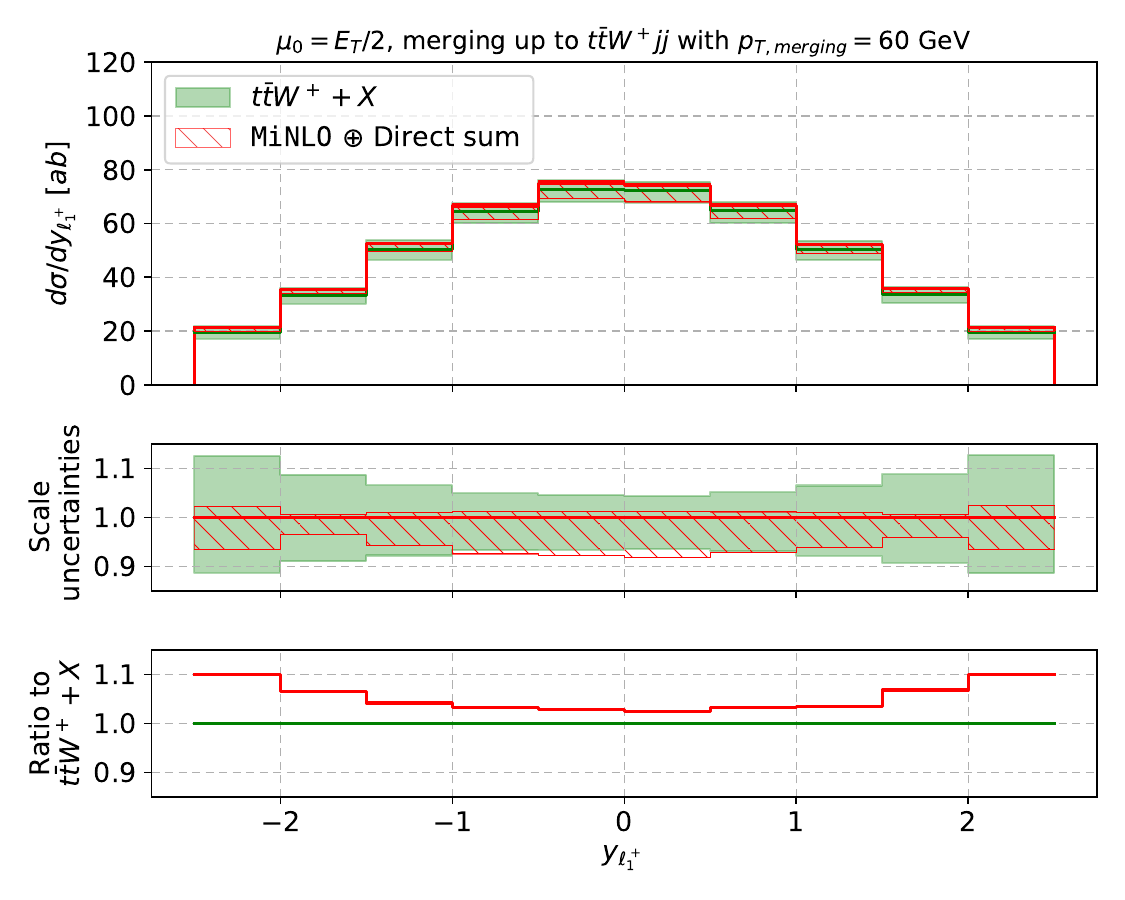}
        \includegraphics[width=0.49\linewidth]{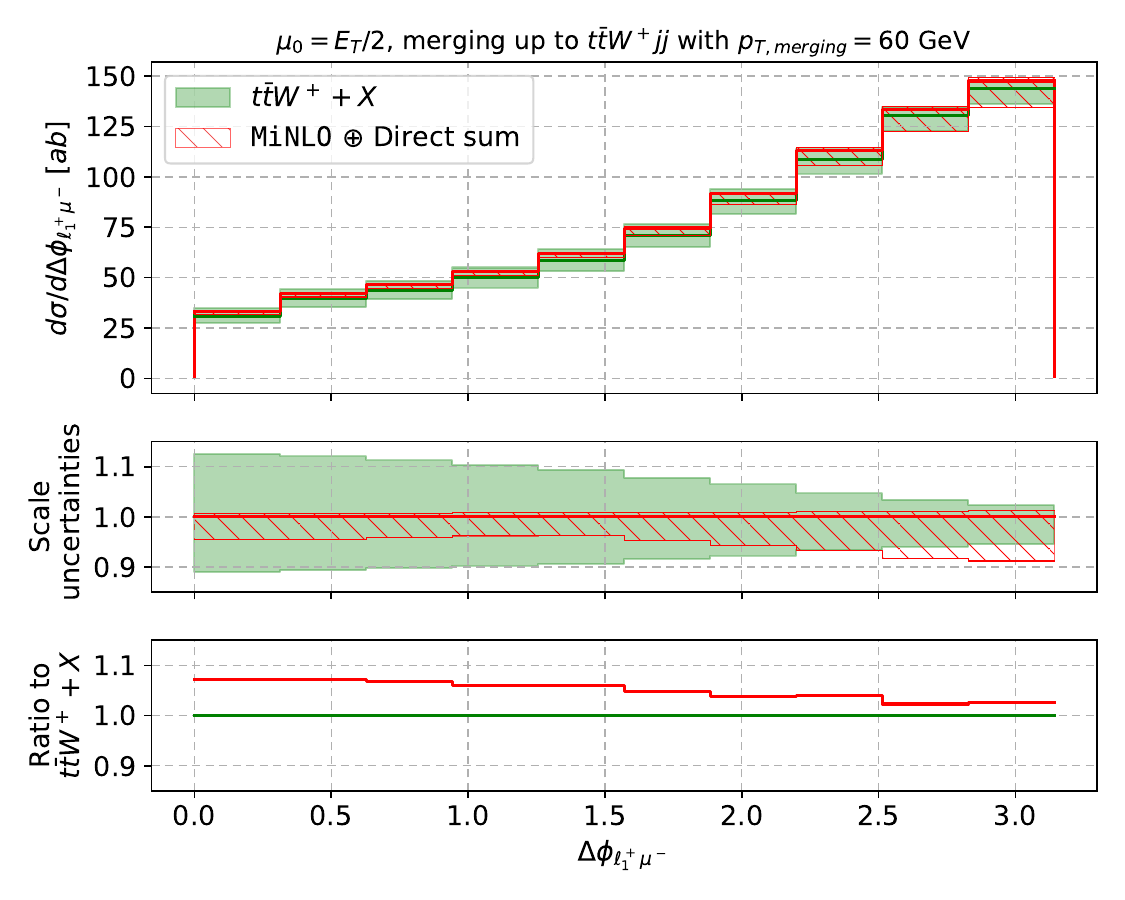}
        \includegraphics[width=0.49\linewidth]{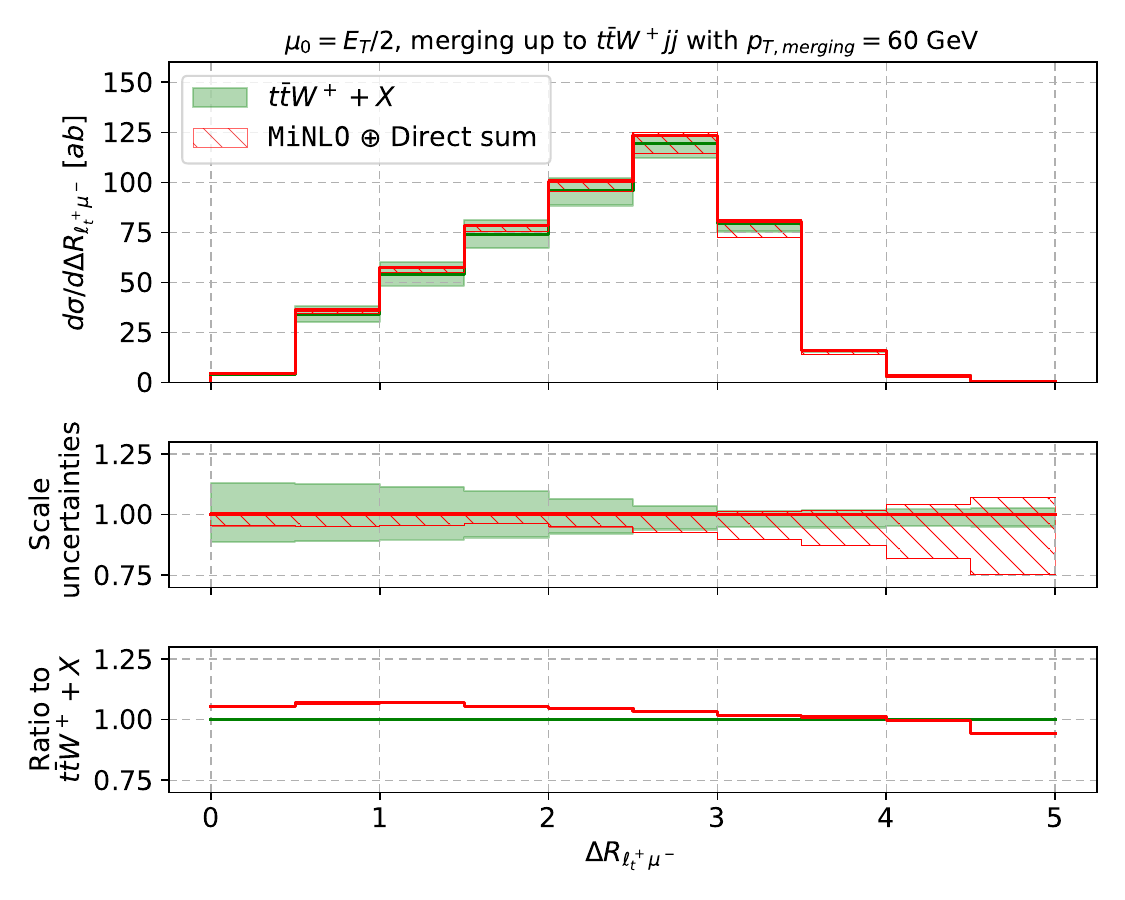}
        \includegraphics[width=0.49\linewidth]{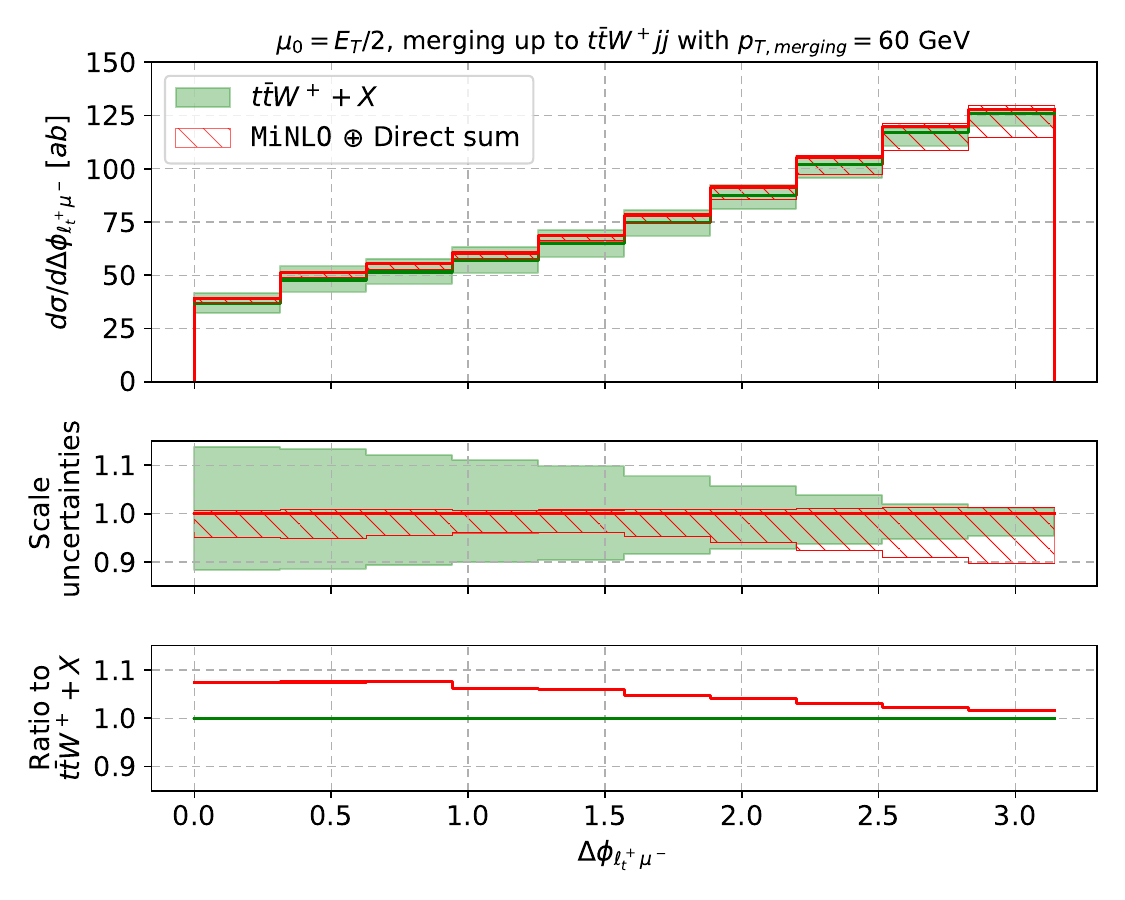}
        \caption{\textit{Same as Figure \ref{fig:diff_merged_1}, but for $y_{\ell_1^+}$, $\Delta \phi_{\ell_1^+ \mu^-}$, $\Delta R_{\ell_t^+ \mu^-}$ and $\Delta \phi_{\ell_t^+ \mu^-}$.}}
         \label{fig:diff_merged_3}
\end{figure}

Figure \ref{fig:merged_integrated} presents the integrated fiducial cross-section results after merging for various values of $p_{T,\,merging}$. These predictions are obtained using the dynamic scale choice $\mu_0 = E_T/2$ as the core scale within the \textsc{MiNLO} framework. However, it should be emphasized that the same behavior is not observed for all other scale choices. Some of them lead to significant theoretical uncertainties. This indicates that the results obtained after merging remain highly sensitive to the scale choice, even after applying the \textsc{MiNLO} procedure. This is not entirely unexpected, as the presence of the merging scale leads to large uncertainties arising from the veto condition imposed on additional radiation. As a result, the impact of Sudakov form factors is reduced, and the results obtained using the \textsc{MiNLO} method are very close to those from the standard NLO calculations. The upper panels of Figure \ref{fig:merged_integrated} show absolute predictions for the fixed-order NLO case and the merged results, labeled as $\textsc{MiNLO} \oplus \mathrm{Direct\ sum}$, incorporating up to one and up to two additional jets, in accordance with the definitions given in Eq. \eqref{merge1jet} and Eq. \eqref{merge2jet}, respectively. Furthermore, the middle panels illustrate the magnitude of scale uncertainties, while the bottom panels also show the ratio to the fixed-order  NLO QCD result.

Firstly, we can notice that the ratio to the fixed-order NLO calculation is remarkably flat across various $p_{T,\,merging}$ values when merging up to one jet is performed. Indeed, the two results differ by only $5\%-6\%$. Conversely, when the  $pp \to t\bar{t}W^+ \, jj$ process at LO is additionally included in the merging procedure, the obtained result becomes more sensitive to the choice of the merging scale. In particular, for low values of $p_{T,\, merging}$, the merged cross-section prediction can be up to $3\%$ smaller than the inclusive result for the $pp \to t\bar{t}W^+ + X$ process, whereas for higher values of $p_{T,\, merging}$ it gradually increases, eventually stabilizing at the $4\%$ level. Secondly, as shown in the middle panels, the associated scale uncertainties are consistently smaller than those estimated for the fixed-order NLO result, starting from $p_{T,\, merging} > 45 \, (60)$ GeV for merging with up to 1 (2) additional jets. For smaller values of the merging scale, the scale uncertainties increase and can exceed the fixed-order NLO ones, particularly when the second leading jet is included in the merging procedure. For example, for $p_{T,\, merging} = 25 \; \rm GeV$, the scale uncertainties are
approximately $20\%$ and $10\%$ for merging  with up to 2 and 1 additional jet(s), respectively, whereas for the 
 $pp\to t\bar{t}W^+  +X$ inclusive prediction at NLO in QCD they are of the order of  $8\%$. This behavior is expected, because the veto values close to the threshold of $p_T = 25 \; \rm GeV$ adopted during the generation of the inclusive results lead to large uncertainties in the exclusive predictions.  Finally, we stress that reducing the theoretical uncertainty in the case when 2 additional jets are added to the merging procedure requires the use of the  NLO  calculations for the full off-shell $pp \to t\bar{t}W^+\,jj$ process, rather than the LO description employed here. We plan to perform such a study in a future work.

Next, we focus on the differential cross-section results, plotting various dimensionful and dimensionless observables. All figures follow the same pattern as Figure \ref{fig:merged_integrated}, namely the top panels show the absolute predictions, the middle ones give the magnitude of scale uncertainties, and the bottom ones show the ratio to the full off-shell result for the $pp\to t\bar{t}W^+ + X$ process at NLO in QCD. We restrict our analysis to the case of the merging with up to two jets and adopt a merging scale of $p_{T,\, merging} = 60$ GeV. As shown in Figure \ref{fig:merged_integrated}, lower values of the merging scale lead to larger theoretical uncertainties, which we wish to avoid. On the other hand, choosing a larger merging scale would increase the dependence on lower-multiplicity contributions when modeling jet dynamics. Therefore, such values should also be avoided to ensure more reliable modeling of the $pp\to \WWW +X$ process. We investigated the effect of varying the $p_{T,\, merging}$ value. In particular, we verified that adopting a value of $p_{T,\, merging} = 40 \, (50)$ GeV leads to only minor changes, and any observed differences are within the estimated theoretical uncertainties. Furthermore, we confirmed that the differential predictions obtained by merging predictions up to one additional jet only are very close to those presented below, and that any observed discrepancies are well within the theoretical uncertainties.

In Figure \ref{fig:diff_merged_1}, we present the transverse momentum of the hardest $b$-jet ($p_{T,\,b_1}$), the transverse momentum of the system of the two hardest $b$-jets ($p_{T,\,b_1b_2}$), the sum of the transverse momenta of the two hardest $b$-jets ($H_{T,\, b}$), and the missing transverse momentum from invisible neutrinos ($p_{T,\,miss}$). A common feature of all these distributions is the reduced scale uncertainties associated with the merged predictions, observed across the entire kinematic range presented. This reduction is particularly pronounced for the observables $p_{T,\,b_1}$, $p_{T,\,b_1b_2}$, and $p_{T,\,miss}$ in the high-$p_T$ tails, where the NLO scale uncertainties for the $pp \to t\bar{t}W^+ + X$ process reach approximately $12\%$ for $p_{T,\,b_1}$ and up to $30\%$ for $p_{T,\,b_1b_2}$ and $p_{T,\,miss}$. By comparison, the corresponding uncertainties for the merged predictions are at most $7\%$ for $p_{T,\,b_1}$ and approximately $10\%-12\%$ for $p_{T,\,b_1b_2}$ and $p_{T,\,miss}$. When comparing central predictions, the differences between the $pp\to \WWW+X$ process and the fully inclusive merged sample are around $3\%-5\%$ in the low-$p_T$ regions, that is, in the phase-space regions that will become accessible at the High-Luminosity Large Hadron Collider (HL-LHC) experiment, where an integrated luminosity of the order of $(3-4)$ ab${}^{-1}$ per experiment is expected. These differences increase at high transverse momenta, reaching about $10\%$ for $p_{T,\,b_1}$ and $H_{T,\,b}$, whereas they are of the order of  $20\%-25\%$ for $p_{T,\,b_1b_2}$ and $p_{T,\,miss}$. 

In Figure \ref{fig:diff_merged_2}, we focus on the kinematics of the positively charged leptons in the $pp\to t\bar{t}W^+ +X$ process. In particular, we present the transverse momentum of the hardest positively charged lepton $(p_{T,\,\ell_1^+})$, the invariant mass of the system of the two charged leptons $(M_{\ell_1^+ \ell_2^+})$, as well as their angular separation and relative azimuthal angle, denoted as $\Delta R_{\ell_1^+ \ell_2^+}$ and $\Delta \phi_{\ell_1^+ \ell_2^+}$, respectively. For the $p_{T,\,\ell_1^+}$ distribution, theoretical uncertainties associated with the merged result may exceed those estimated for the $pp \to t\bar{t}W^+ + X$ process in the high-$p_T$ tails, while remaining smaller in the phase-space regions where the majority of events are concentrated. For the $M_{\ell_1^+ \ell_2^+}$ observable, on the other hand, the scale uncertainties for the merged result remain remarkably constant across the entire kinematic range shown. They are at around $7\%$ even at high invariant masses, where the standard NLO QCD uncertainties reach $20\%$. For the dimensionless observables, scale uncertainties for the merged predictions are generally smaller. Finally, it is worth noting that for all the distributions shown in Figure \ref{fig:diff_merged_2}, the ratios of the merged predictions to the result for the  $pp \to t\bar{t}W^+ + X$ process fall within the range of $(1.02-1.15)$. However, these differences always lie within the substantial theoretical uncertainties associated with the latter prediction.

Figure \ref{fig:diff_merged_3} shows the rapidity of the hardest positively charged lepton $(y_{\ell_1^+})$, the azimuthal angle difference between this lepton and the muon $(\Delta \phi_{\ell_1^+ \mu^-})$, as well as both the angular separation 
and the azimuthal angle difference between the muon and the positively charged lepton most likely originating from the top-quark decay $(\ell_t^+)$, denoted as $\Delta R_{\ell_t^+ \mu^-}$ and $\Delta \phi_{\ell_t^+ \mu^-}$, respectively.  The lepton $\ell_t^+$ is identified as the positively charged lepton for which the invariant mass of the system comprising that lepton, its associated $b$-jet, the corresponding neutrino, and any additional jets, is closest to the top-quark mass. On the one hand, for the $y_{\ell_1^+}$ and $\Delta R_{\ell_t^+ \mu^-}$ distributions, the scale uncertainties for the merged predictions are only slightly larger, in the central rapidity regions and for the back-to-back configurations. On the other hand, for $\Delta \phi_{\ell_1^+ \mu^-}$ and $\Delta \phi_{\ell_t^+ \mu^-}$, the uncertainty bands for the merged predictions become significantly larger in the vicinity of $\Delta \phi \simeq \pi$, reaching values of about $10\%$, compared to $5\%$ for the inclusive $pp \to t\bar{t}W^+ + X$ calculation.  In the remaining phase-space regions, the theoretical uncertainties of the merged predictions are substantially reduced. The only significant exception occurs for $\Delta R_{\ell_t^+ \mu^-} > 3$, in the phase-space regions where few events are generated and large statistical uncertainties arise. Finally, it is worth noting that for all the observables shown in Figure \ref{fig:diff_merged_3}, the normalization differences between the two approaches remain small and do not exceed $10\%$.

%
\section{Summary}
\label{sec:summary}
%

In this paper, we have compared the \textsc{MiNLO} method with standard fixed-order NLO QCD calculations for the full off-shell $pp \to \WWW \,(jj)$ process at the LHC Run II centre-of-mass energy of $\sqrt{s} = 13 \, \rm TeV$. Unlike conventional NLO calculations, where the renormalization and factorization scales are fixed by predefined scale choices, the \textsc{MiNLO} method assigns these scales dynamically based on the most probable branching history of the additional jet radiation. The calculation is further augmented by Sudakov form factors, which encode no-emission probabilities and improve the accuracy of the prediction by resumming large logarithms associated with widely separated scales. However, it should be noted that, even within the \textsc{MiNLO} method, the definition of the core scale is not uniquely prescribed, since only the scale settings associated with the additional jets are determined by the \textsc{MiNLO} algorithm.  Although this paper focuses on the full off-shell  $pp \to \WWW +X$ process, we have presented the implementation of the method within the \textsc{Helac-NLO} framework in a fully generic form, making it applicable to arbitrary processes. 

For the $pp \to \WWW \, j (j)$  process, we have found that the fixed-order NLO QCD and \textsc{MiNLO} predictions agree within their theoretical uncertainties for all scale choices considered. At the same time, the \textsc{MiNLO} predictions have shown a reduced sensitivity to scale variations with increasing jet multiplicity relative to the core process. The effect was most pronounced for the full off-shell $pp \to \WWW\,jj$ process at LO, for which the scale uncertainties were significantly smaller than in the standard LO calculation. For the $pp \to \WWW \, j+X$ process, a significant reduction in scale uncertainties was observed primarily for suboptimal scale choices and when larger transverse-momentum thresholds were imposed on the additional jet.  We have also investigated additional sources of uncertainties within the \textsc{MiNLO} framework, including alternative scale-variation prescriptions and different choices of the resolution parameter $R$ in the inverse $k_T$-clustering algorithm. In both cases, the impact was negligible, with deviations of at most $1\%$ relative to the default setup. At the level of differential cross sections, \textsc{MiNLO} exhibited improved behavior in the tails of selected dimensionful observables associated with light jets when a fixed scale was used, resulting in substantially smaller theoretical uncertainties than the standard fixed-order NLO predictions. When appropriate dynamical scale choices were used in the fixed-order calculation, however, the \textsc{MiNLO} and NLO predictions showed very similar behavior.

In addition, we have presented merged predictions designed to improve the modeling of the full off-shell $pp \to \WWW+X$ process at NLO. This was achieved by introducing a suitable merging scale, $p_{T,\,merging}$, to consistently combine results with different jet multiplicities. The merging was performed up to one and two jets, yielding predictions that were consistent within the estimated theoretical uncertainties. Indeed, the differences in the central predictions did not exceed $8\%$. However, the size of the scale uncertainties in the merged cross-section predictions depended noticeably on the choice of $p_{T,\,merging}$.  Compared to the fixed-order NLO prediction for $pp \to \WWW+X$, reduced scale uncertainties were obtained for $p_{T,\,merging} \ge 35$ GeV when merging up to one jet, and for $p_{T,\,merging} \ge 60$ GeV when merging up to two jets.  With suboptimal scale choices in the merging procedure, the scale uncertainties increased and could exceed those of the fixed-order prediction for all considered values of $p_{T,\,merging}$.  This indicates that the use of \textsc{MiNLO} in the merging procedure does not, by itself, reduce the scale sensitivity of the predictions. However, with an appropriate choice of both $q_{core}$ and $p_{T,\,merging}$, improved predictions were obtained at the differential cross-section level for most of the observables studied. Although the differences between the merged and the fixed-order predictions remained within the corresponding uncertainty bands, the reduced uncertainties obtained after merging indicate a substantial improvement in the modeling of the $pp \to \WWW \, +X$ process.

It is important to stress that the inclusion of the $pp \to \WWW\,jj+X$ sample at NLO in the merging procedure would be essential, as a LO description of the second-leading jet is not sufficient for an accurate modeling of the underlying dynamics of this process. In the present study, the $pp\to \WWW\, jj$ contribution is kept at LO, reflecting the perturbative accuracy with which this final state would enter an NNLO prediction, were such a prediction available in the literature. In this context, a full off-shell NLO QCD calculation for the $pp \to \WWW\, jj+X$ process would be highly valuable for further improving the theoretical description of the underlying process.  Moreover, it would be interesting to investigate the impact of applying the \textsc{MiNLO} method to  $pp\to \WWW \, jj +X$  at NLO in QCD. In line with our previous conclusions, the inclusion of additional jets at higher perturbative accuracy further strengthens the motivation for employing the \textsc{MiNLO} method. We plan to address this aspect in a future study.

\acknowledgments{
We thank Johannes Welter for cross-checking parts of the results presented in this work. 

This work was supported by the Deutsche Forschungsgemeinschaft (DFG) under the following grants:  TRR 257 -  {\it P3H - Particle Physics Phenomenology after the Higgs Discovery}, GRK 2497 -  {\it The Physics of the Heaviest Particles at the LHC.}

Support by a grant of the Bundesministerium f\"ur Forschung, Technologie und Raumfahrt (BMFTR) is additionally acknowledged.

The authors gratefully acknowledge the computing time provided to them at the NHR Center NHR4CES at RWTH Aachen University (project number \texttt{p0020216}). This is funded by the Federal Ministry of Education and Research, and the state governments participating on the basis of the resolutions of the GWK for national high performance computing at universities.}

\appendix

%
\section{Differential cross-section predictions for 
the $\boldsymbol{pp \to t\bar{t}W^+ jj}$ process}
\label{appendix:a}
%

%
\begin{figure}[h!]
        \centering        
        \includegraphics[width=0.49\linewidth]{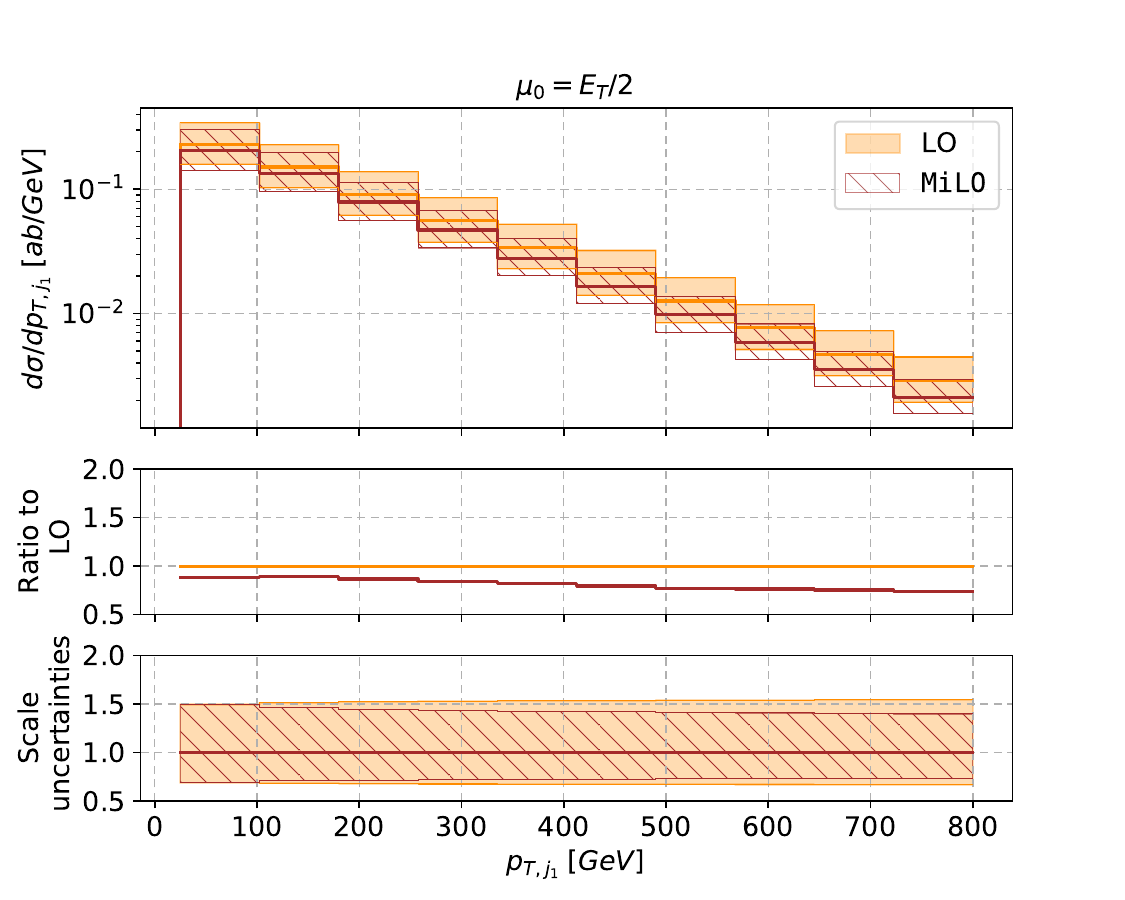}
        \includegraphics[width=0.49\linewidth]{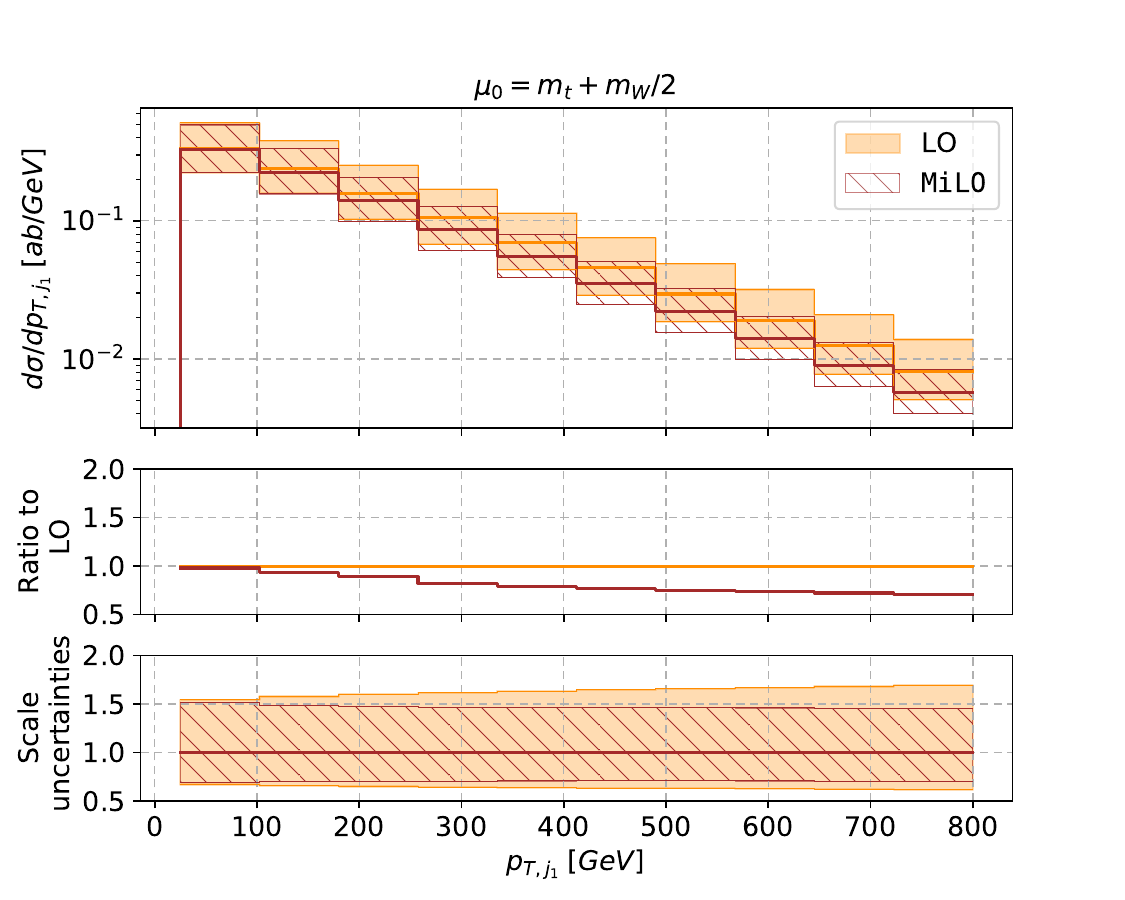}
        \includegraphics[width=0.49\linewidth]{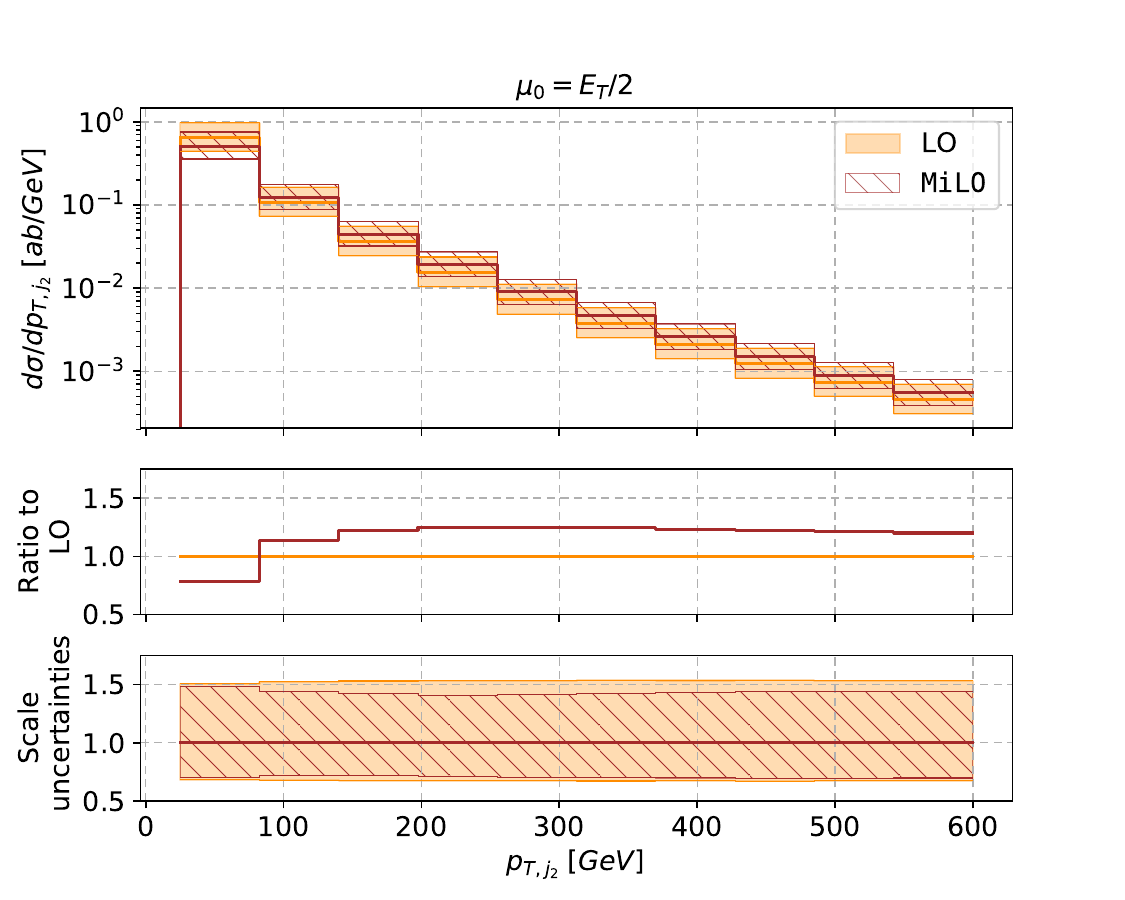}
        \includegraphics[width=0.49\linewidth]{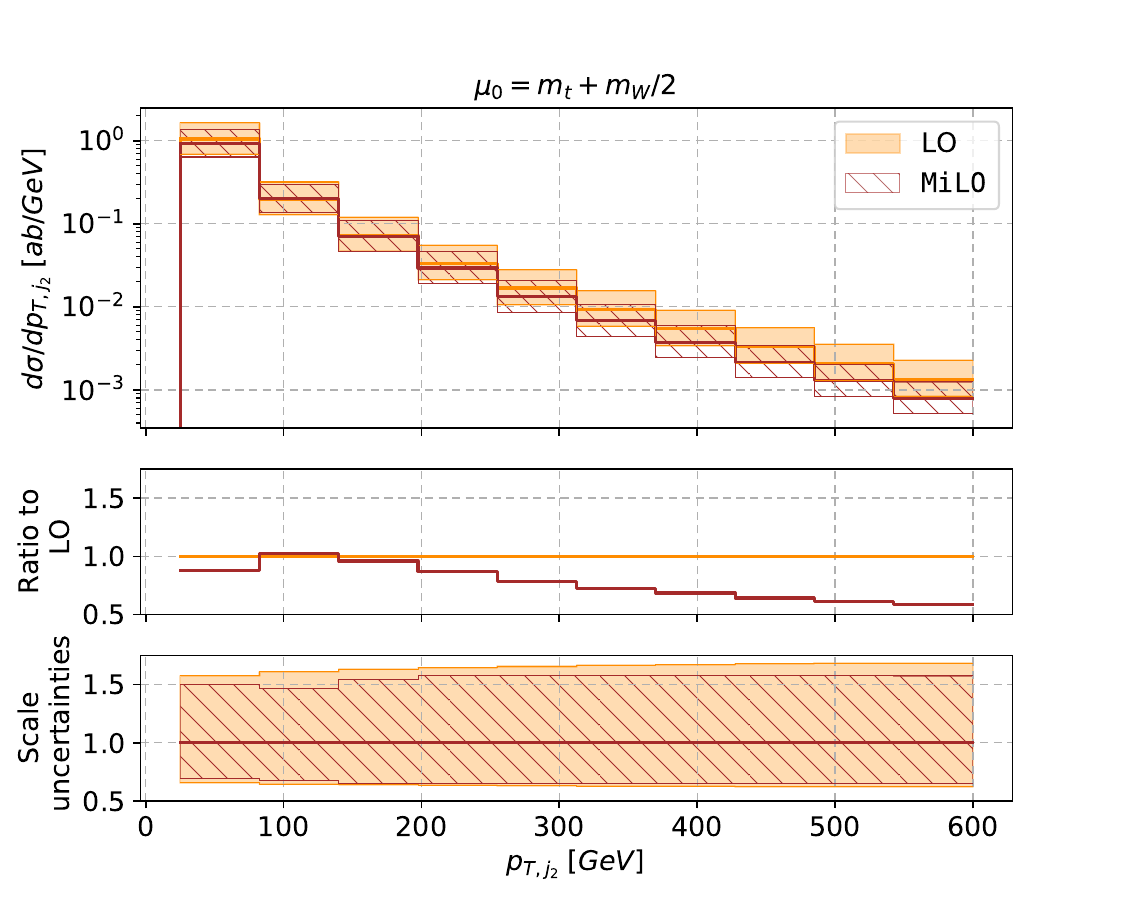}
        \caption{\textit{Differential cross-section distributions for the full off-shell $pp \rightarrow \WWW\,j j $ process at the LHC with $\sqrt{s} = 13 \; \rm TeV$, for the $p_{T, \,j_1}$ and $p_{T, \, j_2}$ observables. Results are shown for the dynamical scale choice $\mu_0 = E_T/2$  and the fixed scale setting $\mu_0 = m_t + m_W/2$, using the standard LO  and the \textsc{MiLO} method. The upper panels show absolute predictions, the middle panels display the ratio to LO results, while the bottom panels illustrate the magnitude of scale uncertainties normalized to the corresponding results.}}
         \label{fig:milovslo_ptj}
\end{figure}
\begin{figure}[h!]
        \centering        
        \includegraphics[width=0.49\linewidth]{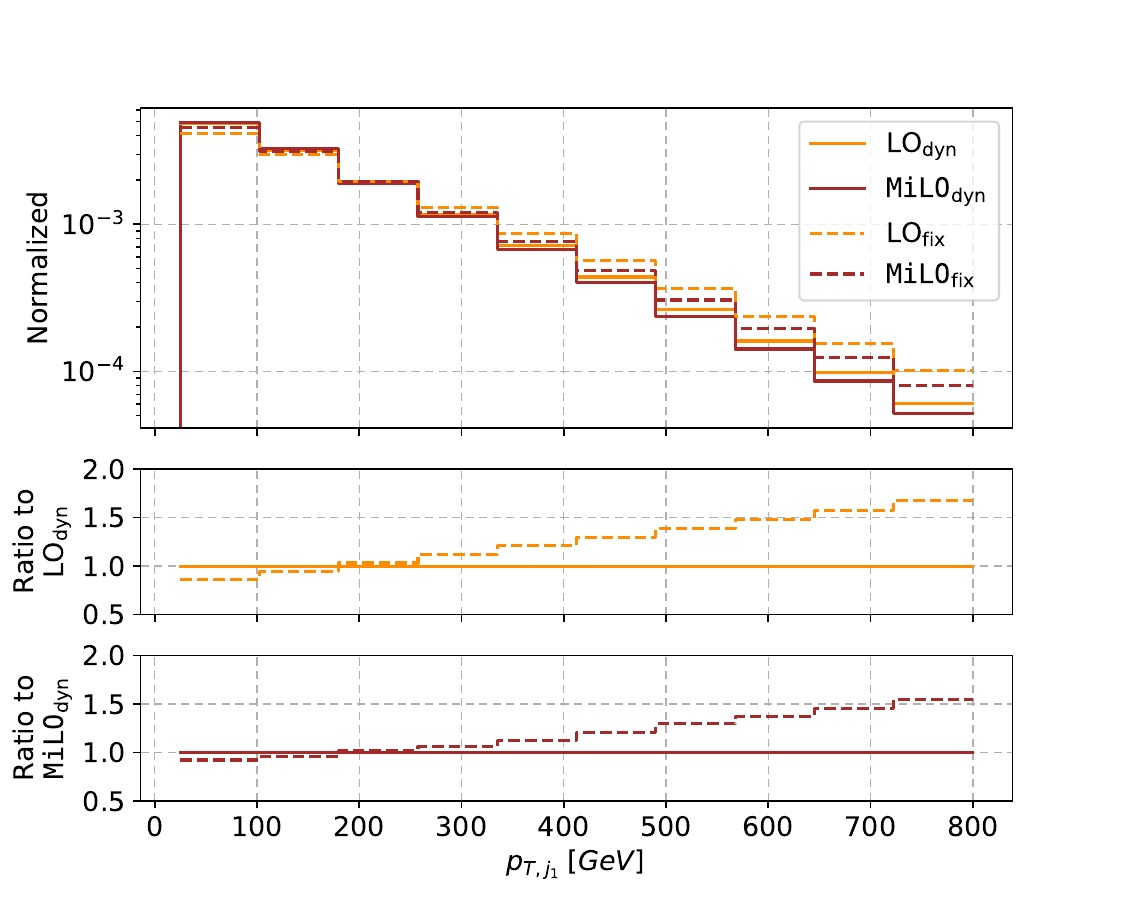}
        \includegraphics[width=0.49\linewidth]{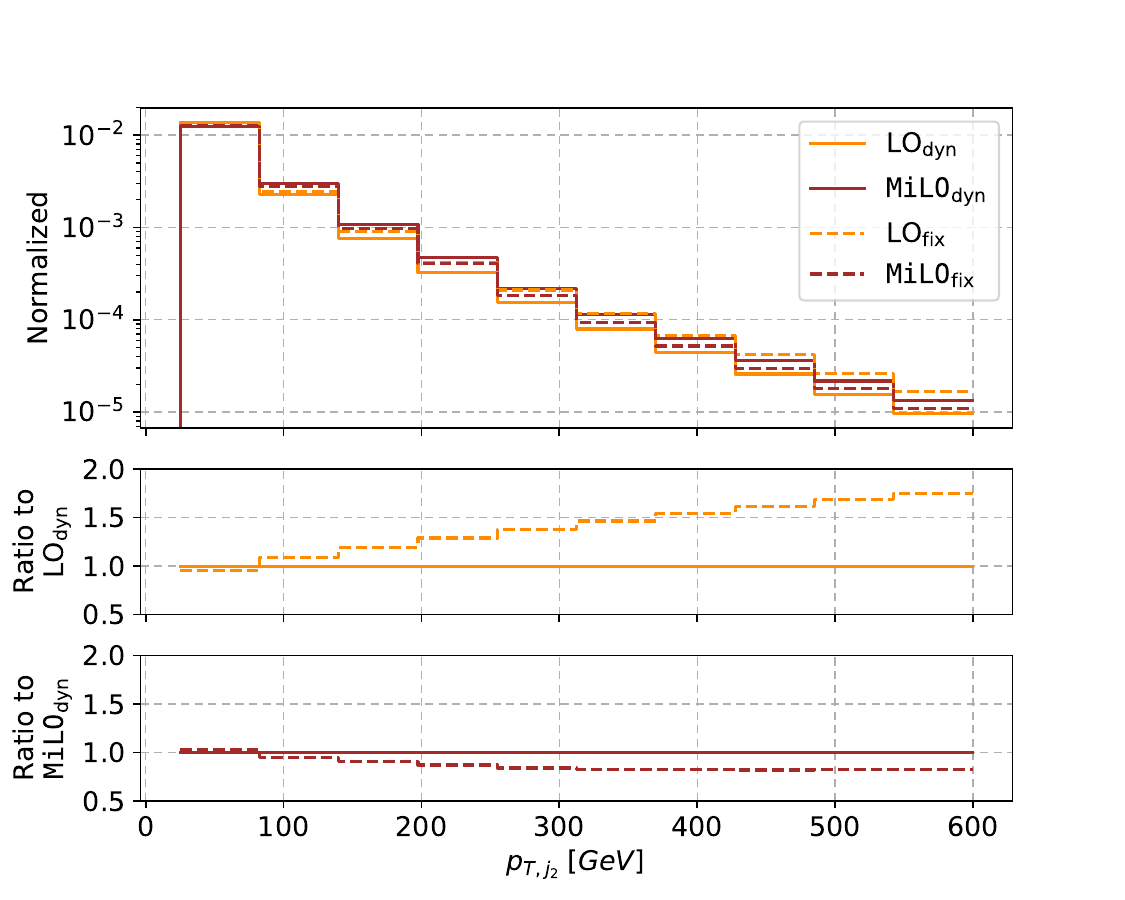}
        \caption{\textit{Normalized differential cross-section distributions for the full off-shell $pp \rightarrow \WWW\,j j $ process at the LHC with $\sqrt{s} = 13$ TeV, for the $p_{T,\,j_1}$ and $p_{T,\,j_2}$ observables. Results are shown for the dynamical scale choice $\mu_0 = E_T/2$ (dyn) and the fixed scale setting $\mu_0 = m_t + m_W/2$ (fix), using the standard LO and \textsc{MiLO} approaches. The bottom panels display the ratios to the results obtained with $\mu_0 = E_T/2$.}}
         \label{fig:milovslo_ptj1_ptj2_norm}
\end{figure}
\begin{figure}[h!]
        \centering        
        \includegraphics[width=0.49\linewidth]{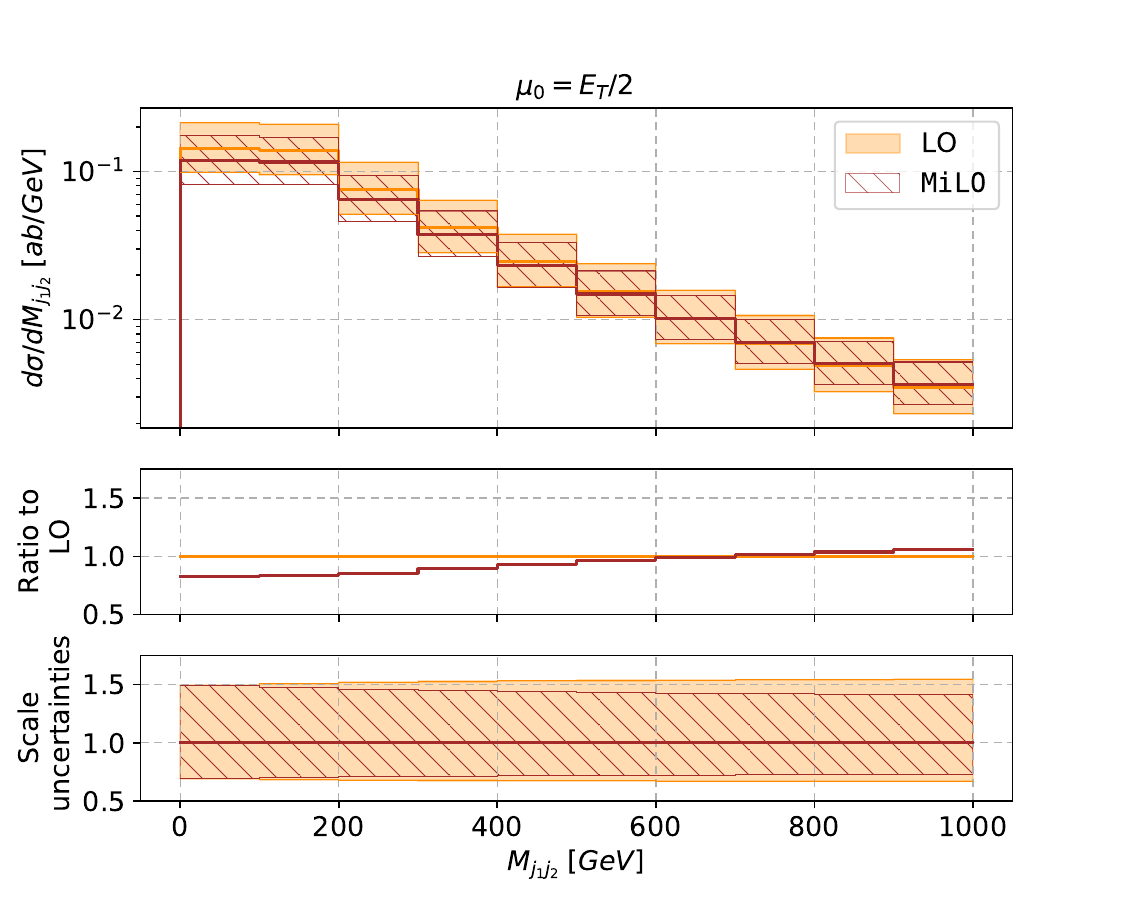}
        \includegraphics[width=0.49\linewidth]{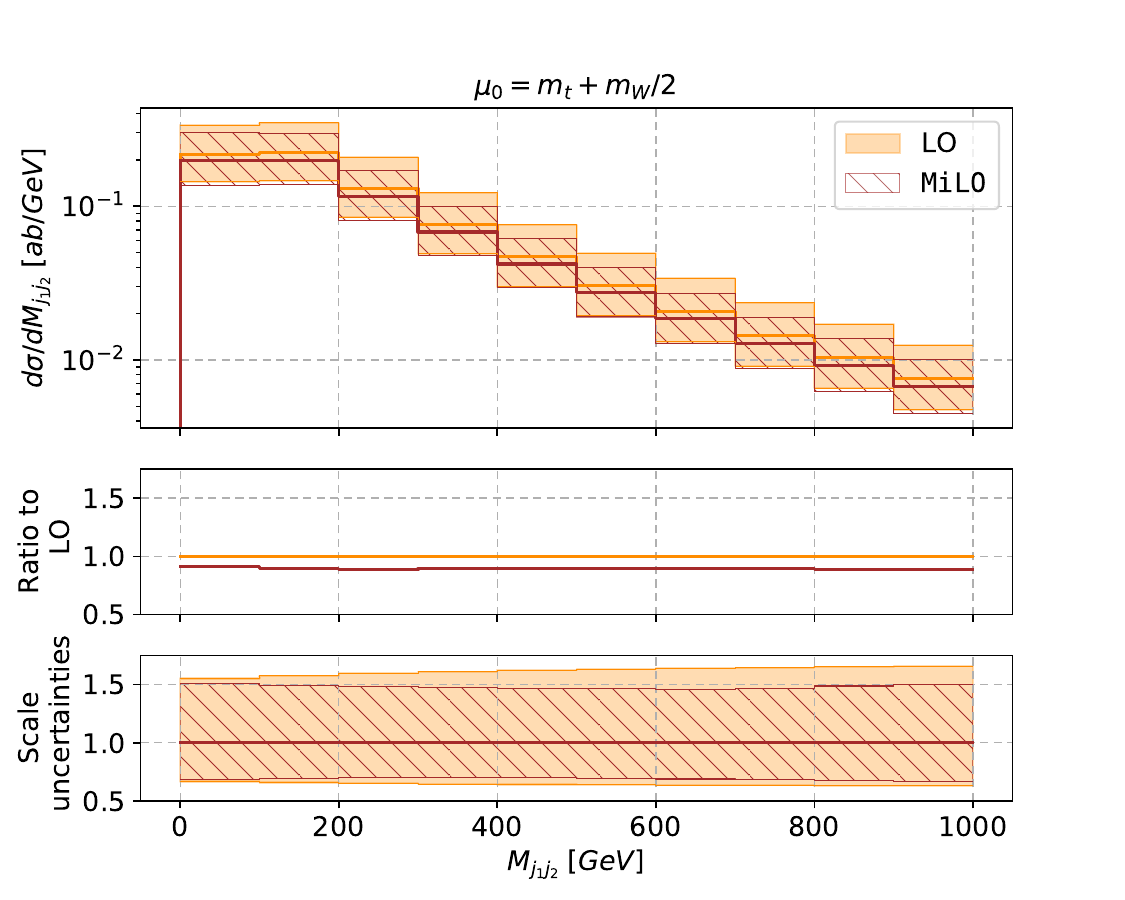}
        \includegraphics[width=0.49\linewidth]{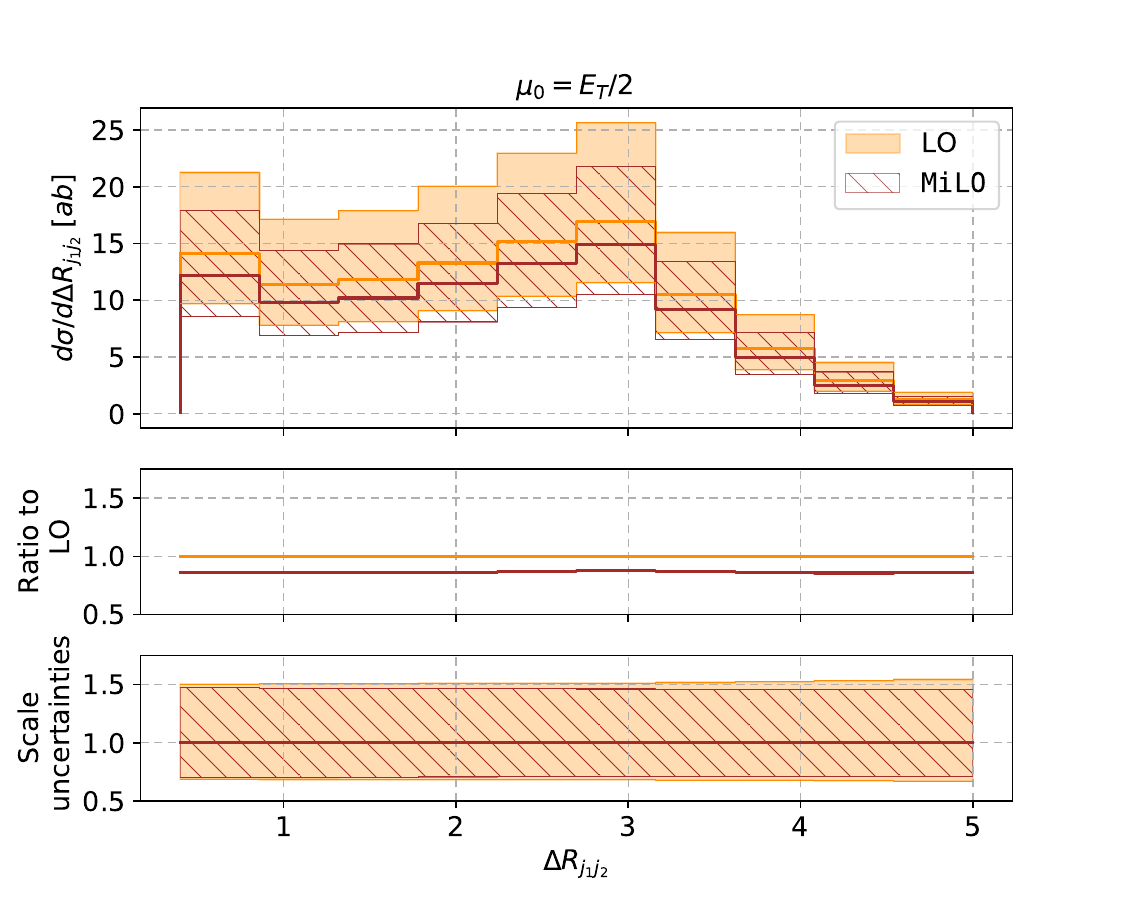}
        \includegraphics[width=0.49\linewidth]{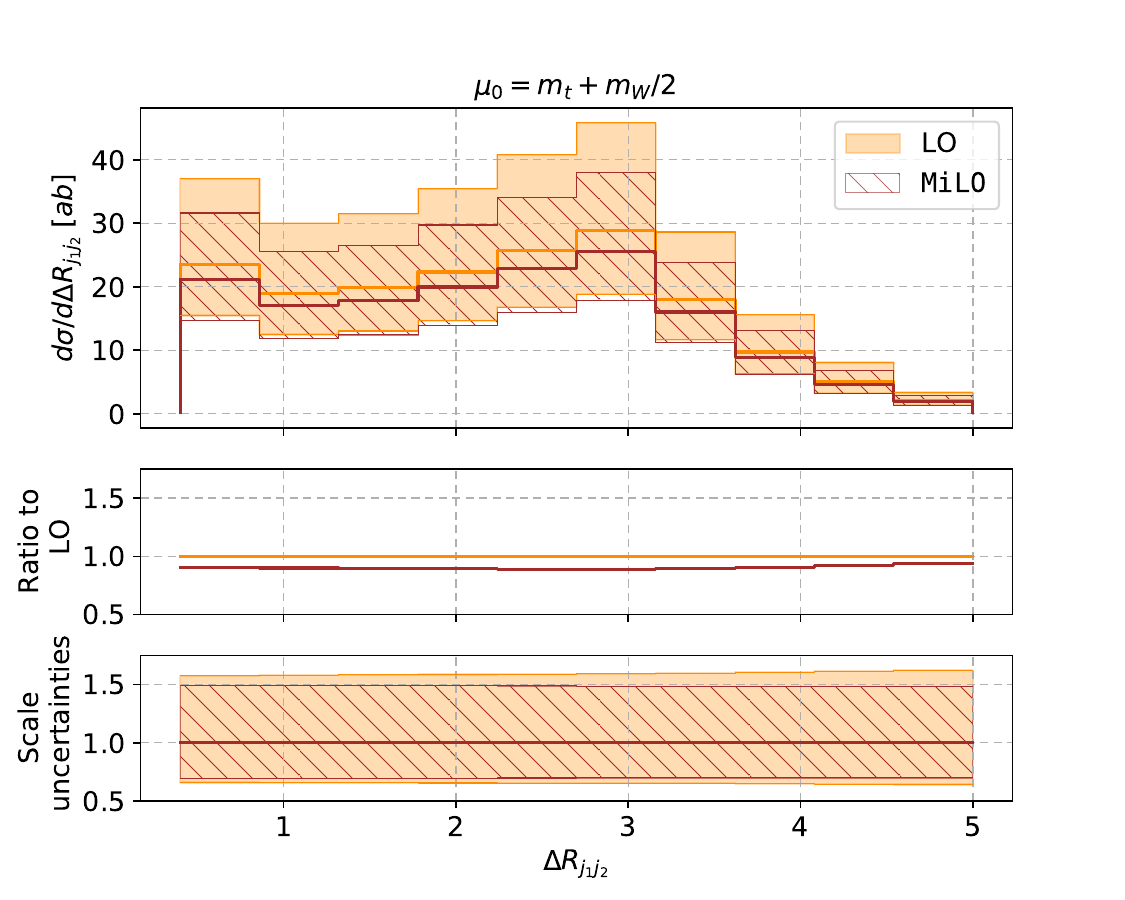}
        \caption{\textit{Same as Figure \ref{fig:milovslo_ptj}, but for  $M_{j_1j_2}$  and $\Delta R_{j_1j_2}$.}}
         \label{fig:milovslo_mjjdrjj}
\end{figure}
In this Appendix, we present differential predictions for the full off-shell $pp \to \WWW \,jj$ process at LO,  comparing the \textsc{MiLO} method with the standard LO calculation. In particular, we show observables related exclusively to the light-jet kinematics.

In Figure \ref{fig:milovslo_ptj} we display the transverse momentum of the first and second hardest jet, $p_{T,\,j_1}$ and $p_{T,\,j_2}$, respectively.  In the tail of the $p_{T,\,j_1}$ distribution, the \textsc{MiLO} predictions differ from the LO results by approximately $25\%-30\%$. For both scale choices, the uncertainties associated with the \textsc{MiLO} results are smaller than those obtained for the corresponding LO predictions. The differences are especially pronounced for the fixed scale $\mu_0 = m_t + m_W/2$, for which the LO scale uncertainties reach $70\%$ in the distribution tails. By contrast, the corresponding \textsc{MiLO} uncertainties are significantly smaller, amounting to $46\%$. In the case of $p_{T,\,j_2}$, the ratio with respect to the LO prediction is particularly sensitive to the scale choice. For the dynamic scale setting, the \textsc{MiLO} prediction is approximately $22\%$ below the LO result at the beginning of the distribution. The ratio then increases gradually and eventually reaches an approximately constant plateau $1.20 - 1.25$. By contrast, for the fixed-scale choice, the \textsc{MiLO} prediction remains consistently below the LO result throughout the entire plotted range, with differences reaching up to $50\%$ in the high-$p_T$ tail.

Nevertheless, the differences between the fixed and dynamical scale choices are more pronounced for the standard LO predictions. This can be seen in Figure \ref{fig:milovslo_ptj1_ptj2_norm}, which shows the same distribution for both LO and \textsc{MiLO}, using both scale choices, with each prediction normalized to its corresponding integrated cross-section prediction. The bottom panels show the ratio of the fixed-scale result to the dynamical-scale result. In the case of $p_{T,\,j_2}$, the corresponding ratio obtained with the \textsc{MiLO} method deviates only moderately from unity, reaching values of approximately $0.8$. By contrast, for the standard LO predictions, the ratio increases up to approximately $1.75$. This demonstrates that the sizable shape distortions induced by the scale choice in the tail of the $p_{T,\,j_2}$ distribution are substantially reduced by applying the \textsc{MiLO} prescription. For $p_{T,\,j_1}$,  the same pattern persists, however, the differences are less pronounced, as expected from the limited additional jet activity.

Finally, in Figure \ref{fig:milovslo_mjjdrjj}, we show the invariant mass, $M_{j_1j_2}$, and the angular separation, $\Delta R_{j_1j_2}$, of the two light jets.  The shape differences observed for $M_{j_1j_2}$ depend strongly on the scale choice. For $\mu_0 = E_T/2$, they vary between $0.8 -1.1$, whereas for $\mu_0 = m_t + m_W/2$, they  remain approximately constant at the level of $0.9$.  For the $\Delta R_{j_1j_2}$ distribution, as well as for all dimensionless observables considered, the \textsc{MiLO}/{\rm LO}  ratios are essentially flat. This indicates that the differences between the two predictions are driven mainly by the overall normalization rather than by distortions in the shapes of the distributions. Moreover, for both scale choices, the \textsc{MiLO} scale uncertainties are systematically smaller than those obtained in the standard LO calculation, particularly in the tails of the dimensionful observables examined.


\bibliographystyle{JHEP}

\providecommand{\href}[2]{#2}\begingroup\raggedright\endgroup

\end{document}